\documentclass[journal]{rmaa}
\usepackage{amsmath}

\usepackage[caption=false]{subfig}

\usepackage{paralist}

\usepackage{psfrag,color}

\usepackage[latin1]{inputenc}

\usepackage{graphicx}
\usepackage{lscape}
\usepackage{txfonts}
\usepackage{natbib}
\usepackage{url}
\usepackage[T1]{fontenc}

\usepackage{url}
\usepackage{xcolor}

\DeclareRobustCommand{\ion}[2]{%
\relax\ifmmode
\ifx\testbx\f@series
{\mathbf{#1\,\sc{#2}}}\else
{\mathrm{#1\,\sc{#2}}}\fi
\else\textup{#1\,{\mdseries\textsc{#2}}}%
\fi}

\DeclareRobustCommand{\ION}[2]{%
\relax\ifmmode
\ifx\testbx\f@series
{\mathbf{#1\,\mathsc{#2}}}\else
{\mathrm{#1\,\mathsc{#2}}}\fi
\else\textup{#1\,{\mdseries\textsc{#2}}}%
\fi}
\newcommand{\lam}{$\lambda$}

\newcommand{\hii}{\ION{H}{ii}}

\newcommand{\nii}{[\ION{N}{ii}]}
\newcommand{\oi}{[\ION{O}{i}]}
\newcommand{\oii}{[\ION{O}{ii}]}
\newcommand{\oiii}{[\ION{O}{iii}]}
\newcommand{\sii}{[\ION{S}{ii}]}

\newcommand{\Ha}{$\rm{H}\alpha$}
\newcommand{\Hb}{$\rm{H}\beta$}

\newcommand{\update}[1]{{ #1}}

\title{From global to spatially resolved in low-redshift galaxies}

\author{S. F. S\'anchez$^1$, C. J. Walcher$^2$, C. Lopez-Cob\'a$^1$, J. K. Barrera-Ballesteros$^1$, A. Mej\'ia-Narv\'aez$^1$, C. Espinosa-Ponce$^1$, Camps-Fari\~{n}a$^{1}$}
\altaffiltext{1}{Instituto de Astronom\'\i a, Universidad Nacional Auton\'oma de M\'exico, A.P. 70-264, 04510 M\'exico, D.F.,  Mexico.}
\altaffiltext{2}{Leibniz-Institut f\"ur Astrophysik Potsdam (AIP), An der Sternwarte 16, D-14482 Potsdam, Germany}
\shortauthor{S\'anchez et al.}
\shorttitle{resolving low-redshift galaxies}

\listofauthors{S. F. S\'anchez}
\indexauthor{S. F. S\'anchez}

\abstract{Our understanding of the structure, composition and evolution of galaxies has strongly improved in the last decades, mostly due to new results based on large spectroscopic and imaging surveys. In particular, the nature of ionized gas, its ionization mechanisms, its relation with the stellar properties and chemical composition, the existence of scaling relations that describe the cycle between stars and gas, and the corresponding evolution patterns have been widely explored and described. More recently, the introduction of additional techniques, in particular Integral Field Spectroscopy, and their use in large galaxy surveys, have forced us to re-interpret most of those recent results from a spatially resolved perspective. This review is aimed to complement recent efforts to compile and summarize this change of paradigm in the interpretation of galaxy evolution. In particular we cover three particular aspects not fully covered in detail in recent reviews: (i) the spatially resolved nature of the ionization properties in galaxies and the confusion introduced by considering just integrated quantities; (ii) the nature of the global scaling relations and their relations with the spatially resolved ones; and (iii) the dependence of the radial gradients and characteristic properties of the stellar populations and ionized gas on stellar mass and galaxy morphology. To this end we replicate published results, and present novel ones, based on the largest compilation of IFS data of galaxies in the nearby universe to date.}  

\resumen{Nuestro entendimiento de la estructura, composici\'on y evoluci\'on de las galaxias se ha visto ampliamente modificado en las \'ultimas d\'ecadas principalmente debido a la explosi\'on de resultados basados en grandes muestreos en imagen y espectroscop\'\i a. En particular, la naturaleza de la ionziaci\'on observada, su relaci\'on con las propiedades estelares y la composici\'on qu\'\i mica, la existencia de relaciones evolutivas o de escala que regulan el ciclo entre el gas y las estrellas, y los patrones evolutivos se han explorado y descritos en detalle. M\'as recientemente, la introducci\'on de t\'ecnicas adicionales, en particular la espectroscop\'\i a de campo integral, ha forzado a una reinterpretaci\'\i on de estos resultados recientes desde una perspetiva espacialmente resuelta. Este art\'\i culo de revisi\'on tienen como objetivo complementar esfuerzos recientes enfocados en la compilaci\'on y resumen de este cambio de paradigma en la interpretaci\'on de la evoluci\'on de las galaxias. En particular aborda tres aspectos no cubiertos completamente en detalle por revisiones anteriores: (i) la naturaleza espacialmente resuelta de la ionizaci\'on en las galaxias y la confusi\'on introducida al considerara solo cantidades integradas; (ii) la naturaleza de las relaciones de escala globales y su relaci\'on con las espacialmente resueltas; y (iii) la dependencia de los gradientes radiales y los valores caracter\'\i sticos de las propiedades de las poblaciones estelares y del gas ionizado en las galaxias con la masa y la morfolog\'\i a de las galaxias. Con este proposito hemos replicado resultados publicados y presentado igualmente resultados nuevos bas\'andonos en la mayor compilaci\'on de datos IFS de galaxias en el universo local hasta la fecha. }

\addkeyword{galaxies: evolution}
\addkeyword{galaxies: star-formation}
\addkeyword{galaxies: resolved properties}
\addkeyword{galaxies: fundamental parameters}
\addkeyword{techniques: imaging spectroscopy}

\begin{document}






\maketitle



\section{INTRODUCTION}
\label{sec:intro}

The evolution of galaxies during cosmic time is the story of 
the cycle of transformation of gas into stars, the production of
metals inside these stars, the release of metals during stellar life and death, and
the interaction between these processes and their environment, i.e their host galaxies' 
dynamics and overall structure. All this evolution
leaves signatures in the observed properties of galaxies that we can 
analyse to reconstruct it. The analysis of this 
fossil record is a key tool to understand how
galaxies in the nearby universe evolved. In combination with the massive
acquisition of spectroscopic data, both integrated  
\citep[e.g., Sloan Digital Sky Survey, SDSS, Galaxy and Mass Assembly survey, GAMA,][respectively]{york2000,gamma}, and
spatially resolved \citep[e.g., Calar Alto legacy Integral Field Spectroscopy Area, 
CALIFA, or Mapping Nearby Galaxies at APO, MaNGA][respectively]{sanchez12,manga} 
have increased considerably our
understanding of the processes that govern galaxy evolution.

In a recent review, \citet{ARAA}, the most recent results obtained by the 
analysis of Integral Field Spectroscopy (IFS) Galaxy Surveys (GS) were 
summarized. Among the results reviewed there were the following: 
(i) the sources of ionization across the optical extent 
of galaxies; (ii) the interplay among the global (i.e., integrated/characteristic) properties of galaxies, the local (i.e., spatially resolved) ones, and the link between these two kinds of relations; and (iii) the radial distributions
of different properties of the stellar and ionized gas. However, due
to the narrow scope and space limitations of such reviews some important
aspects of those results were not fully addressed. 

In particular it was not possible to include a detailed description
of the adopted dataset (a compilation of publicly accessible IFS data),
and the description of the analysis performed to derive the described
results. We include those details in the current manuscript. Furthermore, we now provide (i) additional detail on the nature of the different ionizing sources and how the ionized gas is observed in galaxies; (ii) we demonstrate analytically how local
and global relations are connected and (iii) provide a quantitative statement on the  gradients described in the \citet{ARAA}.

The main aim of the current article is to provide the details
that were not covered in \citet{ARAA}, presenting more quantitative
results. Even though, the two article are clearly complementary,
the current one presents results not described in detail in that review and
includes new ones.
This review is organized as follows: (i) a description of the
adopted dataset is provided in Sec. \ref{sec:data}; (ii) 
Sec. \ref{sec:analysis} includes a summary of
the performed analysis; (iii) the results of the described 
analysis are presented in Sec. \ref{sec:diag}, including a detailed
description of the different sources of ionization within galaxies and the
diagnostic diagrams widely used to disentangle them through observational 
signatures; (iv) the
analytical description of the connection between local and global
relations is included in Sec. \ref{sec:local}, showing that indeed
both relations are essentially the same; (v) finally, a quantitative
statement on the radial gradients and characteristic values of 
the different resolved properties
explored in \citet{ARAA} is included in Sec. \ref{sec:grad}; we
summarize our results and the main conclusions in Sec. \ref{sec:summary}.

\begin{figure*}[ht]
  \centering
  \includegraphics[width=8cm, clip, trim=1 30 7 5]{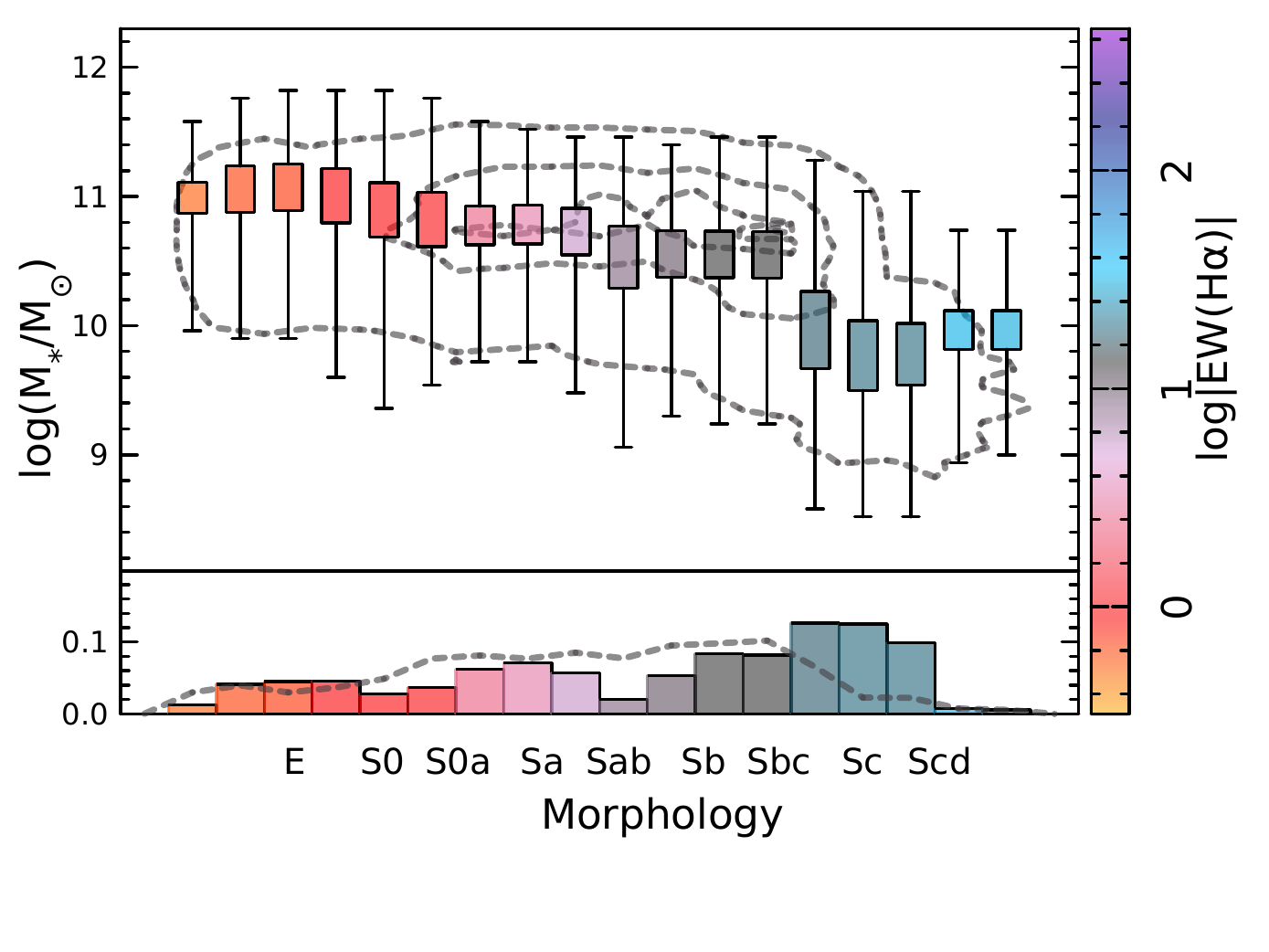}\includegraphics[width=8cm, clip, trim=1 30 7 5]{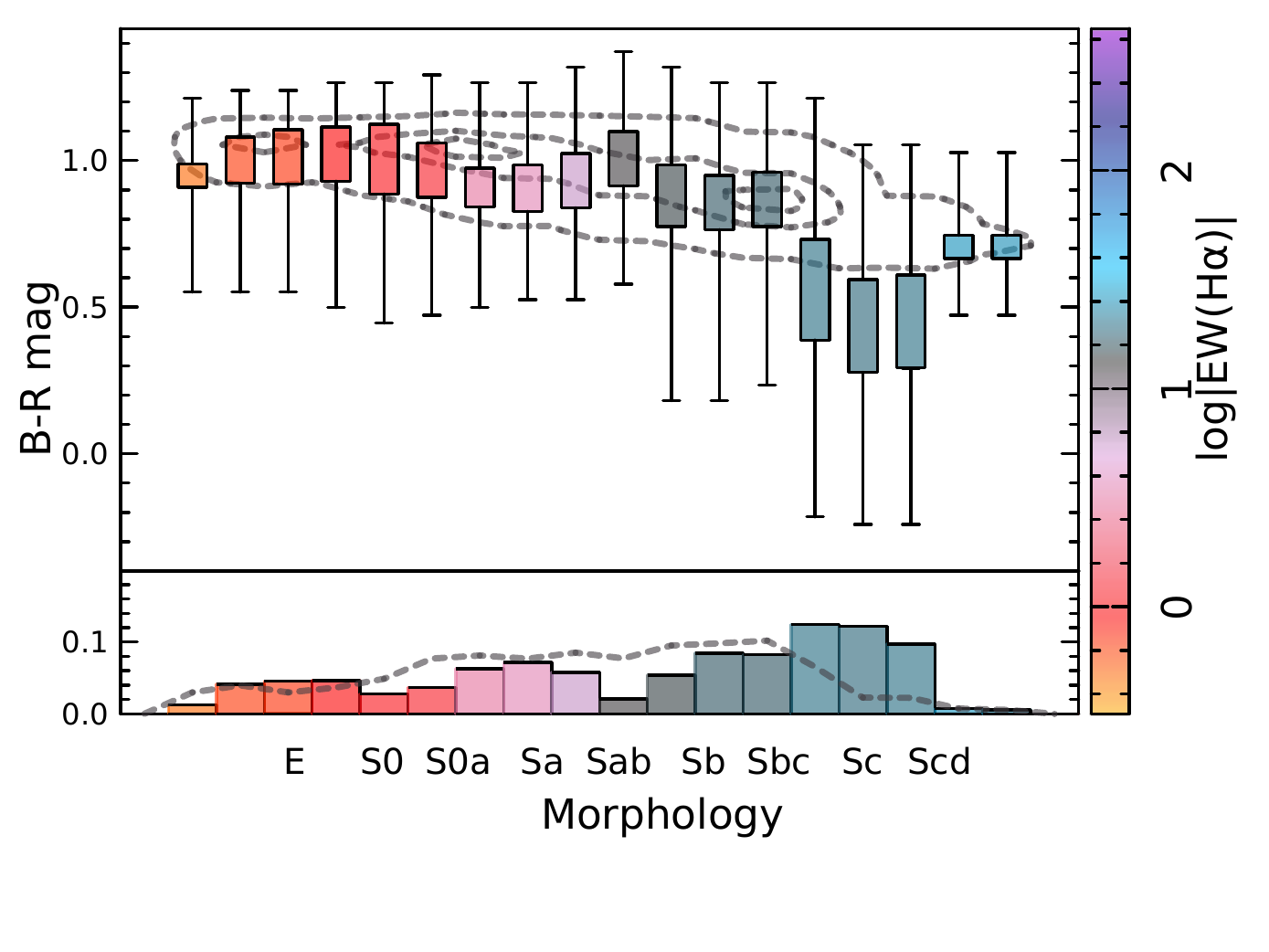}
  \includegraphics[width=8cm, clip, trim=1 30 7 5]{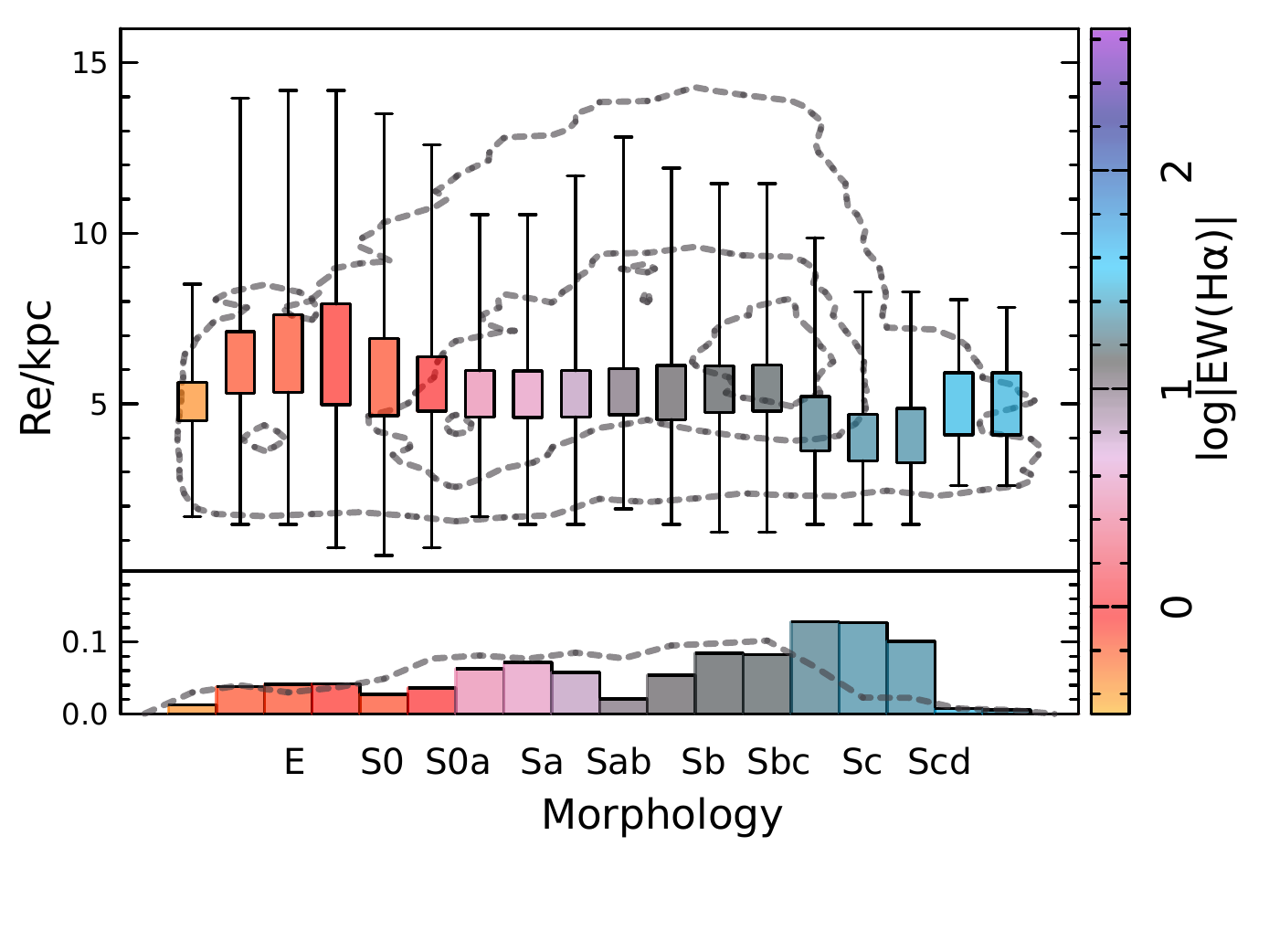}\includegraphics[width=8cm, clip, trim=1 30 7 5]{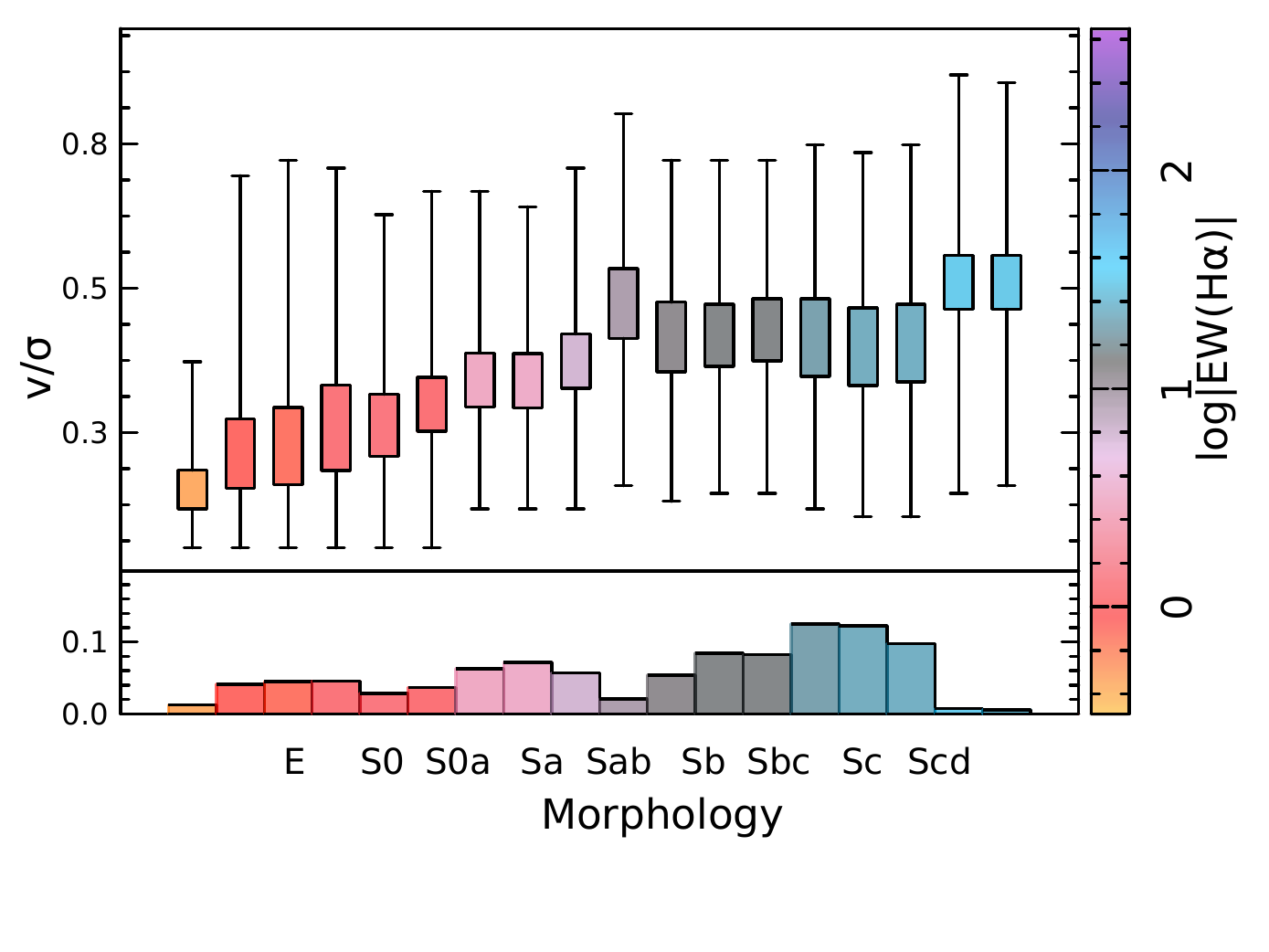}
\caption{Distribution of stellar masses (top-left panel), B-R color (top-right panel), effective radius (bottom-left panel) and v/$\sigma$ ratio within one effective radius (bottom-right panel) versus the morphological type for the full sample of galaxies. Symbols have the same meaning in each panel. Boxes are located at the average value for each morphology bin, with the size in the y-axis corresponding to the standard deviation around this value. Colors represent the mean value of the EW(H$\alpha$) of the galaxies, and errorbars indicate the range of values covered by 98\%\ of the sample, in each bin and for each of the explored parameters. In addition, for the first three panels, we include, as grey-dashed contours, the density distribution reported by \citet{nair2010} for a sub-sample of the galaxies in the SDSS survey, located at a similar redshift range.}
  \label{fig:Morph}
\end{figure*}

\section{Data sample}
\label{sec:data}

We adopted the same dataset already presented in \citet{ARAA}. This
represents a compilation of the publicly accessible IFS data provided
by the most recent IFS-GS and compilations, including: AMUSING++
\citep[447 ][]{galbany16b}, eCALIFA \citep[910][]{sanchez12,pisco},
MaNGA \citep[4,655][]{bundy15} and SAMI
\citep[2,222]{croom12}. Details of the particular characteristics of
each survey and the differences between the provided data are
discussed in detail in Appendix A of \citet{ARAA}. They all provide
spatially resolved spectroscopic information of large samples of
galaxies mostly located at $z\sim$0.01-0.06. After removing a few
cubes with low-S/N, or covering just a fraction of the optical extent
of the target galaxies, the final compilation comprises 8203 galaxies,
5637 with morphological information. However, due to the strong
differences among the different surveys, not all galaxies are sampled
with the same quality. \update {Thus, we select what we consider
  {\it the best quality data}, in terms of the ability to explore the
  spatial variations of galaxy properties in most optimal way, by
  restricting the dataset to those galaxies/cubes that satisfy the
  following criteria}:

\update{

  {\it (i) They should have a reliable morphological classification.} This is
  extremely important since one of the main goals of the current
  exploration is to characterise the physical resolved properties of
  galaxies for different morphological types.

  {\it (ii) Galaxies should be sampled out to 2.5 effective radii
    (R$_e$).} This requirement was included to explore only those
  galaxies which IFS data covers a significant fraction of their
  optical extension. This is particular important for disk-dominated
  late-type objects, which bulge may cover a range up to 0.5-1.0 R$_e$
  \citep[e.g.][]{rosa14}, and the average properties of the disk would
  not be well covered if the FoV of the IFS data is limited to 1-1.5
  R$_e$. Furthermore, it is known that beyond 2 R$_e$ disk galaxies
  may present a different behavior than that of the main disk, showing
  truncations or upturns in their surface-brightness
  \citep[e.g.][]{kruit01,bakos08,kruit11}, and deviations from the
  global oxygen abundance trends \citep[e.g.][]{marino16}. But is is
  also relevant in elliptical galaxies, in particular those that
  present some remnants of star-formation in the outskirts, but no
  where else \citep[e.g.][]{gomes16}. Thus, to cover up to 2.5 R$_e$
  guarantees that we sample the real radial distributions in
  galaxies, not being biased to neither the properties of the very
  central regions nor those of the most outer ones.

  {\it (iii) R$_e$ should be at least two times the
    full-width-at-half-maximum (FWHM) of the point spread function
    (PSF) of the data.} This is indeed a basic requirement to
  guarantee that galaxies are resolved by the data. If the PSF
  FWHM is of the order or larger than R$_e$, the considered galaxy would be unresolved. Thus, even if it is sampled beyond its full optical extension no
  reliable gradient or variation across the FoV could or should be
  derived. Although it may sounds obvious it is sometimes 
  ignored, in particular since sometimes there is a confusion between
  the size of sampling element (e.g., the pixel or spaxel in which the data are
  recorder or stored) and resolution (given by the FWHM of the PSF or
  beam of the instrument or final dataset).

  {\it (iv) We limit the redshift range up to $z<0.02$. } This
  requirement was included to restrict ad maximum the range of
  cosmological distances sampled by the data (D$_{\rm L}<$90 Mpc,
  $t<$380 Myr), but without excluding a significant fraction of
  galaxies of a particular type (mostly morphology) or stellar
  mass. This way there is no space for a strong cosmological evolution
  of the properties between the galaxies sampled at higher and lower
  redshifts, and most of the galaxies would be sampled at a similar
  physical resolution. Some of the original samples from which our
  collection is drawn span through much larger cosmological times (up
  to z$\sim$0.15), with strong correlations between galaxy properties
  and redshift \citep[e.g.][]{wake17}. This creates a complication
  in the exploration of the dependence of the derived observables with
  either the properties of the galaxies and/or their cosmological evolution.

  {\it (v) Highly inclined galaxies are excluded (i.e. we require
    $i<75^o$).} It is known that when a galaxy is observed at high
  inclinations many of its global and spatial resolved properties are
  strongly biased \citep[e.g., the molecular gas derived from the dust attenuation,][that we explore later on]{concas19}. Although this is a general problem, it is obviously
  more evident in disk-dominated galaxies. There is a combined effect of
  (a) dust attenuation, that obscures more the inner than the outer regions
  of inclined galaxies, (b) the intrinsic differences between radial and
  vertical variations, and (c) the difficulty to deproject the observed properties. For instance, disk-galaxies with prominent bulges,
  bars or thick disk would present strong vertical variations
  that should not be assigned to radial differences \citep[e.g.][]{levy18}.
  Although these galaxies are very important laboratories for the exploration
  of some particular and relevant galactic processes, like outflows \citep[e.g.][]{carlos19}, they are not good to provide representative properties
  of galaxies in general \citep[e.g.][]{ibarra19}.

  {\it (vi) The field-of-view (FoV) covered by the IFS data should
    have a diameter of at least 25$\arcsec$. } This requirement is
  introduced to have a good sampling of the radial properties of the
  galaxies. Considering that the PSF FWHM of the collected data range
  between $\sim$1$\arcsec$ (AMUSING data), and 2.5$\arcsec$ (the rest
  of the IFS-GS), and that we have imposed that galaxies are sampled
  at least up to 2.5 R$_e$, with an R$_e$ of at least 2 times the PSF FWHM,
  the current requirement guarantees that we have between 5 and 10 resolution
  elements to explore the considered radial distributions. Below
  this number we consider that the derived gradients would not be
  very reliable.
}

This {\it well resolved} sub-sample contains almost 1,500 
galaxies. In this review we sometimes adopt the full dataset, sometimes 
the {\it well resolved} one, depending on which is more appropriate. 
We clearly indicate which sample is used to derive each result.

Figure \ref{fig:Morph} shows the main properties of the sample of
galaxies compiled, including the morphological distribution against
stellar mass, $B-R$ color, effective radius and the ratio
between velocity and velocity dispersion (within one R$_e$). In
general the compiled sample resembles, in the observed distribution of
these properties, what would be observed for a volume complete
sample in a similar redshift range. For comparison purposes we have
included in Fig. \ref{fig:Morph}, when feasible, the locus of 
the galaxy sample by \citet{nair2010} from SDSS DR4 \citep{abazajian09} 
at a redshift range similar to ours.
Our compilation is dominated by
late-type galaxies ($\sim$70\%), with a clear peak at Sb/Sbc, and a 
decline towards earlier types, in particular concerning elliptical
galaxies. As expected, there is an increase of the average stellar
mass from late- to early-types, from $\sim$10$^{8.5}$ M$_\odot$ (Sd) to
$\sim$10$^{11.5}$ M$_\odot$ (E). This mild trend is smooth, and for each
particular morphological bin a considerable range of stellar
masses is covered. The main difference with respect to the SDSS 
distribution is an under-representation of Sab galaxies and an
over-representation of Sc/Sd galaxies.

However, despite these differences, the overall trend between morphology
and stellar mass is very similar. 
Both in \citet{ARAA} and the current exploration
we try to separate the effects of stellar mass and morphology dividing
galaxies in mass/morphology bins. However, it is worth noticing
that in this type of segregation there are intrinsic biases. In particular, groups of early-type galaxies (E/S0) and
low stellar mass ($<$10$^{9.5}$M$_\odot$) and late-type galaxies
(Sc/Sd) and high stellar mass ($>$10$^{11}$M$_\odot$) have so few
galaxies that the results are not statistically robust. Although we have included (and discussed) these bins in the
present publication, we advise the reader to treat the results obtained for 
these bins with due caution.

A similar trend as the one observed between morphology and mass 
is observed between $B-R$ color and mass \citep[e.g.][]{balogh04}. Early-type galaxies are red, 
covering a very narrow range of colors (i.e., defining a clear 
red sequence at $B-R\sim$1.2 mag). Later type galaxies have 
bluer colors, covering a wider range of colors (i.e., a cloud
rather than a sequence). The most relevant difference between the two
groups is that there is a deficit of blue/early-type galaxies (E/S0,
$B-R<$0.5 mag) and a corresponding deficit of red/late-type galaxies
(Sc/Sd, $B-R>$0.8 mag). Again, this distribution resembles what 
would be observed in a volume limited sample at a similar
cosmological distance.

Regarding the sizes of the galaxies, characterized by the effective
radius (R$_e$), there is also a relation with morphology. However, this 
seems a secondary correlation arising because of the well known 
relation of R$_e$ with stellar mass \citep[e.g.~][Fig.~20]{conse06,conse12,ARAA}. 
However, this trend is more shallow than the one with M$_*$. 
On average, early-type galaxies are slightly larger and cover 
a wider size range than late-type galaxis. This trend has a large
dispersion, and for a given morphology galaxies present a wide range
of sizes, although in general there is a lack of spiral galaxies
larger than R$_e>$10 kpc (for galaxies later than Sb). This figure
shows that size is primarily dependent on galaxy mass, rather than
morphology. When comparing with the literature results, there are 
clear differences. The data presented by \citet{nair2010} present a 
sharper increase in size from Scd to Sab galaxies, with a drop 
for earlier type galaxies. This distribution is not expected naively, 
and could be related to a size bias in that sample than a real effect. 
In any case, the average values are not very different between the 
two samples. 

Finally, we present the distribution of the ratio between stellar
rotation velocity and velocity dispersion within one R$_e$ over
morphology. This ratio is a proxy for the fraction of ordered rotation in
these galaxies \citep[e.g.][]{cappellari16}. Large values correspond to
galaxies with stars following well ordered orbits distributed in a
plane or disk, i.e., cold, rotationally supported orbits. On the
contrary, low values correspond to galaxies with stars on
hot/warm orbits (pressure supported), that present a triaxial
structure, including galaxies with strong bulges \citep[e.g.][]{zhu18a}. 
As expected
early-type galaxies present the lowest values for the v/$\sigma$
ratio, with a mode $\sim$0.1 and a deficit of galaxies with
v/$\sigma >$0.5 for pure Ellipticals. On the other hand, late-type
galaxies present the largest values, with a mode $\sim$0.5 (for Sc/Sd)
galaxies. It is interesting to highlight that the range of values
covered by this parameter for late-type galaxies is also wider, which is 
a consequence of projection effects. Like in the case of the
previous figures, the observed distribution for the compiled sample 
agree with the expected one for galaxies in the nearby Universe \citep[e.g.][]{davies83}.

The symbols in the different panels of Fig.~\ref{fig:Morph} are 
color-coded by the average distribution of the equivalent width (EW) 
of the H$\alpha$ emission line. As extensively discussed in
different studies \citep[e.g.][]{sta08,cid-fernandes10,sanchez18} and
reviewed in \citet{ARAA}, this parameter segregates well
between star-forming and retired galaxies (SFGs/RGs). It separates
equally well between star-forming and retired areas (SFAs/RAs) within
galaxies \citep[e.g.][]{mariana19}. The distributions shown in
Fig.\ref{fig:Morph} illustrate clearly the connection between the different
global properties and the star-formation activity of
galaxies. Early-type, massive, red, large and pressure supported
galaxies are mostly RGs, with little or no star-formation. On the
contrary, late-type, less massive, bluer, smaller and rotationally 
supported galaxies are mostly SFGs. Thus, those galaxies that contribute most
to the star formation budget in the nearby universe. There are clear, 
continuous trends from SFGs to RGs, just as from late to early type 
galaxies, with most galaxies in the transition region between the two groups 
corresponding to early spirals (Sa/Sb), i.e.~spirals with prominent bulges.

In summary, the compiled sample of galaxies covers the space of
explored parameters just as well as a well defined, statistically 
significant sample at the same cosmological distance. In general, the
distributions are similar to those reported for this kind of samples
\citep[e.g.][]{blanton09}. Therefore, although our sample was assembled 
in an ill-defined manner, the properties and results extracted from 
the analysis of this sample can be considered a good representation 
of the average population of galaxies in the nearby Universe 
(i.e.~within a few hundreds Mpc). 

\section{Analysis}
\label{sec:analysis}

To provide a homogeneous analysis of this somewhat heterogeneous dataset 
we analysed all cubes using the same tool, the {\sc Pipe3D} pipeline
\citep{pipe3d_ii}. This pipeline was designed for IFS datacubes to 
(i) fit the stellar continuum with spectra from stellar population 
models and (ii) extract the information about the emission lines of 
ionized gas. {\sc Pipe3D} uses {\sc FIT3D } algorithms as basic fitting
routines 
\citep{pipe3d}\footnote{\url{http://www.astroscu.unam.mx/~sfsanchez/FIT3D/}}. 
We include here a brief description
of the fitting procedure \citep[extensively described in][]{pipe3d,pipe3d_ii},
and a more detailed description of how the different 
parameters used in this review \citep[and in][]{ARAA} were derived.

\subsection{Stellar Population analysis}
\label{sec:ana_ssp}

The fitting of the stellar continuum requires a minimum signal-to-noise 
ratio (S/N) to provide reliable results. Since it is not guaranteed 
that this S/N is reached throughout the entire FoV (or optical extent 
of the galaxy), as a first step, a spatial binning is performed in 
each datacube to increase the S/N above this limit by co-adding 
adjacent spectra. This limit was selected to be 50 for most of 
the compiled data (CALIFA, MaNGA and AMUSING++), and 20 for the 
SAMI data (that have slightly lower S/N in the continuum). The actual 
value of this S/N limit was derived based on simulations for the 
considered spectral resolutions and wavelength ranges covered 
by the data \citep{pipe3d}.

Then, the stellar continuum of the co-added spectra corresponding to each 
spatial bin was fitted with a stellar population model, taking into 
account a model for the line-of-sight velocity distribution (LOSVD), 
and the dust attenuation. The stellar population model consists of 
a linear combination of a set of simple stellar populations (SSP) 
taken from a particular library. Therefore, the model spectrum is 
described by the following equation:

\begin{flalign}\label{eq:dec}
  \begin{aligned}
S_{obs}(\lambda) \approx S_{mod}(\lambda) = \\ \left[ \Sigma_{ssp} w_{ssp} S_{ssp}(\lambda) \right]
10^{-0.4\ A_{V}\ E(\lambda)} \ast G(v,\sigma) \\
  \end{aligned}
\end{flalign}

\noindent where $S_{obs}(\lambda)$ is the observed intensity of the spectrum at the
wavelength $\lambda$ for a particular bin; $S_{mod}(\lambda)$ is the overall 
model, that is derived by minimizing the difference with respect to 
$S_{obs}(\lambda)$ (by means of $\chi^2$ minimization); 
$w_{ssp}$ is the normalization of each contributing model SSP spectrum 
$S_{ssp}(\lambda)$; $A_{V}$ is the dust attenuation in the V-band 
(in magnitudes), and E($\lambda$) is the adopted extinction curve 
\citep[in this particular case][]{cardelli89}. This unbroadened 
model spectrum is convolved with the LOSVD, $G(v,\sigma)$, modelled 
by a Gaussian function of two parameters (the velocity, $v$, and 
velocity dispersion, $\sigma$). Thus, the best fitting model comprises 
three non-linear parameters ($A_v$, $v$ and $\sigma$), and a set of 
linear parameters, $w_{ssp}$, one for each SSP in the considered 
library. Note that Eq. \ref{eq:dec} assumes that the kinematics of 
all stars is described with a single LOSVD, and they are all 
affected by a single dust attenuation. These are simplifications 
of the problem. Young and old stars are known to follow different 
orbits within galaxies, and they may be affected by different 
dust attenuation. Experiments with more complex decomposition 
procedures are described in \citet{vale16}.

Each SSP in the considered library is represented by a single spectrum 
that is the result of co-adding all the spectra of the surviving 
stars \citep[i.e., considering the mass-loss with time, e.g.][]{court13}
created by a single burst that happens a certain time in the past 
(i.e.~the age of the SSP) from gas with a certain chemical composition 
(i.e.~a certain metallicity). SSPs are created by stellar population 
synthesis codes \citep[e.g.][]{bc03}, using as basic ingredients: 
(i) an Initial Mass Function (IMF) of stars 
\citep[e.g.][]{salpeter55,chab03}; (ii) a model for the evolution of 
the stars, described by isochrones in the Hertzsprung-Russell diagram; 
and (iii) a synthetic \citep[e.g.][]{coelho07, maraston10} or observational 
\citep{miles} stellar library of spectra for each star with a 
particular set of physical parameters (i.e.~at each location 
within the HR diagram and for each metallicity). The different 
ingredients included in the generation of the SSP, and differences in the
model algorithms produce subtle differences in the synthetic stellar 
population spectra for the same physical parameters. Therefore, any 
inversion method/stellar decomposition like the one described before 
(Eq.~\ref{eq:dec}) may produce different quantitative results depending 
on the adopted SSP library. The limitations of this method have been 
described in more detail elsewhere \citep{walcher11, conroy13}.

The adopted implementation of {\sc Pipe3D} uses the GSD156
SSP-library.  This library, first described in
\citet{cid-fernandes13}, comprises 156 SSP templates, that sample 39
ages (1 Myr to 14 Gyr, on an almost logarithmic scale), and 4
different metallicities (Z/Z$\odot$=0.2, 0.4, 1, and 1.5), adopting
the Salpeter IMF \citep[][]{Salpeter:1955p3438}.  It is assumed that
the number of stars in each selected resolution element (typically
$\sim$1 kpc) is large enough ($\sim>$10$^{4}$M$_\odot$), to have a
complete statistical coverage of the IMF \footnote{Note that for
  smaller apertures or very low-surface brightness galaxies, this may
  not be the case}. These templates have been extensively used in
previous studies
\citep[e.g.~][]{perez13,rosa14,ibarra16,sanchez18,sanchez18b}.

\update{We should stress that this particular library is not a priori
  better than other adopted ones to derive the properties of the
  stellar populations. Detailed comparisons and simulations presented
  in different studies
  \citep[e.g.][]{cidfernandes:2014,rosa14,pipe3d,pipe3d_ii,rosa16},
  demonstrate that as far as the space of parameters is fairly
  covered (mostly expected ages and metallicities), the explored
  quantities are well recovered and/or the values are consistent at
  least qualitatively when using different SSP templates. The main
  reason why the GSD156 library is best placed for the current study so far is because
  it is the only one used to explore in common the different dataset
  comprised in this study and, at the same time, confronted against
  mock IFS observations of galaxies created from hydrodynamical simulated
  \citep{ibarra19}. Therefore we understand the systematics associated
  with the derivation of parameters in a better way that with other
  SSP templates.}

Once the best model for the stellar population in each bin was derived, 
the model is adapted for each spaxel. This is done by re-scaling the 
model spectrum in each bin to the continuum flux intensity at the 
considered spaxel, as described in \citet{cid-fernandes13} and 
\citet{pipe3d} (we say that the model is \textit{``dezonified''}). 
Finally, based on the results of the fitting, it is possible to 
derive different physical quantities in each parameter $P$, both 
light-weighted (LW) and mass-weighted (MW), using the formulae:

\begin{equation}
{\rm log} P_{LW}  = \frac{\Sigma_{ssp} w_{ssp} P_{ssp}}{\Sigma_{ssp} w_{ssp}}
\label{eq:LW}    
\end{equation}

\noindent and

\begin{equation}
{\rm log} P_{MW}  = \frac{\Sigma_{ssp} w_{ssp} M/L_{ssp} P_{ssp}}{\Sigma_{ssp} w_{ssp} M/L_{ssp}}
\label{eq:MW}    
\end{equation}

\noindent where: (i) $P$ is the considered parameter, (ii) $w_{ssp}$ are the 
light weights (normalizations) described in Eq. \ref{eq:dec}, and (iii) $M/L_{ssp}$ is 
the mass-to-light ratio of the considered SSP. The parameters can be 
derived both in a spatially resolved way (spaxel-by-spaxel or bin-by-bin) 
or integrated (on coadded spectra or averaged across the FoV). 
These equations are also used in \citet{ARAA} and the current review to 
obtain further quantities of interest. Among them the most relevant are: (i) the average 
light-weighted Mass-to-light ratio ($M/L$), 
obtained by substituting {\it P} by $M/L_{ssp}$ in Eq. \ref{eq:LW}; 
(ii) the stellar mass surface density ($\Sigma_*$), by multiplying 
the average $M/L$ derived before with the surface brightness ($\mu$), i.e.:

\begin{flalign}\label{eq:mu}
  \begin{aligned}
\Sigma_* = \mu M/L \ \ \ \ \ \ \ \ \ \ \ \\
\mu = \frac{4\pi D_L^2 I_{obs,V}}{A_{spax}} 10^{0.4 A_{V,*}}
  \end{aligned}
\end{flalign}

\noindent where $D_L$ is the luminosity distance and $A_spax$ is the area of each 
spaxel (in the corresponding units, pc$^2$ in the present review). 
By integration over the FoV it is possible to derive the integrated 
stellar Mass (M$_*$); (iii) in a similar way, if instead of co-adding 
all the ages included in the SSP library, both quantities are added 
from the beginning of star formation in the universe up to a certain 
look-back time (and with additional corrections for the mass-loss), 
it is possible to derive $M_{*,t}$ and $\Sigma_{*,t}$ at a certain 
look-back time ($t$). $M_{*,t}$ and $\Sigma_{*,t}$ are the Mass (Density) 
Assembly Histories (MAH) of a galaxy \citep[or a region in a
galaxy][]{perez13,ibarra16}; (iv) the derivative of this MAH over 
time is the Star Formation History (SFH), which is nothing else 
than the Star Formation Rate as a function of time ($\mathrm{SFR}(t)$). 
$SFR(t)$ is indeed defined as

\begin{equation}\label{eq:sfr}
    \mathrm{SFR}(t)=\frac{d M_{*,t}}{d t} \approx \frac{\Delta M_{*,t}}{\Delta t} \equiv \frac{M_{*,t_1}-M_{*,t_0}}{|t_1-t_0|}
\end{equation}

\noindent where $t_1$ and $t_0$ are two look-back times (where $t_1<t_0$ 
and $t_0-t_l$ is relatively small). This way it is possible to 
estimate the most recent SFR from the stellar population analysis, 
usually defined as SFR$_{ssp}=$SFR$_{32{\rm Myr}}$ \citep{rosa16} 
(although other time ranges, in general below 100 Myr are considered 
too) \footnote{a more recent one could be derived using the H$\alpha$ dust corrected luminosity, that would corresponds to $<$10 Myr}; (v) the LW and MW  Age ($Age_{LW|MW}$) and metallicity 
($[Z/H]_{LW|MW}$), derived substituting $P$ by the age and 
metallicity of each SSP included in the library (i.e., $Age_{ssp}$ 
and $[Z/H]_{ssp}$), either adding over all ages (i.e., the current 
LW and MW ages and metallicities) or up to an age corresponding 
to a certain look-back time, $t$ ($Age_{LW|MW,t}$ and $[Z/H]_{LW|MW,t}$). 
This way it is possible to derive the Chemical Enrichment History of 
a galaxy (or a region within a galaxy) by exploring $[Z/H]_{LW,MW,t}$ 
over time \citep[e.g.][Camps-Fari\~na et al. in prep.]{vale09,walcher15,rosa16}.
Errors in the different parameters are derived based on the uncertainties in 
the individual analyzed spectra propagated through a Monte-Carlo 
procedure included in the {\sc FIT3D} code.

\subsection{Analysis of the ionized gas}
\label{sec:ana_gas}

In conjunction with the analysis of the stellar population 
we explore the properties of the ionized gas (both resolved an 
integrated) by deriving a set of emission line parameters, including
the flux intensity, equivalent width and kinematic properties. 
To that end we create a cube that contains 
just the information from these emission lines by subtracting the best fitting 
stellar population model, spaxel-by-spaxel, from the original cube. This 
{\it gas-pure} cube inherits a variance vector from the original cube that is 
made from two components: the original noise associated with the observations 
and the standard deviation of the residuals obtained from a Monte Carlo 
run of the continuum fitting with stellar population models. 
Finally, the parameters for each individual 
emission line within each spectrum at each spaxel of this cube are extracted 
using a weighted momentum analysis as described in \citet{pipe3d_ii}. 
More than 50 emission lines are included in the analysis, in the case 
of the CALIFA, MaNGA and SAMI datasets, and around 20 lines in the 
case of MUSE. Among them we include the strongest emission lines within 
the optical wavelength range: \Ha, \Hb, \oii\ \lam3727\footnote{not 
covered by MUSE data at the considered redshift},
\oiii\ \lam4959, \oiii\ \lam5007, \nii\ \lam6548,
\nii\ \lam6583, \sii \lam6717 and \sii \lam6731. The final 
product of this analysis is a set of maps showing the spatial distributions 
of the emission line flux intensities and equivalent widths. Integrated 
(or averaged) quantities across the optical extent of galaxies (or 
across the FoV of the instrument) are then easily derived. 

Finally, as in the case of the stellar mass, we derive the spatial 
distribution of different physical quantities used in the present 
review \citep[and throughout][]{ARAA}: (i) The attenuation of ionized 
gas emission ($A_V$),  derived from the spatial distribution of the 
\Ha/\Hb~ratio. We adopt the canonical value of 2.86 \citep{osterbrock89} 
for the non-attenuated ratio and use a Milky Way like extinction law 
\citep[][]{cardelli89} with R$_{V}$=3.1. (ii) The SFR, both resolved 
(i.e.~the SFR surface density, $\Sigma_{SFR}$) and integrated is 
derived from the H$\alpha$ dust-corrected luminosity (and surface 
brightness), following a prescription described by \citet{catalan15}. 
To calculate the attenuation-corrected H$\alpha$ luminosity L$_{H\alpha}$ 
we use a formula similar to Eq. \ref{eq:mu}, 
substituting the H$\alpha$ intensity for $I_{obs}$, and the gas
attenuation, i.e.~$A_{V}$ as defined before, for $A_{V,*}$. 
We then use the scaling derived by \citet{kennicutt89}, 
SFR$=8\ 10^{42}$ L$_{H\alpha}$, but apply it to each 
spaxel. Consistency tests of the SFR derived using H$\alpha$ and the stellar population analysis can be found in the literature 
\citep[e.g.][]{rosa16,sanchez19}. (iii) The molecular gas mass (M$_{gas}$), 
and its resolved version, the molecular gas mass surface density 
($\Sigma_{gas}$), is derived from both the dust-to-gas calibrator described 
by \citet{jkbb20}, defined as $\Sigma_{gas} = 23 A_V$, and an updated
calibration with the functional form $\Sigma_{gas} = a\ A_V^b$, presented 
in Barrera-Ballesteros et al. (in prep.). We will refer to the latter as
$\Sigma_{gas}'$ hereafter. (iv) The Oxygen abundance, 12+log(O/H), is 
derived from strong emission line calibrators, using those spaxels 
compatible with ionization related to star-formation (i.e.~young, 
massive OB stars). To select those spaxels we follow the prescriptions
described in Sec. \ref{sec:class}. A list of possible calibrations 
was included in \citet{sanchez19}. However, for the present contribution, 
we limited ourselves to the O3N2 calibrator by
\citet{marino13}. 

Based on these primary parameters, we can also derive some additional
parameters discussed within this review: (i) Star-formation efficiency, 
defined as the ratio between the $SFR$ and M$_{gas}$ (or between 
$\Sigma_{SFR}$ and $\Sigma_{gas}$, for their spatially resolved 
version). This parameter is just the inverse of the depletion time 
($\tau_{dep} \equiv \frac{1}{SFE}$). (ii) The specific star-formation 
rate (sSFR), defined as the ratio between $SFR$ and M$_*$ (or 
between $\Sigma_{SFR}$ and $\Sigma_{*}$, for their spatially resolved version).

All the physical parameters derived from emission lines are not 
directly calculated by {\sc Pipe3D}, although they are part of an 
analysis post-processing that is performed by the same algorithms 
for the different analyzed datasets. Finally, we should strongly 
stress that {\sc Pipe3D} is just one of several different 
pipelines/tools developed in the last years with the goal of 
analysing IFS data \citep[e.g.,
PyCASSO, LZIFU, MaNGA DAP][]{amorim17,ho16,dap}. Most
of the performed comparisons demonstrate a remarkable agreement in the
results \citep{pipe3d_ii,dap,sanchez19}.




\section{Ionizing sources in galaxies}
\label{sec:diag}

Star forming galaxies (SFGs) and star forming areas within galaxies (SFAs) are
frequently identified based on the observational properties of the ionized
gas. In our current understanding of the star-formation process when a
molecular gas cloud reaches the conditions of the Jeans instability
\citep{jeans02,bonnor57} it fragments and collapses, igniting
eventually star formation activity \citep{low76,true97}. This
process creates thousands of stars at the typical scale of a molecular
cloud, in general. These stars are not equally distributed in 
mass: as expected from a fragmented cloud \citep{bate05},
there is a larger number of less massive stars than of more massive
ones \citep[][]{salpeter55,chab03}. These later massive, short-lived,
young stars (classified spectroscopically as O and B stars) have a
blue spectrum, with a significant contribution of photons below the limit
required to ionize not only Hydrogen (E$>$13.6 \update{eV},
$\lambda<$912\AA), but also other more heavy elements, in particular
Oxygen, Nitrogen and Sulfur. Therefore, they ionize the gas
distributed around the recently formed stellar cluster, producing a
large number of emission lines observed in the optical spectrum. These
emission lines arise due to the recombination of ions with electrons, 
and the subsequent cascade of lines as the electron drops onto low energy
levels, or by the radiative de-excitation of electrons in levels
previously excited by collisions between ions and
electrons\citep[e.g.][]{osterbrock89}. These ionized gas clouds 
are the classical
\ION{H}{ii} regions \citep[e.g.][]{sharp59,peim67}. 

The ratios between emission lines originating from ionized metals and 
those from Hydrogen depend on the physical conditions inside the nebulae. 
In general, classical \ION{H}{ii} regions show 
typical values for certain line ratios (like \oii/\Hb, \oiii/\Hb,
\oi/\Ha, \nii/\Ha and \sii/\Ha):
(i) $<$1 dex in the case of \oiii/Hb; (ii) $<-$0.1 dex, in the case of
\nii/Hb and \sii/\Ha; and (iii) $<-$1 dex, in the case of \oi/Ha \citep[e.g.][]{osterbrock89}. The reason for these relatively low values 
is the shape of the ionizing spectra, that, although hard enough to cause 
some ionization, are not hard enough to provide enough high
energy photons to strongly ionize heavy elements like Oxygen or
Nitrogen. These ratios are then further modulated by the Oxygen and 
Nitrogen abundances, the ionization parameter 
(the ratio between the available ionizing photons and the hydrogen
content), the electron density, dust content and even the geometry of
the nebulae with respect to the ionizing source
\citep[e.g.~][]{BPT1981,evans85,dopita86,veilleux87,veil95,dopita00,kewley01,kewley02,sanchez15,mori16}.

However, young stars resulting from recent SF are not the only
ionizing sources in galaxies (although they are in general the
dominant one). Other sources, in order of importance (strength and
frequency) are: (i) The hard and intense ionizing spectra associated
with non-thermal and thermal emission of active galactic nuclei, which 
are observed in a limited fraction of galaxies. This ionization is
particularly important in the central regions of galaxies, and for
those galaxies in the AGN phase \citep[$\sim$10\% of galaxies in
the nearby Universe, or even less, e.g.~][although their relative importance
increases at high-z]{schawinski+2010,lacerda20}. (ii) The hard but 
weak ionizing radiation from hot evolved stars \citep[HOLMES,
post-AGB stars][]{binette94,flor11}, that could significantly contribute to
the excitation of the so-called diffuse ionized gas in galaxies \citep[DIG,
e.g.~][]{sign13,lacerda18}. (iii) Shocks associated with galactic
winds, either induced by high-velocity galactic-scale winds due to the
kinetic energy introduced by central starbursts
\citep[e.g.][]{heckman90} or AGN, or low-velocity winds associated with gas
cooling processes or internal movements in triaxial galaxies
\citep{dopita96}. This ionization may contribute significantly 
to the DIG too. (iv) Supernova remnants, associated with past (but 
relatively recent, $<$100 Myr) SF processes, also present an
expanding shock wave, but with very different geometry than the
previous ones. Similar in shape to \hii\ regions, and frequently
miss-classified or mixed with them, they could explain a fraction of
nitrogen enhanced regions described in the literature \citep[][Cid
Fernandes et al., submitted]{ho97,sanchez12b}. In general all those
ionizing sources present harder ionizing spectra than the ones observed
in \hii\ regions. Therefore, they emit a relatively larger
fraction of their flux in high energy photons and thus produce 
larger values for the  line ratios described above.

When using emission lines as diagnostics of the ionized ISM,
it is important to keep in mind a second concept, which we will call the 
"ionization conditions" in the remainder of this review. Indeed, it is 
customary to sub-sum all gas that does not reside in a  
spatial region clearly associated with star formation 
(i.e.~\ION{H}{ii} regions) or clearly associated with AGN (narrow line 
regions NLR) into the diffuse ionized gas (DIG). The DIG is in general 
all emission that has no clear peaky structure, but is rather smooth 
in its surface brightness distribution. DIG may still show structure 
(filaments, spiral arms), but to a lesser degree than typical
for \ION{H}{ii} regions.

A clear cut case where the ionization source has to be distinguished 
from the ionization conditions is the leaking of ionizing photons 
from \ION{H}{ii} regions. Indeed, these photons may present a somewhat 
harder ionizing spectrum than the original source \citep[hot, massive 
stars e.g.][]{weil2018}, and therefore present a much wider range of line 
ratios than the \ION{H}{ii} regions directly associated with the 
ionizing source. As we will see later, in galaxies that are primarily 
ionized by SF, leaked photons can contribute significantly to the DIG.

\begin{figure*}[ht]
\begin{centering}
\includegraphics[height=10.25cm, clip, trim= 15 25 82 1]{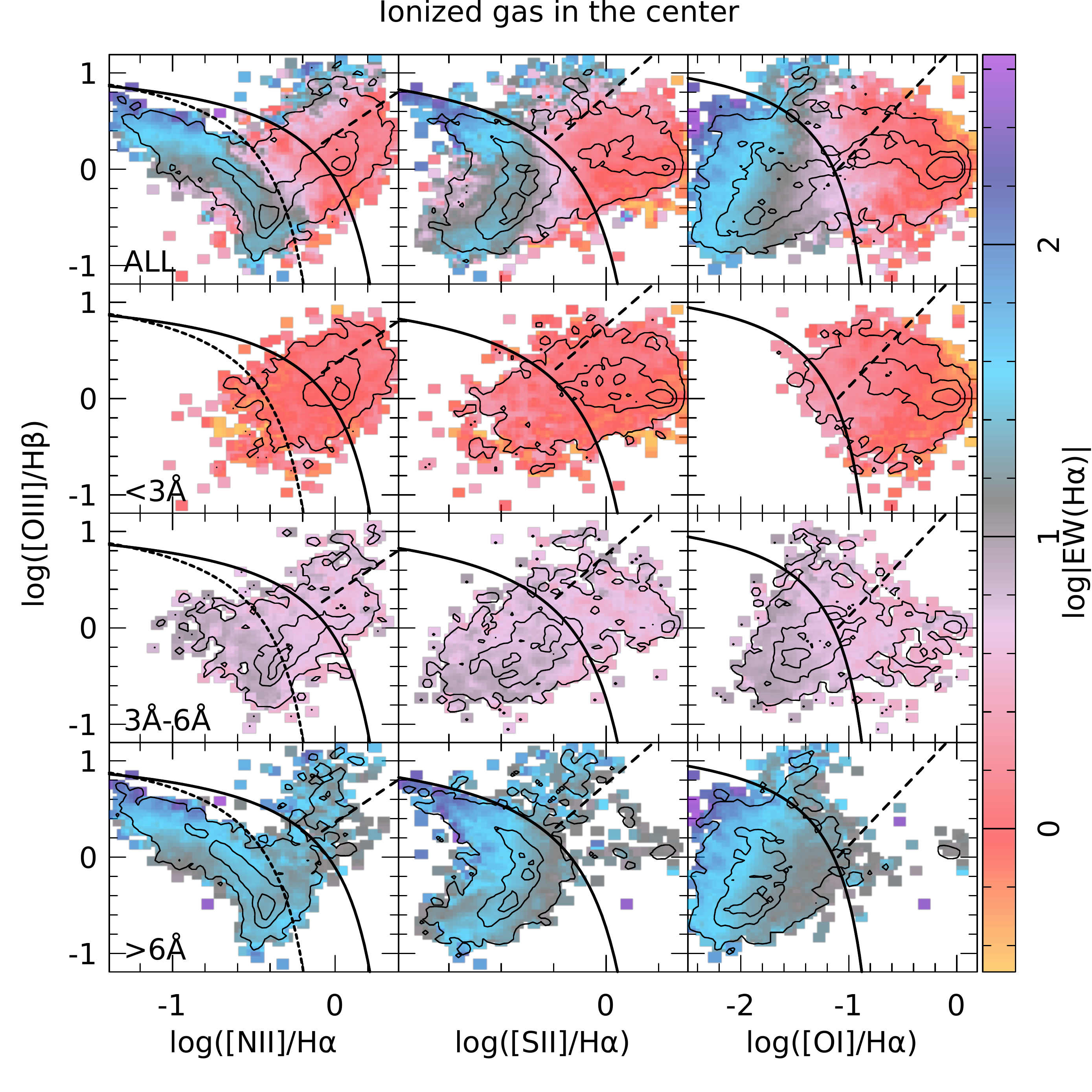}
\includegraphics[height=10.25cm, clip, trim= 80 25 1 1]{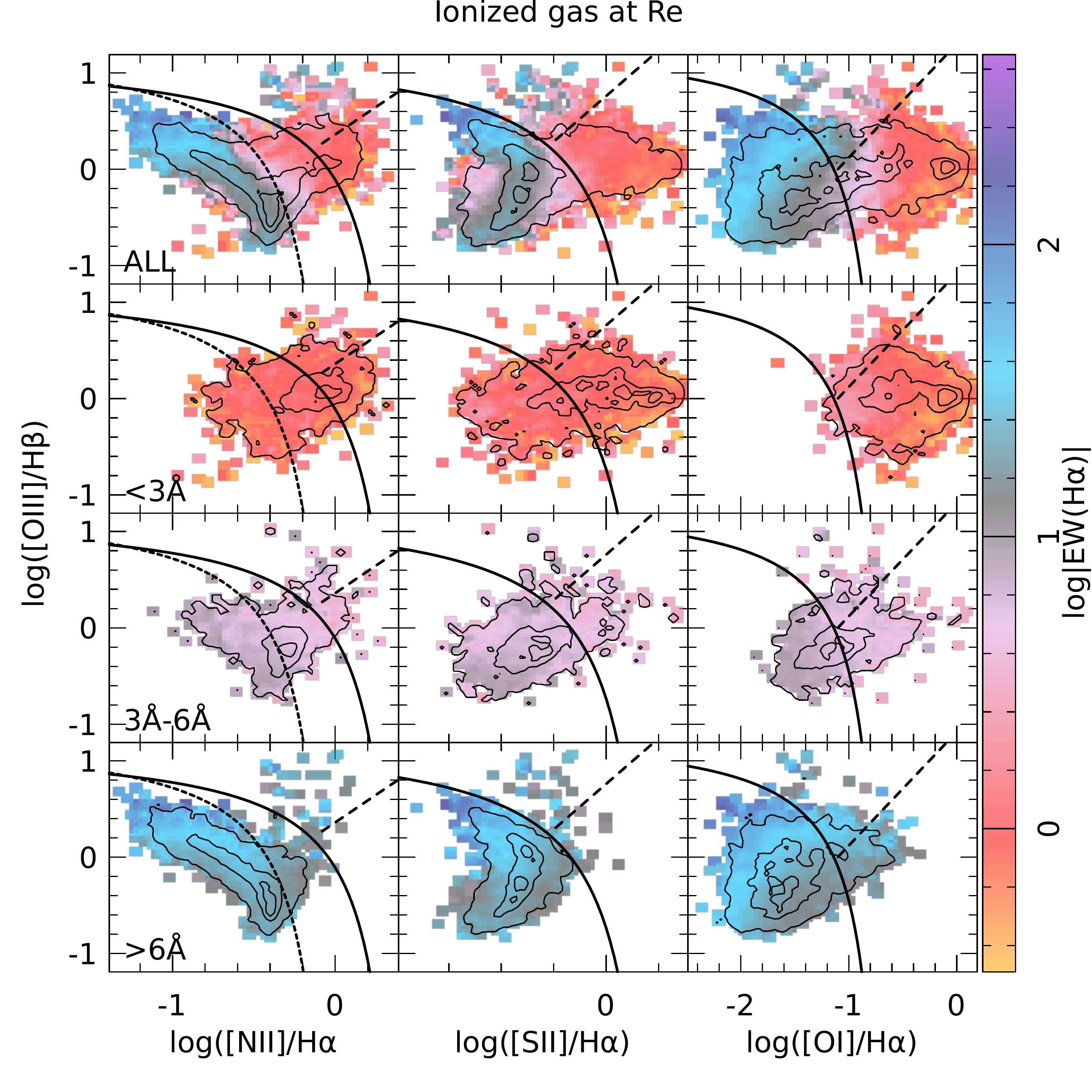}
\end{centering}
\caption{{\it Left panels}: Diagnostic diagrams for ionized gas built from 
the central ($\sim$3$\arcsec$ diameter) apertures of the galaxies in the 
sample, including the distributions of the [\ION{O}{iii}/H$\beta$] vs.
[\ION{N}{ii}]/H$\alpha$ line ratio (left panel), [\ION{O}{iii}/H$\beta$] 
vs. [\ION{S}{ii}]/H$\alpha$ (central panel), and [\ION{O}{iii}/H$\beta$] 
vs. [\ION{O}{i}]/H$\alpha$ (right panel). Each galaxy 
contributes a single point in the distributions, which in turn are shown 
as contours representing the density of objects. Each contour encompasses 
95\%, 50\% and 10\% of the points, respectively. The color code shows the 
average EW(H$\alpha$), on a logarithmic scale, of all galaxies at each 
point in the diagrams. In all panels the solid line represents the 
locus of the \citet{kewley01} boundary lines, with the proposed 
separation between Seyferts and LINERs indicated with a dashed-line. 
Finally, in the left-most diagram the dotted-line represents the 
locus of the \citet{kauff03} demarcation line. The upper panels 
show the distributions for all galaxies, irrespective of their 
EW(H$\alpha$) values. Then, form top to bottom, galaxies are 
separated by EW(H$\alpha$), comprising low values ($<$3\AA, 
panels in the 2nd row), intermediate values ( $3-6$\AA, panels in 
the 3rd row), and high values ($>$6\AA, bottom-panels). 
{\it Right panels:} Similar plots for the average ionized gas 
at the effective radius in each galaxy. }
\label{fig:BPT_cen}
\end{figure*}

The physical differences between the possible ionizing sources listed 
before (mostly the hardness of their spectra), and the ionized gas 
(mostly their metal content) have been used to define
demarcation lines in diagrams comparing pairs of metallic to 
hydrogen line ratios, with the intent of distinguishing between ionizing 
sources directly. These are the so-called diagnostic diagrams
\citep[e.g.][]{BPT1981,osterbrock89,veil01}. This concept is quite 
successful when it comes to identifying gas ionized by star forming regions, which shows 
a nearly one-to-one correspondence between the location of the 
spatial regions within the diagnostic diagrams and the ionizing 
source \citep[e.g.~][]{kewley01}. Unfortunately, 
the distinction among the different physical processes in the second 
group is less clear \citep[e.g.][]{cid-fernandes10}.
Some of the complication arises because of the mixing between different
ionizing sources. This is particularly important for those ionizing 
sources that contribute to the DIG, where shocks, the contribution of
ionization by old stars, and even photons leaked
from \hii\ regions could be spatially co-existing \citep[e.g.][]{della20}.
We will try to shed some light on this complex problem in the next section.

\subsection{Complexity and myths of diagnostic diagrams}
\label{sec:complex}

Figure \ref{fig:BPT_cen} shows the distribution of galaxies 
in three of the most frequently adopted diagnostic diagrams
\citep[][]{veil95}. The left panels show the
distributions for the central regions of galaxies,
corresponding to $\sim$1 kpc at the average redshift of our compilation
(3$\arcsec \times$ 3$\arcsec$ aperture). This region corresponds to
the one most affected by any ionization associated with an AGN, a
shock induced by a central outflow, or ionization due to hot evolved stars
\citep[more frequently present in the bulge of galaxies, 
the so-called cLIERs, e.g.~][]{belfiore17a}. The right panels present
similar distributions for a ring at one
effective radius of each galaxy.  This region is far enough from the
center to be clearly less affected by central ionizing sources, 
and therefore the ionization is more related to \hii\ regions in 
the case of SFGs. In all these plots each galaxy contributes as a 
single point in the considered distributions. The distributions are 
color-coded by the average value of EW(H$\alpha$) for all the
galaxies at a particular locus within the diagram. As indicated
before, several studies have demonstrated that the EW(H$\alpha$), in
combination with the described line ratios, is a good discriminator
between ionization conditions: (i) AGN and high-velocity shocks
present in general high values of the EW(H$\alpha$), $>$6\AA\
\citep[e.g.~][]{sta08,cid-fernandes10}. (ii) HOLMES/post-AGBs and
low-velocity shocks present in general low values of the
EW(H$\alpha$), $<$3\AA\  \citep[e.g.][]{binette94,sarzi10,lacerda18,carlos20}.  
(iii) Weak AGN could present EW values between 3-6\AA\
\citep[e.g.][]{cid-fernandes10}, and sometimes even lower 
(for very weak ones). It is important to note here that \hii\
regions are expected to present a high value of the EW(H$\alpha$) too. 
In our sample they present values larger than 6\AA\
\citep[e.g.,][]{sanchez14,lacerda18,espi20}. However,
they are located in a different region in the diagnostic diagrams, 
as indicated before.

The described trends can also be seen in the average distributions 
shown in the top panels of both right and left figures. Ionized 
regions with high EW are mostly located in the left part of the
diagrams, i.e., at the classical location of \hii\ regions. 
On the other hand, ionized regions with low EW are found in the
right parts of the three diagrams. We will show in upcoming
sections that indeed low EW is in general associated with DIG. 
Such regions are more numerous in the left panels (central 
regions of galaxies) than at one R$_e$. 

There are a few galaxies with high EW regions 
in the upper right area of the diagram. They are clearly less 
numerous than the low EW ones and therefore do not show up 
well in our representation. Their influence can, however, be seen 
in the three diagrams in the last row on the left. A comparison with 
the three panels on the right in the same row also shows that they 
are more numerous in the
center than at one R$_e$. Those correspond to either strong 
AGN ionization or the contribution of shocks created by 
high velocity galactic
outflows \citep[e.g.~][]{bland95,veil01, ho14, carlos19, carlos20}.

As mentioned before, based on the described average distributions,
different demarcation lines have been proposed for these diagrams 
to separate between the different ionizing sources. 
The most popular ones are the
\citet{kauff03} (K03) and \citet{kewley01} (K01) curves (included in
Fig. \ref{fig:BPT_cen} as a dashed and a solid line). They are usually
invoked to distinguish between star-forming regions (below the K03
curve) and AGN (above the K01 curve). The location between both
curves is generally assigned to a mixture of different sources of
ionization, being refereed to as the {\it composite} region
\citep[e.g.][]{cid-fernandes10,davies16}. However, as we will 
discuss later, this is only
one of multiple possibilities to populate that area of the
diagnostic diagrams. The demarcations lines have very different
origin. The K03 line is a purely empirical boundary traced by hand as
an envelope of the star-forming galaxies detected in the SDSS
spectroscopic survey.
The second demarcation line was derived based on a set of 
photo-ionization models, as the envelope of the largest 
values for the considered line ratios that can be produced 
by ionization due to young stars and a continuous star formation 
\citep[similar curves were derived by][
using other photoionization models and star-formation
histories]{dopita00,stas06}. In essence, only these later demarcation
lines are physically driven, indicating which region of the
diagrams cannot be populated by ionization due to
star-formation. 

A first exploration of the distributions shown in
Fig.~\ref{fig:BPT_cen} seems to demonstrate that the proposed
demarcation lines do a good job in segregating at least the harder
(AGN, shocks, post-AGBs) from the softer (\hii) ionization. When
exploring the upper panels in both figures it seems that all \hii\
regions (high-EW, left-size) are well constrained by the K03
demarcation line while most of the hard ionized regions are above 
the K01 one. This seems to be particularly true for AGN, that 
correspond to the hard ionized regions (upper right in each diagram) 
in the bottom panels (i.e., with high EWs). 

However, while it is certainly true that low spatial resolution or 
single aperture data may mix different spatial components of 
galaxies with different ionization mechanisms, and thus may 
populate the region between the two demarcation lines, it is not true
that only mixed ionization can be found there. A detailed
inspection of the distributions segregated by the EW(H$\alpha$)
clearly demonstrates so. High EW regions with mixed ionization
could only result from the mixing of \hii\ and AGN/shock 
ionization. However, such a case could be present only in a very
limited fraction of galaxies ($<$10\% or so, at the considered
redshift), and only in the central regions of galaxies (left panels). 
So, in general, mixed ionization resulting from a mixture of SF and 
DIG should be identifiable by intermediate EW values. This 
corresponds to the third row of panels in both figures. While 
a substantial fraction of spatial regions with intermediate EW 
are located in the intermediate region, a considerable fraction 
of these are well below and above the K03 and K01 curves. 
{\bf Thus, there are no features in the EW-augmented diagnostic 
diagrams that would allow to define a unique locus of mixed ionization.}

Furthermore, exploring the diagrams corresponding to the high and low
EWs (2nd and 4th rows), the complete continuity in the distributions 
over the K01 and K03 lines seems to indicate that this 
intermediate region is populated by a non-negligible fraction 
of both DIG and/or \hii\ regions \citep[the so-called nitrogen 
enhanced regions][that could be polluted by SNR]{ho97}. {\bf Thus, 
the region between both demarcations lines is not exclusively 
populated by spatial regions with a mixed ionization.} Finally, 
an additional complication are the low EW regions well 
below both the K01 and the K03 demarcation
line (in particular at one R$_e$ and for K01). Some authors have even claimed that
low-metallicity AGNs could also populate a region below the K01 (or
the K03) line \citep[e.g.][]{sta17pue}. This indeed does not 
contradict the nature of the K01 line, that was defined
as a maximum envelope for the \hii/SF regions (which seems to be a valid
interpretation), and not as a definitive boundary between soft and
hard ionization (as frequently and wrongly interpreted).

In summary: (i) classifying the ionizing source based only on the
distribution in the so-called diagnostic diagrams may be valid only at
a statistical level. Thus, for individual targets in 
boundary/intermediate regions, the use of these diagrams may lead 
to important mistakes; (ii) considering the additional information
provided by the EW(H$\alpha$) may mitigate the miss-classifications
induced by a pure selection based on the loci within these diagrams; 
and (iii) interpreting the location in a diagram as a combination 
of mixing of ionizing sources may be largely miss-leading. 
However, as we will see in the next section,
additional information provided by the morphology/shape of the ionized
structures, the underlying stellar population and even the kinematics
of the gas may shed some further light on the ionization conditions.

\subsection{Spatial distribution of the ionized gas}
\label{sec:spatial}

\begin{figure*}
\includegraphics[width=8cm, clip]{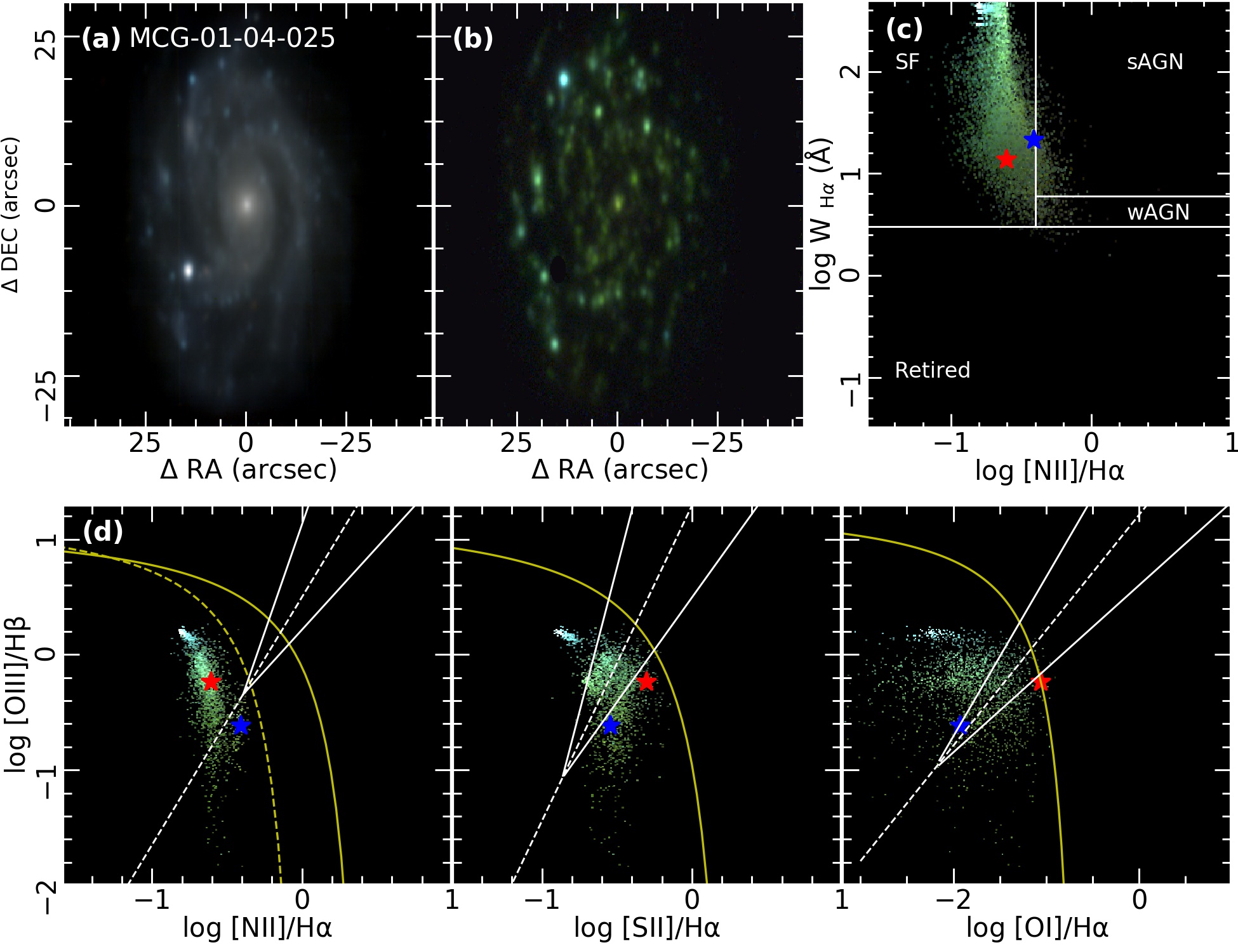}\includegraphics[width=8cm, clip]{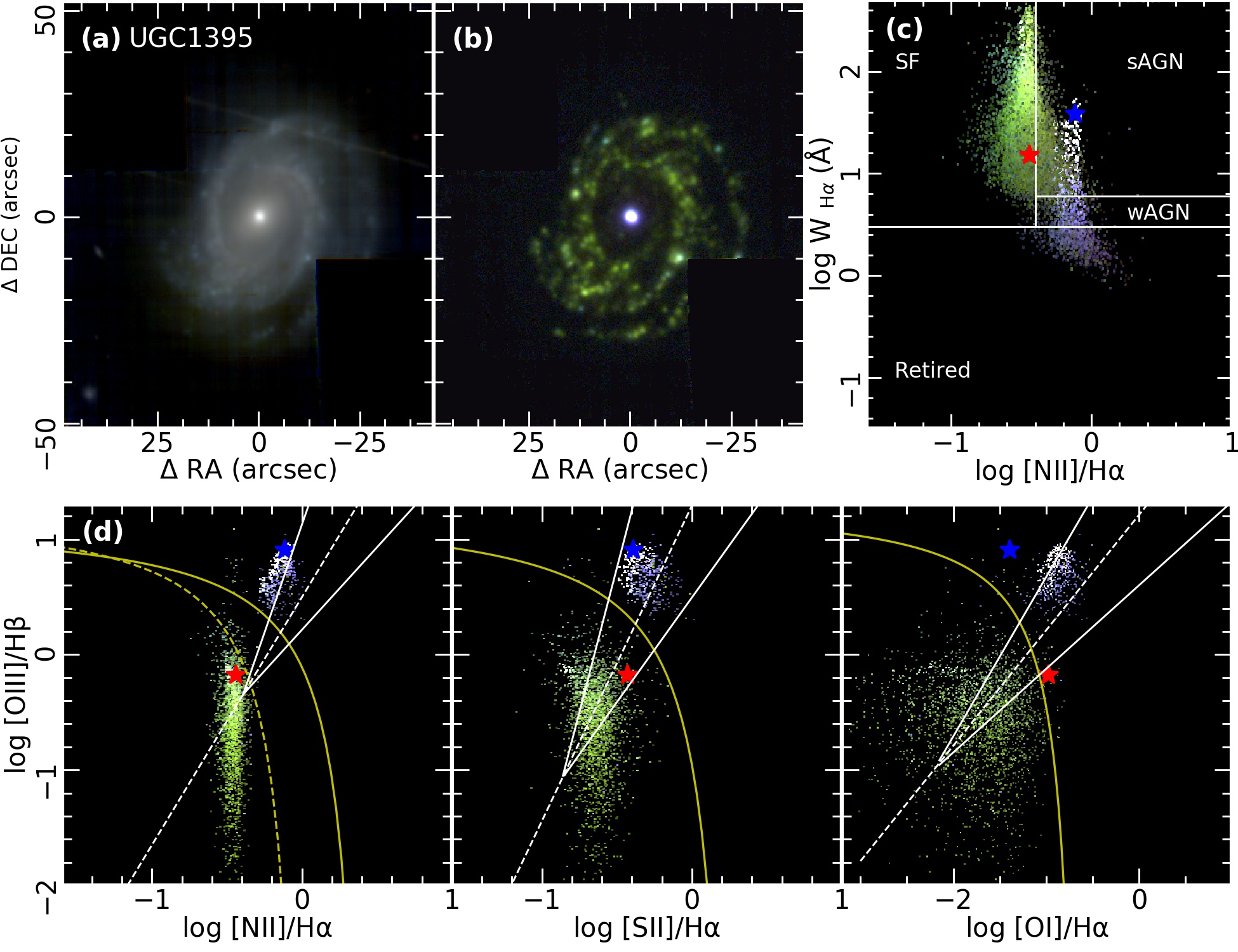}
\includegraphics[width=8cm, clip]{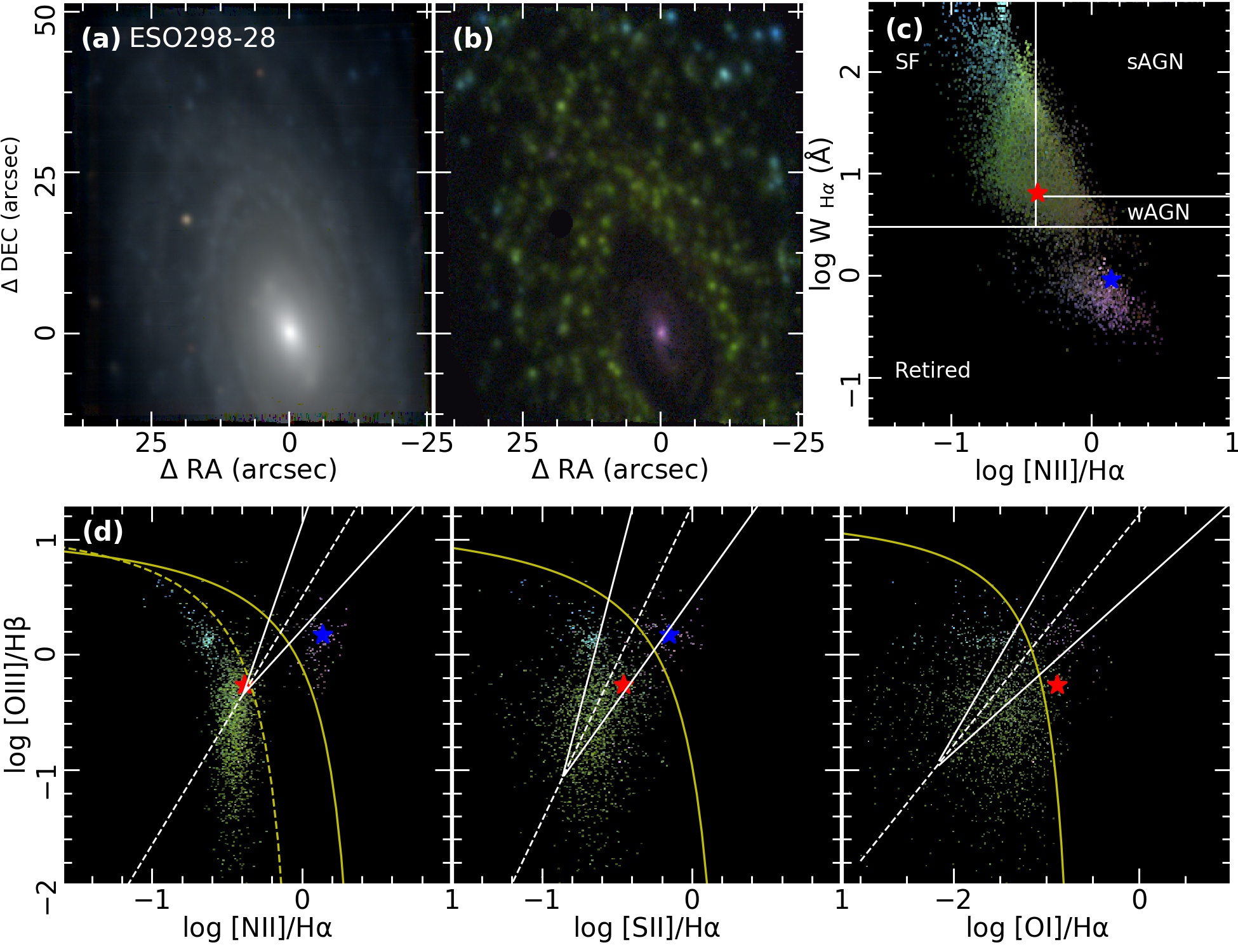}\includegraphics[width=8cm]{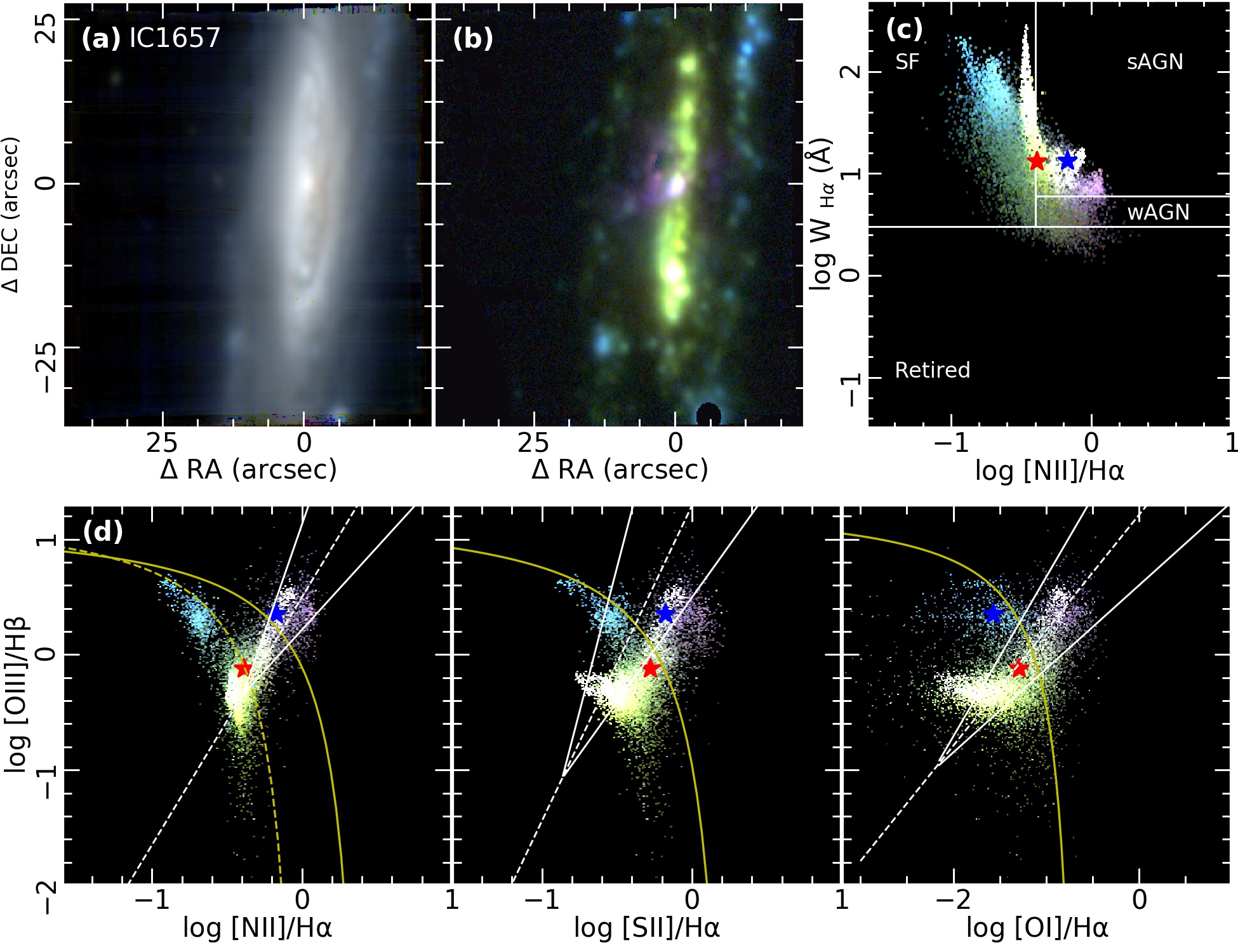}
\caption{Each panel shows, for different galaxies: (a) the continuum image created using $g$ (blue),$r$ (green) and $i$-band (red) images extracted from a MUSE datacube, by convolving the individual spectrum in each spaxel with the corresponding filter response curve (left-panel); (b) the emission line image created using the [\ION{O}{iii}] (blue), H$\alpha$ (green) and [\ION{N}{ii}] (red) emission line maps extracted from the same datacube using the {\sc Pipe3D} pipeline (central-panel); (c) the WHAN diagnostic diagram presented by \citet{cid-fernandes10}, showing the distribution of EW(H$\alpha$) versus the [\ION{N}{ii}]/H$\alpha$ line ration; and finally, (d) the classical diagnostic diagrams involving the [\ION{O}{iii}]/H$\beta$ line ratios vs. [\ION{N}{ii}]/H$\alpha$ (left), [\ION{S}{ii}]/H$\alpha$ (middle) and
[\ION{O}{i}]/H$\alpha$ (right), respectively \citep{baldwin81,veil95}. The four diagnostic diagrams included in panels (c) and (d) are color-coded by the values shown in the emission line image shown in panel (b). The average value of the parameters shown in panels (c) and (d) across the entire FoV of the IFU data is shown as a red star in each diagnostic diagram, while the central value is marked with a blue star. The solid and dashed lines represent the location of the \citet{kewley01} and \citet{kauff03} demarcation lines, respectively. The name of each galaxy shown
in each panel is included in the figures, comprising from top to bottom MCG-01-04-025, UGC1395, ESO0298-28 and IC1657.}
\label{fig:bpt_map_SF1}
\end{figure*}

\begin{figure*}
\includegraphics[width=8cm]{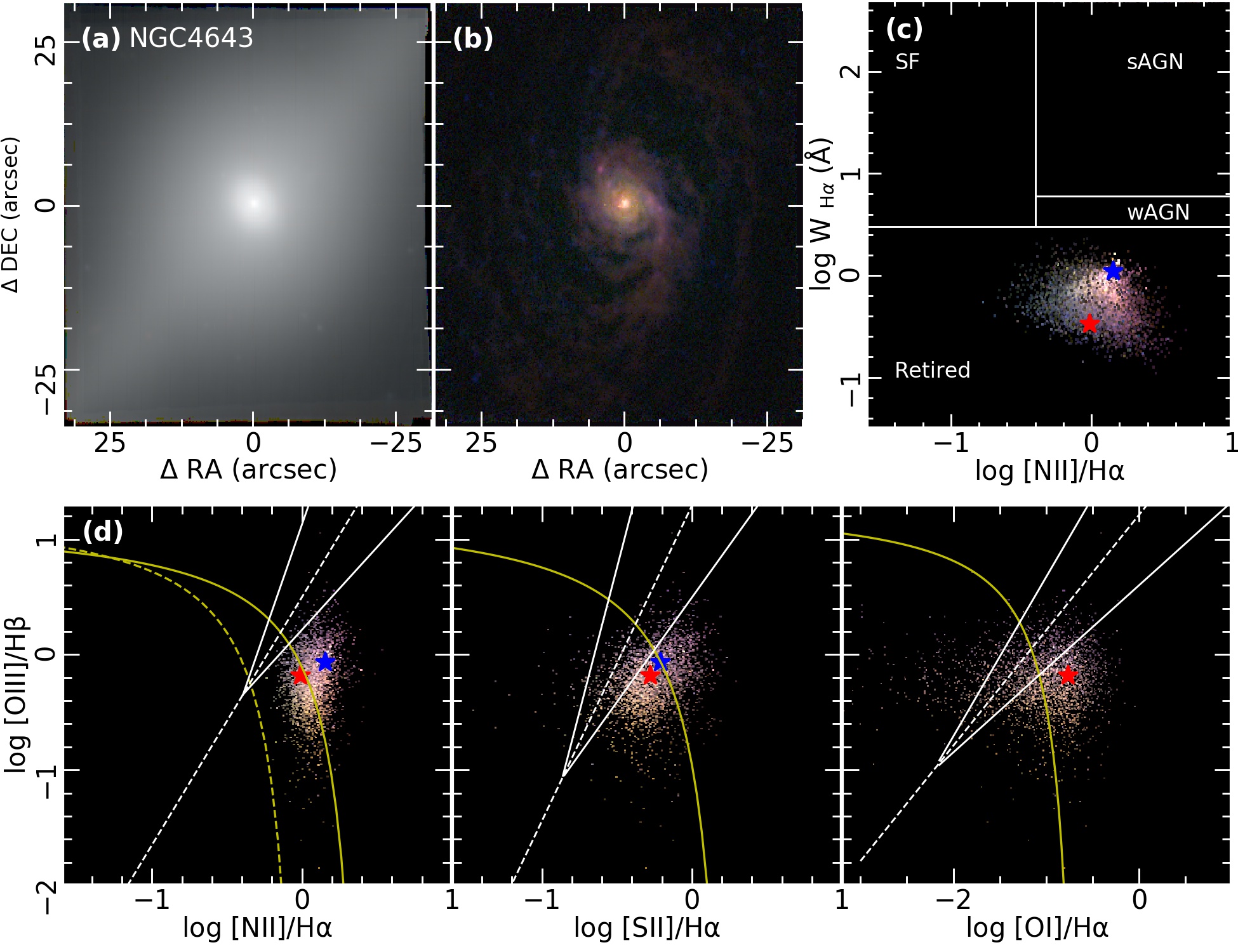}\includegraphics[width=8cm]{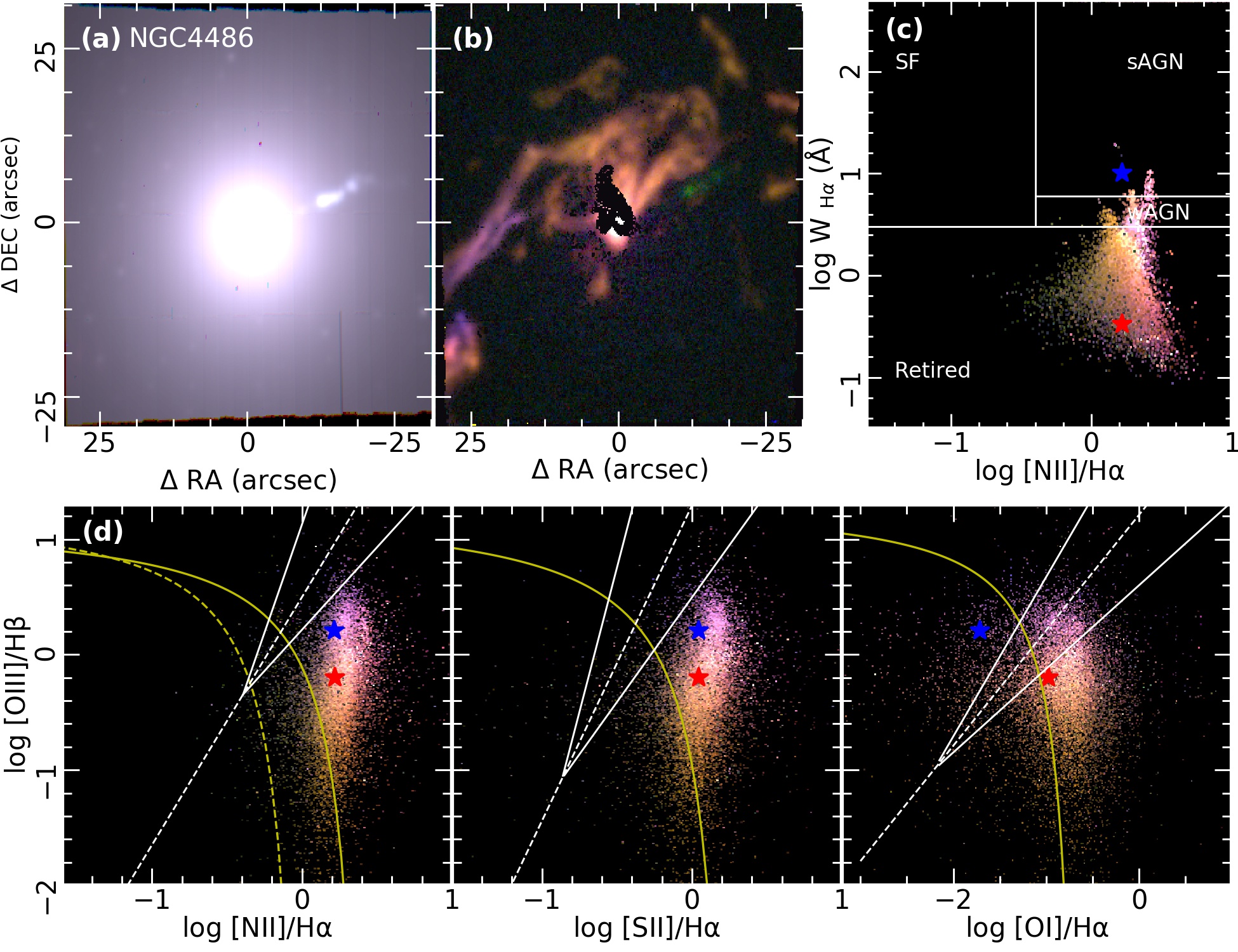}
\caption{Similar figure as Fig. \ref{fig:bpt_map_SF1}, for galaxies NGC4643 and NGC4486}
\label{fig:bpt_map_SF2}
\end{figure*}

\begin{figure}
\includegraphics[width=8cm]{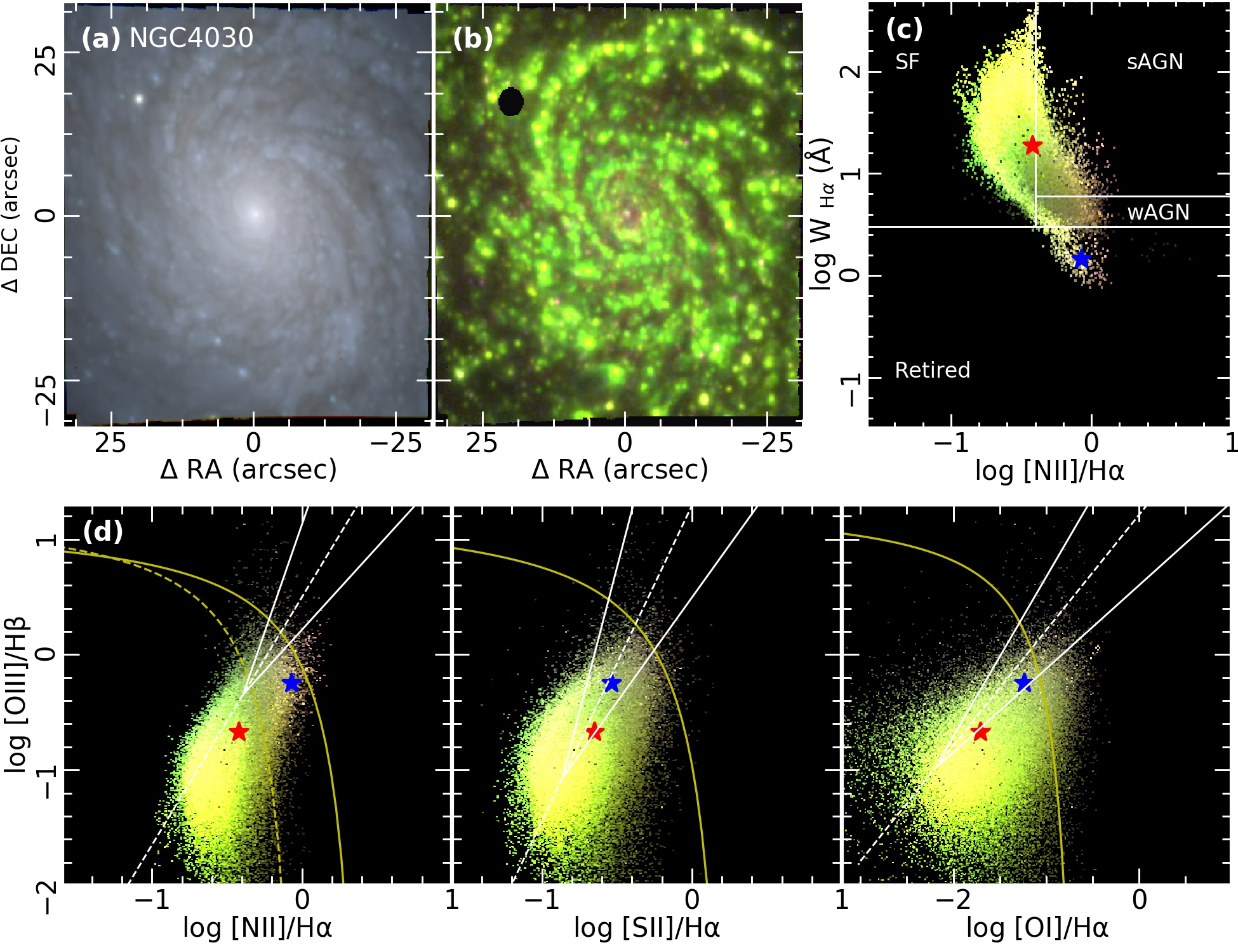}
  \caption{Similar figure as Fig. \ref{fig:bpt_map_SF1}, for galaxy NGC4030.}
\label{fig:bpt_map_SF3}
\end{figure}

In this section we explore the spatial distribution of the ionized gas
and their line ratios in some prototype galaxies included in our
galaxy sample. The main aim of this section is to reinforce
the results highlighted in the previous section and to demonstrate how
the shape/morphology and general spatial distribution of the ionized
gas may help to disentangle among different ionizing sources beyond
the use of just diagnostic diagrams (and EWs). We adopted just 
data taken with the best spatial resolution (MUSE data), although 
similar conclusions could be extracted from other datasets 
(strongly affected by resolution effects, in some cases). 
Figures \ref{fig:bpt_map_SF1}, \ref{fig:bpt_map_SF2} and
\ref{fig:bpt_map_SF3} show, for each of the considered galaxies, 
in the top-left panel, a true color continuum image.
This image is reconstructed from the datacubes by convolving 
the individual spectra
at each spaxel with the response curve of the $g$, $r$ and $i$-band
filters. Then, the three images are combined into one, 
with each filter corresponding to the blue, green and red color
respectively. In addition it shows, in the top-middle panel, a color
image created using the [\ION{O}{iii}] (blue), H$\alpha$ (green) and
[\ION{N}{ii}] (red) emission line maps extracted from the 
{\it gas-pure} datacube, as described in Sec. \ref{sec:ana_gas}. 
These two images (continuum and emission line one) allow us to 
explore the distribution of the ionized gas structures across 
the optical extent of galaxies, and its association with the 
different morphological sub-structures (such as bulges, disks, 
arms, bars..). In addition to these two maps we include, for 
each galaxy, four different diagnostic diagrams, including 
the ones shown in Fig. \ref{fig:BPT_cen}, and the WHAM diagram
\citep[][]{cid-fernandes10}, that compares the \nii\ ratio with the
EW(H$\alpha$). Thus, this figure takes into account the main 
conclusion of Section \ref{sec:complex} in the classification of 
the ionizing sources, mitigating the segregation problems by 
considering the EW(H$\alpha$) in addition to the classical 
diagnostic diagrams.

Each pixel shown in the top-middle image is mapped with the same color
in the four diagnostic diagrams, showing clearly the association
between the loci in those diagrams and the spatial distribution of the
ionized gas.  Figure \ref{fig:bpt_map_SF1} shows four examples of 
galaxies with a considerable number of \hii\ regions, that 
dominate the ionization across the galaxy disk. Those are seen as
clumpy/peaked ionized structures, almost circular, at the 
spatial resolution of these data ($\sim$0.1-0.5 kpc), as described 
by \citet{laura18}. They are clearly distinguished in the 
emission-line maps, tracing the spiral arm structure. Furthermore, 
they are mostly located in the lower-left region of the diagnostic 
diagrams, forming an arc where the classical \hii\ regions are 
found \citep[e.g.][]{osterbrock89}. They are easily identified in 
the case of the four considered galaxies (MCG-01-04-025, UGC\,1395, ESO\,0298-28, and IC1657).

In the case of MCG-01-04-025 all the ionization seems to be 
produced directly by \hii\ regions (clumpy structures), with a 
possible component of DIG (not clumpy, smooth distribution). In 
this galaxy DIG is most probably due to photon leaking from those
regions too, being the most frequent or dominant ionizing source 
for the diffuse gas in late-type spirals as reported in the 
literature \citep[e.g.][]{zur00,rela12}. The color change
in the ionized gas map for the \hii\ regions from the center 
(more green) to the outer parts (blueish),
reflects both the well known negative abundance gradient in these
galaxies \citep[e.g.][]{sear71,vila92,sanchez14,laura18}, and 
most probably a positive gradient in the ionization parameter
\citep[e.g.][]{sanchez12b,sanchez15}. 

The other three galaxies present additional ionizing conditions in their
central regions. In the case of UGC\,1395, the almost point-like, 
strong and hard ionization, in the very center of the
galaxy is a clear indication of the presence of an AGN. Indeed, 
the line ratios in this central region are located well above the 
K01 curve, with an EW(H$\alpha$) that in most of the cases is 
well above the 6\AA\ cut proposed by \citet{cid-fernandes10} 
for strong AGN. Despite the clear presence of an AGN, we should 
notice that a strict cut in the EW is not fully valid for this 
resolved spectroscopic data. At the edge of the distribution 
dominated by the AGN the EW drops below 6\AA\ and even 3\AA, 
just because the PSF size and/or the strong radial decline 
expected for this kind of ionization \citep{singh13,papa13}. 
Thus, the inclusion of the EW(H$\alpha$) helps for the 
discrimination of the nature of the ionization, but without 
spatial information it may provide an incomplete picture.

In the central region of the next example galaxy, ESO\,0298-28, 
a hard ionization is clearly present as well. However, contrary to the 
previous case, although the line ratios are very similar, both 
the EW(H$\alpha$) \citep[clearly below the 3\AA\ cut proposed 
by][for retired galaxies]{cid-fernandes10} and
the spatial distribution (smoother, following the stellar 
light distribution), indicate that the ionized gas has
a completely different nature in the central regions of this galaxy. 
We suggest that this is a good example of diffuse ionized gas 
associated with ionization by hot evolved stars. 
Different theoretical explorations have demonstrated that those stars can 
produce the required ionizing photons to explain the observed ionized gas 
\citep[e.g.][]{binette94,flor11}. The exploration of the properties of the 
underlying stellar population and their compatibility with the observed ionized
gas properties (i.e., the fraction of young and old stars able to ionize the 
gas), is becoming an important tool to identify this ionizing source 
\citep{gomes16,mori16,espi20}. Based on this kind of analysis, it is expected 
that these specific ionizing sources are ubiquitous in galaxies, 
albeit more evident in structures associated with old stellar populations
\citep[e.g.][]{sign13,belfiore17a}. Indeed, stars in galaxies were formed
mostly a long time ago \citep{panter07,perez13}, meaning that old stars, the 
progenitor population of hot evolved stars are available everywhere. 
However, it is obvious that this ionization is observed most frequently 
in the absence of \hii\ regions, as its characteristic line ratios would
otherwise be swamped by the more luminous ionizing sources. 

Even photons leaked from \hii\ regions may blur the signature of 
ionization by hot evolved stars, that is in general very weak, with a 
typical EW(H$\alpha$)$\sim$1\AA\ \citep[e.g.][as indicated before]{binette94}.
This ionization is also difficult to distinguish from a weak AGN, and 
in general it is not feasible to fully discard the presence of those 
faint central sources. However, the spatial association with the 
stellar continuum and the lack of a central peak (although weak) 
in both the flux intensity and EWs of H$\alpha$ is a guidance to 
discard (or at least not confirm) the presence of an AGN. The 
selection of AGN candidates in optical spectroscopic surveys without
considering the EW(H$\alpha$) is a general mistake, that was discussed 
in detail in the literature \citep[e.g.][]{cid-fernandes10}, but it 
is still not fully abandoned by the community.

The last galaxy shown in Fig. \ref{fig:bpt_map_SF2}, IC\,1657, is 
a disk galaxy (Sab) that has been classified as a Seyfert-2 
based on its emission line ratios \citep[e.g.][]{gu06}. It 
shows strong X-ray emission, a hallmark of the presence of 
nuclear activity. However, its X-ray-to-IR properties are 
somehow atypical, indicating a heavily obscured AGN \citep{lanz19}. 
The observed ionization throughout the FoV of the current 
data was explored in detail by \citet{carlos20}. They demonstrate 
the presence of an outflow at galactic scales that produces shock
ionization in a bi-conical structure emanating from the center 
of this galaxy. Fig. \ref{fig:bpt_map_SF2} shows this structure, 
as a {\it pinky} triangle (in projection), over-imposed on top of the 
ionization associated with \hii\ regions located in the heavily 
inclined disk of this galaxy (clumpy ionized regions, greenish 
and blueish). Contrary to previous claims,
we consider that the presence of an AGN cannot be fully confirmed 
or discarded by the optical data. It is true
that the central ionization is compatible with the presence of 
a nuclear source: it presents a hard ionization, with line ratios 
above the K01 curve and an EW(H$\alpha$) larger (but only 
marginally larger) than 6\AA. If a single fiber observation 
was taken on this central region, this galaxy would be clearly
classified as an AGN candidate. However, there is a lack of such
ionization outside the very central region, and even this one is 
clearly associated with the cone defining the shock ionization
associated with the outflow. Everywhere else, the ionization is
dominated by \hii\ regions, as indicated before. Indeed, the line 
ratios in the very center are at the edge of the K01 curve (for 
two of the diagnostic diagrams) and below it for one of them 
(the one involving the [\ION{O}{i}]/H$\alpha$ ratio. Those line 
ratios could be a consequence of the mix between shock ionization 
and the underlying ionization due to young hot stars. As a matter of fact, 
the conclusion by \citet{carlos20} was that this galaxy hosts 
an outflow most probably due to strong star-formation 
activity in the central regions. 

Independently of the final conclusion on the presence or not 
of an AGN, the observed line ratios across the central region 
are fully compatible with those usually considered as evidence 
of an AGN, and that would most probably be the conclusion from
single aperture spectroscopic data. In reality, the ionization 
structure of this galaxy is far more complex, 
showing (i) clear evidence of strong SF activity across its 
entire disk and towards the very center, (ii) a conical 
outflow, and (iii) maybe (or not) the presence of an AGN. 
Lacking spatially resolved information, despite the combined 
use of diagnostic diagrams and EW(H$\alpha$), the description 
of the ionizing source would be limited and miss-leading.

Figure \ref{fig:bpt_map_SF2} presents similar plots for two more 
galaxies: NGC\,4643 and the well known NGC\,4486 (M87). The first
galaxy,  NGC\,4643, lacks any trace of ionization associated with 
SF activity. No evident \hii\ region is detected in the emission 
line maps, which show no greenish clumpy ionized structure that is
evident in the four previously explored galaxies. NGC4643 is an 
S0 ring galaxy observed by MUSE as part of the TIMER survey
\citep{gado19}, it was also observed within the Atlas3D survey
\citep{cappellari11}. It presents a strong bar, and there is no 
evidence of SF in the optical images \citep[although there are 
known cases of SF in the outer regions of early-type
galaxies][]{gomes16}. If there is remnant SF activity, the 
\hii\ regions are not observed, implying that they would be less 
luminous and smaller than typical \hii\ regions in spiral galaxies. 
This would imply a strong variability in the H$\alpha$ luminosity
function of \hii\ regions, that so far is not observed
\citep[e.g.][]{brad2006}. Thus, all evidence indicates that there 
is no SF activity in this galaxy. Therefore, the observed 
ionization is most probably due to ionization by hot evolved 
stars: it is diffuse, following the stellar continuum, hard 
on average, and with low EW(H$\alpha$). However, low velocity 
shocks cannot be excluded \citep[e.g.,][]{dopita96}, although 
they would be expected more in the presence of weak AGN outflows 
or cooling flows in elliptical galaxies in cluster cores  
\citep[e.g.][]{Balmaverde2018,roy18,Olivares2019,carlos20}. 
Additionally, slow shocks are expected to show a filamentary 
structure and there is not reason their flux distributions should 
follow the stellar continuum emission 
\citep[and kinematics, e.g.~][]{kehrig12,lin17,cheung18}. Therefore, we 
consider this object as a clear candidate of DIG due to HOLMES 
or post-AGB stars. Nevertheless, not all spatial regions are 
located above the K01 curves in the diagnostic diagrams, 
although the ionization is on average clearly harder than for
ionization associated with \hii\ regions. Indeed, in the 
classical BPT diagram half of the observed regions are 
located in the so-called {\it intermediate} region between 
the K01 and K03 demarcation lines. This is a clear example 
of a possible misleading use of the diagnostic diagrams. 
Without considering the EW(H$\alpha$) this galaxy would be 
classified as an AGN (and in fact in the literature has been 
reported as a weak AGN or LINER). Moreover, a considerable 
fraction of the region covered by the FoV of the IFU data 
would be classified as showing {\it mixed} ionization, 
suggesting the presence of clearly unobserved SF activity. 

The diversity of ionization conditions observed in the previously 
explored galaxies can only be appropriately explored when using the 
diagnostic diagrams together with the spatial shape of the 
ionized structures. A similar situation is observed in 
NGC\,4486 (M87). This extremely massive galaxy in the center 
of the Virgo-A cluster hosts a super-massive black-hole 
without any doubt \citep[as recently demonstrated by the ][]{M87_EHT}. Known to be a 
radio-galaxy for decades \citep[e.g.][]{meis96}, it was one 
of the optically detected counterparts of a radio-jet 
\citep[indeed, the optical emission was reported in the 
first decades of the 20th century, e.g.][]{curt1918}. 
The counter-part is so strong that it is clearly seen in 
the continuum emission color-maps in the upper-left panel 
of  Fig.~\ref{fig:bpt_map_SF2}. It is also appreciated in 
the emission line maps as a series of greenish knots placed 
along the continuum counterpart, as already described in 
the literature \citep[e.g.][]{jarvis90}. More interesting 
to us is the filamentary emission structure observed across the 
north-eastern half of the MUSE FoV. This ionization was previously 
reported as a disky ionized gas structure based on narrow-band 
images by \citet{ford1994}. However, neither the shape of 
the ionization (filamentary, not clumpy) nor the spatial 
distribution (not following the shape of a disk), neither 
the distribution across the diagnostic diagrams, together 
with the distorted kinematics \citep{carlos20}, support 
this interpretation. The line ratios and the EW(H$\alpha$) 
indicate that the ionization is most probably due to either 
hot evolved stars or shocks of moderate velocity. However, 
considering the morphology, we are more inclined to suggest 
that this ionization is due to shocks. The nature of this 
gas is clearly under debate. However, recent results suggest 
that galaxies in the centers of clusters may present ionized 
gas originating from cooling flows that may be connected with cluster-wide 
flows \citep[e.g.,][]{Balmaverde2018,Olivares2019}. Remnants 
of past wet mergers may also be an alternative origin of that 
gas. In the particular case of M87 this inflow of gas could 
be the feeding mechanism of the AGN. Whatever is the ultimate 
origin of the gas, it is clear that the presence of the AGN 
is not easily uncovered by the optical line ratios, and the 
properties of the ionized gas can easily lead to confusion with 
post-AGB ionization. Furthermore, a fraction of the line 
ratios are located below the K01 curve, and like in the case 
of NGC\,4643, SF that may induce a {\it mixed} ionization is 
not observed. Like in the previous case, the complexity of 
the ionization would be impossible to uncover without a detailed 
exploration of the shape, distribution, and location in the 
different diagnostic diagrams, together with the use of the
EW(H$\alpha$) and the comparison with the spatial distribution 
of the continuum emission (either stellar or the radio-jet). 
Additional information, like the one provided by the gas 
and stellar kinematics (and the comparison between them), 
and an analysis of the velocity dispersion and asymmetry 
of the lines helps disentangle the real nature of such 
ionizing sources \citep{Agostino2019,carlos20}.

Figure \ref{fig:bpt_map_SF3} presents the last showcase of our
illustration of the complexity of deriving the nature of the 
ionization in galaxies. NGC\,4030 is a grand design Sbc galaxy 
observed almost face-on. Its disk shows hundreds of \hii\ regions 
easily identified in the emission-line image (middle-top panel 
in Fig. \ref{fig:bpt_map_SF3}) as green clumpy ionized regions. 
As expected, most of the ionization is located below the K01 
demarcation line (with a large fraction below the K03). 
However, contrary to previous examples discussed before, as
MCG-01-04-025, UGC\,1395 and ESO298-28, the distribution does 
not follow the classical location of \hii\ regions. Certainly, 
a substantial fraction of the line ratios are located in 
between the K03 and K01 region, an even above the K01 curve 
in the classical BPT diagram. Following the usual and broadly 
accepted interpretation of this diagram, this shift should be 
due to {\it mixing} of the ionization produced by the 
overlap of the \hii\ regions with other ionizing source 
that pollute those line ratios \citep[either by a central 
AGN or DIG due to hot evolved stars, e.g.][]{davies16,lacerda18}.
However, a more detailed exploration indicates that the polluting 
sources correspond to clumpy ionized structures, morphologically 
similar to \hii\ regions, but with line ratios corresponding 
to the presence of a harder ionization. Those regions clearly
correspond to those regions with higher [\ION{N}{ii}]/H$\alpha$ 
line ratios, previously detected in the central regions of 
galaxies \citep{kennicutt89,sanchez12b}, usually refereed as 
Nitrogen enhanced regions \citep{ho97,sanchez15}. Recent 
explorations have shown that they are compatible with 
supernova remnants combined in some cases with ionization 
by young hot stars (Cid-Fernandes et al., submitted). 
{\bf Thus, again, the intermediate region can be populated 
without invoking an AGN to explain the mix of ionization.}

In summary, gas in galaxies could be ionized by many different
physical processes, like in the case of our own galaxy. Therefore,
associate a single ionization to the observed emission across an
entire galaxy is a first order approximation that can lead to
considerable errors, in particular in the derivation of the physical
parameters of the ionized gas, like dust attenuation or 
oxygen abundance (due to the non linearity of the
combination of line ratios). Even more, our ability to distinguish
between the different ionization conditions should not rely on the
classical diagnostic diagrams only. They should be combined with
morphological and kinematic information about the emission line
structures and complemented with the study of the properties of the
underlying stellar populations to determine their potential ionizing
sources. This highlights the fundamental importance of integral field
spectroscopy in the study of the ionized gas in extragalactic sources.
However, these data are also limited by their spectral and in particular
their spatial resolution, which can produce a mix of ionization
\citep[e.g.][]{davies16} and limit our understanding of the derived
properties \citep[e.g.][]{rupke10,mast14}.

\subsection{Ionized gas: A practical classification scheme}
\label{sec:class}

\begin{figure*}
\includegraphics[width=17cm, clip, trim=10 30 0 0]{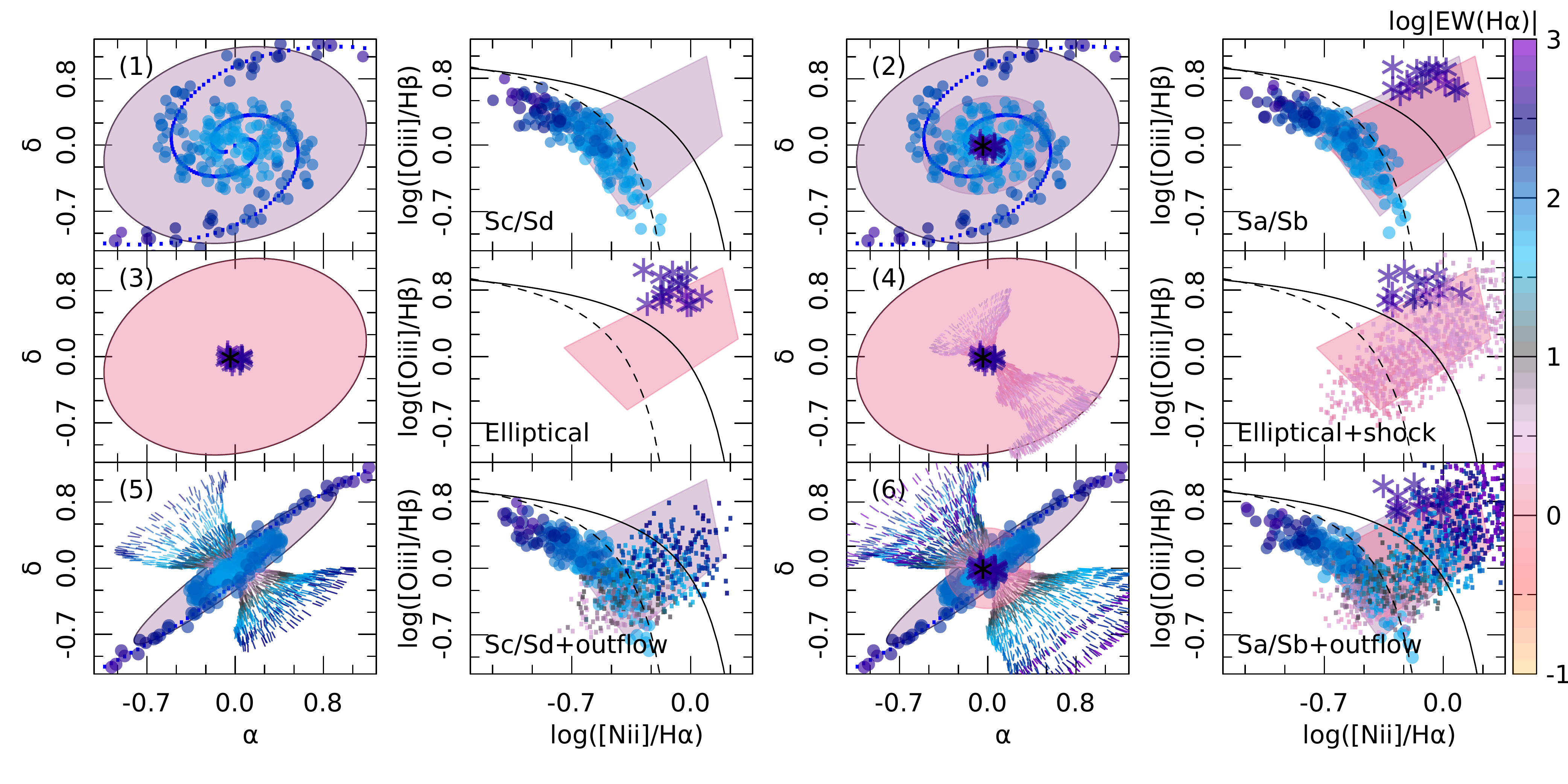}
  \caption{Scheme of the main ionizing conditions typically observed in galaxies of different morphological types, including both the distribution across the optical extent (left) and the classical BPT diagram (right): (1) late-type spirals (Sc/Sd), without a prominent (or without) bulge. Their main morphological feature is a thin disk (ellipse) together with the spiral arms (dotted line). The main components of the ionized ISM are the \hii\ regions distributed mostly in the disk and following more or less the spiral structure (blue solid-circles), together with some diffuse ionized gas that in this case is dominated by photon leaked from those regions (represented as a pale pinky ellipse). (2) Early-type spirals (Sa/Sb), with a prominent and well defined bulge, in addition to the disk and spiral arms. They present \hii\ regions and diffuse ionized gas in the disk too, like late spirals. However, the DIG is ionized  by a mix of photon leaking and ionization by hot evolved stars, mostly associated with the presence of old stellar populations (i.e., more clearly observed in the bulge, shown as pink central ellipse). Some also show ionization due to AGN, mostly located in the central regions (violet stars). (3) Early-type galaxies (E/S0), with very weak or no disk.  Ionization across their optical extent is dominated by hot evolved stars (pink ellipse), that ionize the diffuse gas, with the possible presence of a central AGN. (4) Early-type galaxies may present shock-ionized gas, observed mostly in central galaxies in clusters, radio-galaxies, and weak AGN \update{\citep[e.g., like in the case of Geyser galaxies,][]{geysers18}}. The structure of these relatively low-velocity shocks is filamentary, with a velocity dispersions that is slightly larger than the one observed in the diffuse ionized gas), and with kinematics largely decoupled from those of the stellar populations in these galaxies. (5) Late-type spirals may present galaxy scale shock ionization associated with galactic winds, that can usually be seen in edge-on or highly inclined galaxies. They present a patchy filamentary distribution, following a conical or biconical structure, emanating from the central regions and escaping to high altitudes with respect to the disk height in some case. (6) Early-type spirals may present the same kind of galactic winds, but in this case they can be produced by the kinetic energy injected by an AGN too. When observed at high inclination the ionization in the vertical direction can be associated with old stellar populations either in the bulge or a thick disk (not shown in the figure). This latter ionization is less patchy, more diffuse. In all panels, colors represent the typical EW(H$\alpha$) associated with the different ionization conditions.}
\label{fig:bpt_sch}
\end{figure*}

Based on the results outlined before, we propose a new procedure 
to classify the components of the ionized ISM. To illustrate it 
we present in Figure \ref{fig:bpt_sch} a scheme of the 
distribution of the most dominant ionization conditions for 
different galaxy types. We consider this classification procedure 
is valid for spatially resolved spectroscopic data between a 
few hundred pc to a few kpc scales. However, it may not be 
valid for smaller scales, where the ionization structure is 
resolved for the different conditions described in the explored data.

\begin{itemize}
\item {\bf A star-forming region} is observed as (i) a clumpy/peaked
region (clustered) in the ionized gas maps of a galaxy with (ii) line
ratios below the \citep{kewley01} demarcation line in at least 
one of the classical diagnostic diagrams 
shown in Fig. \ref{fig:BPT_cen} (and Fig. \ref{fig:bpt_map_SF1},
\ref{fig:bpt_map_SF2}, and \ref{fig:bpt_map_SF3}), with (iii)
EW(H$\alpha$) above 6\AA\ (see Fig. \ref{fig:bpt_sch}, in 
particular panels 1 and 2), and with (iv) a fraction of light 
assigned to young stars ($Age<100$Myr), in the $V-$band of at 
least 4-10\%. This definition is valid for Giant \hii\ regions 
and \hii\ region clusters at the indicated resolution. Small 
\hii\ regions, like the Milky Way's Orion nebula, would be 
spatially and spectroscopically diluted, being confused with 
the diffuse gas. In other words, this classification scheme 
guarantees that the selected regions are indeed ionized by young
stars, but it cannot guarantee a complete selection of all 
regions with the presence of young ionizing stars. Higher 
spatial resolution data, covering bluer wavelength ranges 
(UV) would improve this selection process.

\item {\bf An AGN ionized region} is observed as (i) a central ionized
region (almost unresolved to the considered resolution), well 
above the intensity of the diffuse ionized gas (Fig.
\ref{fig:bpt_sch}, panel 2 and 3, and  Fig. \ref{fig:bpt_map_SF1},
top-right panel). (ii) Its emission line ratios are above the 
K01 demarcation lines in the three diagrams discussed before, 
and (iii) its EW(H$\alpha$) is above 3\AA\ \citep[6\AA for strong
AGN][]{cid-fernandes10}. Below that limit it is not possible 
to determine if the ionization is due to an AGN or to other 
processes (post-AGBs/HOLMES, low-velocity shocks, photon leaked 
from \hii\ regions). They (iv) present a decrease of the considered
line ratios with respect to the central values in the galaxy, and 
(v) show a steep decay in the flux intensity, which is never 
shallower than a $r^{-2}$ distribution \citep[e.g.,][]{sign13}. 

\item {\bf Diffuse gas ionized by hot, evolved stars} is seen as (i) 
a smooth (not clumpy or filamentary) ionized structure that 
follows the light distribution 
of the old stellar populations in galaxies. (ii) The EW(H$\alpha$) 
in these regions is clearly below 3\AA , with (iii) the fraction 
of young stars never larger than 4\%. (iv) The distributions in the
diagnostic diagrams cover a wide range of values from the 
LINER-like area towards the range covered by metal-rich \hii\ 
regions (see Fig. \ref{fig:bpt_sch}, panel 3). (v) This component 
is observed in galaxies with old stellar populations (massive, 
early types, e.g., Fig. \ref{fig:bpt_map_SF2}, left panel) or 
regions in galaxies with the same characteristics (bulges, 
Fig.~\ref{fig:bpt_map_SF1}, bottom-left panel). (vi) The 
kinematics of this ionized gas does not deviate significantly 
from that of the old stellar population in the galaxy. Two caveats: 
In high spatial resolution data (10-100 pc) this component may 
present, in some cases, clumpy structures associated with individual 
source. This is not visible at the resolutions considered in this 
review. Also, as indicated before, the signal from small-size \hii\ regions is 
diluted at these resolutions by this ubiquitous DIG emission, 
what could alter the observed line ratios. 

\item {\bf Diffuse gas due to photon-leaking by \hii\ regions} 
is observed as (i) a smooth ionized structure present in galaxies 
with young stellar populations (in general, low mass and 
late-type galaxies) or regions in galaxies with the same
characteristics (disks, e.g. Fig.~\ref{fig:bpt_sch}, panel 1 
and Fig.~\ref{fig:bpt_map_SF1}, greenish diffuse ionization shown in 
the emission line image located at the disk of the four galaxies).
This component should (ii) have a fraction of young stars never 
larger than 4\% within the same resolution element and at the
considered resolution \citep[e.g.][]{espi20}. (iii) Its 
kinematics are not fundamentally different from those of the disk.
In the diagnostic diagrams this component may present varying 
line ratios, 
from very similar to \hii\ regions to significantly different from 
them \citep[e.g.][]{weil2018}. It also covers a wide range of
EW(H$\alpha$). Caveats: In high spatial resolution data (10-100 pc) 
it may present some shells or bubble-like structures, not visible 
at the considered resolutions. In late spirals (Sc/Sd), it 
may be the dominant or at least a large fraction of the DIG
\citep[e.g.][]{rela12}. In this case, this component should be
included in the photon budget to derive the SFR in galaxies
\citep[e.g.][]{zur00}. 

\item {\bf A high-velocity shock ionized region} is seen as (i) a
filamentary or bi-conical ionized gas structure with fluxes (and 
EWs) well above those of the diffuse ionized gas. Its emission 
line ratios cover a wide range of values. In general, (ii) the 
line ratios are above the K01 demarcation line in 
the three diagrams, and (iii) in most cases, the lines are 
asymmetrical, with (iv) a clear increase of the line 
ratios (in particular [\ION{O}{i}/H$\alpha$]) with the 
velocity dispersion and the distance from the source of the outflow. In addition (v) they have an EW(H$\alpha$) above 
3\AA. (vi) This component is usually located in the central 
regions (or emanating from this region), for both star-formation driven (Fig. \ref{fig:bpt_sch}, 
panel 5, and Fig. \ref{fig:bpt_map_SF1}, bottom right panel) 
and/or AGN driven outflows (Fig. \ref{fig:bpt_sch}, panel 6). 
Caveats: The line ratios can spread from the area usually 
associated with AGN ionization (top-right region of the diagram), 
to the area covered by \hii\ regions (i.e., below both 
demarcation lines, and spread through the so-called 
{\it mixed} region). A demarcation line has been proposed to 
separate between SF and AGN driven outflows 
\citep[][]{sharp10}, although this should be tested using
large/statistically significant samples \citep[e.g.~][]{carlos20}.

\item {\bf A low-velocity shock ionized region} shares many of 
the characteristics of the DIG by old, evolved stars, and indeed 
it is considered by different authors as part of this diffuse
gas component \citep[e.g.][]{dopita96,monreal10}. However, 
these regions present (i) a clear filamentary structure (Fig.~ 
\ref{fig:bpt_sch}, panel 4), and (ii) a velocity distribution 
not following the general rotational pattern of the galaxy. 
The recently named \update{{\it Geyser-galaxies} \citep{cheung18,geysers18}} or the cooling flows 
observed in ellipticals (in clusters in general) present most 
probably shock ionization corresponding to this type (e.g., Fig.~
\ref{fig:bpt_map_SF2}, right panel).

\item {\bf Supernova remnants} are less frequent than the previous
types, they are observed (i) at this resolution as clumpy/ionized
regions, similar in shape to the Giant \hii\ regions/star-forming 
areas discussed before. However, they (ii) are ionised by a harder 
ionization spectrum and therefore show higher values for the 
line ratios shown in the classical diagnostic diagrams. 
Caveats: At the current resolution this component is hard to see
without considerable contamination by adjacent or superposed 
(through the line of sight) \hii\ regions or photon leaking 
ionized DIG. For this reason they cover a wide range of line 
ratios, from the classical location of \hii\ regions towards the
intermediate region between the K03 and K01 regime. In the emission
line maps presented in this review, this component may appear as
reddish clumpy structures (e.g., Fig. \ref{fig:bpt_map_SF3}). In
general it is required to explore other emission lines, frequently
associated with SN and SNR \citep[e.g.][]{snr_elines}, to detect 
them (e.g., Cid-Fernandes et al. submitted).

\end{itemize}

We present this classification scheme as a practical tool, with the 
hope to improve on current classification methods. However, it is 
by construction incomplete and will require future revisions 
based on the improvement of our understanding of the 
ionization conditions observed in galaxies.


\section{Global and resolved relations}
\label{sec:local}

One of the main results emerging from IFS-GS, as highlighted by
\citet{ARAA}, is that the global relations uncovered in the
exploration of extensive properties of galaxies present 
local/resolved counterparts that are valid at kiloparsec scales.
Among them, the most evident ones are:

\begin{itemize}

\item The star-formation main sequence \citep[SFMS,
e.g.~][]{brin04,renzini15}, that relates the SFR and the 
M$_*$ for SFGs, has a resolved version that relates the 
$\Sigma_{\rm SFR}$ and the $\Sigma_*$ for SFAs \citep[rSFMS, e.g.~][]
{ryder95,sanchez13,mariana16,hsieh17,mariana19}.

\item The Mass-Metallicity relation \citep[MZR][]{tremonti04}, 
that relates the central and/or characteristic gaseous Oxygen abundance,
12+log(O/H), of a galaxy with its M$_*$, is reflected in the 
relation found between the local (spatially resolved) 
Oxygen abundance and the mass surface density $\Sigma_*$ in
star-forming regions \citep[rMZR, e.g.][]{rosalesortega:2012,jkbb16}.

\item The scaling relation between the molecular gas mass and the
stellar one \citep[e.g.~][and references therein]{calette18},
corresponds to the recently reported relation between $\Sigma_{\rm gas}$
and $\Sigma_*$ \citep[e.g.][]{lin19,jkbb20}. For its similitude 
with the SFMS, we will refer to this relation as the {\it mass 
gas main sequence}, or MGMS, and to its resolved version as 
rMGMS, following \citet{lin19}.
  
\end{itemize}

In addition to all these extensive relations and their local/resolved
intensive counterparts, recent IFS-GS have allowed to explore in
detail well-known relations, like the Schmidt-Kennicutt relation
\citep[SK-law,][]{kennicutt98}. The SK-law was formulated as a 
relation between
intensive quantities, connecting the average $\Sigma_{\rm SFR}$ and the
average $\Sigma_{\rm gas}$ across the optical extent of
galaxies. However, in principle it was also a global relation, since it
relates characteristic properties of individual galaxies. Much 
work has gone into investigating the true nature of the SK-law, 
in particular concerning the spatial scales over which it holds 
and the phases of the neutral ISM which provide the tightest 
relation. A consensus has emerged in which the SK-law is tightest 
when relating SFR to molecular hydrogen H$_2$ over scales of a few 
hundred parsec \citep[e.g.~][]{bigiel08,bigiel11}. The advent 
of IFS-GS, in combination with spatially resolved explorations of the
gas content, has allowed to confirm that relation over statistically 
large samples at kiloparsec scales \citep[e.g.][]{bolatto17}. For 
nomenclature consistency we refer to the resolved SK-law as rSK-law.

It is important to note here that these relations are valid either 
for SFGs (global) or star-forming areas (SFAs, local/resolved). 
Retired galaxies (and retired areas) do not follow those relations, 
showing in general much lower values of the SFR ($\Sigma_{\rm SFR}$) 
and M$_{\rm gas}$ ($\Sigma_{\rm gas}$) for a fixed M$_*$ ($\Sigma_*$), 
and slightly lower values of SFR ($\Sigma_{\rm SFR}$) for a 
fixed M$_{\rm gas}$ ($\Sigma_{\rm gas}$). Whether those RGs 
(and RAs) follow well defined trends or are distributed over extended loci
(clouds) in each diagram, is under debate \citep[e.g.][]{hsieh17}. 
Regarding the MZR (rMZR), it is difficult to determine if they 
follow the same trends or not, since the gas-phase metallicity is very 
hard to obtain for RGs (RAs). In the few cases in which a globally 
retired galaxy presents a few SFAs \citep[e.g., ][]{gomes16a}, 
it seems that those regions have a slightly lower oxygen abundance 
than predicted from the relation \citep[][]{ARAA}. However, more 
statistics are required in this regards.

There are two possible ways to look at the evidence connecting 
the resolved/local relations with their global counterparts. 
First, concerning the MZR, theoretical investigations seems to imply that 
chemical evolution is not dominated by local processes only, but 
transport of gas and metals plays a significant role 
in the physical origin of the relation \citep[e.g.~][]{trayford19}. 
Similar arguments could be made invoking radial migration, accretion 
and merging when it comes to the rSFMS. However, the second way to 
look at the relations is purely empirically, to investigate whether 
they are at least interchangeable in terms of their intensive vs.~
extensive nature \citep[e.g.~][]{jkbb16, gao18, salmeida19}. 
Indeed, we will show in the 
next section that the extensive/global relations are in general a 
natural consequence of the intensive/resolved ones, while the reverse 
is not true. While this does not yet make up a proof of which physical 
processes shape those relations, it becomes probable 
that most processes surrounding star formation and metal enrichment 
in galaxies are pre-dominantly local, while global processes only perturb,
but do not fundamentally alter the resulting relations. 

In the following we investigate two topics: (i) we transform the global
extensive relations into intensive ones, by deriving the average surface
densities of the involved quantities \citep[following the exploration
of the SK-law by][]{kennicutt98}. Then we compare the intensive global
relation with the local/resolved one; and  (ii) we explain with simple
calculations why a resolved relation directly leads to an intensive
global one, following \citet{salmeida19}.

\begin{figure*}
  \includegraphics[width=8.4cm]{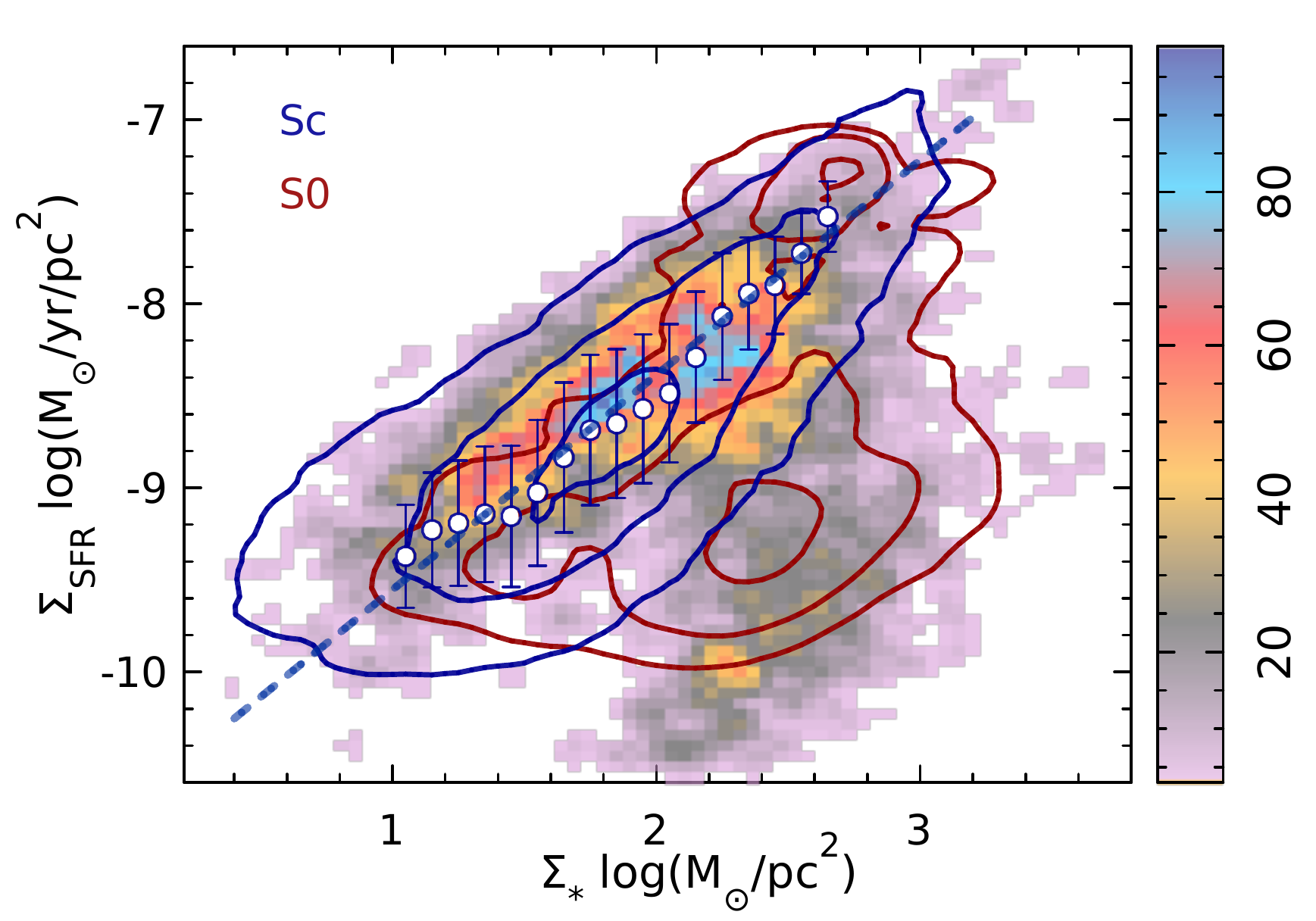}\includegraphics[width=8.5cm]{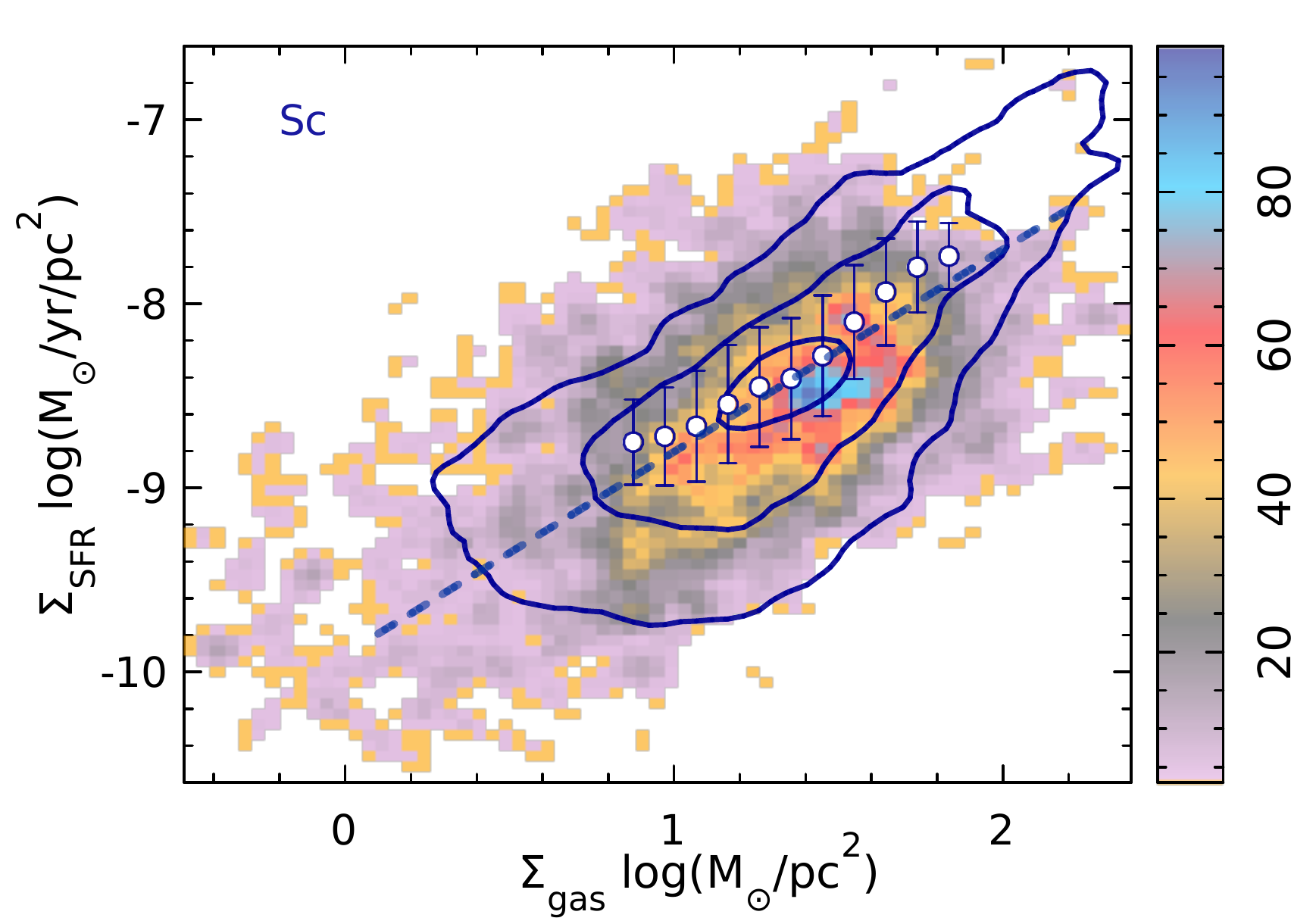}
\includegraphics[width=8.4cm]{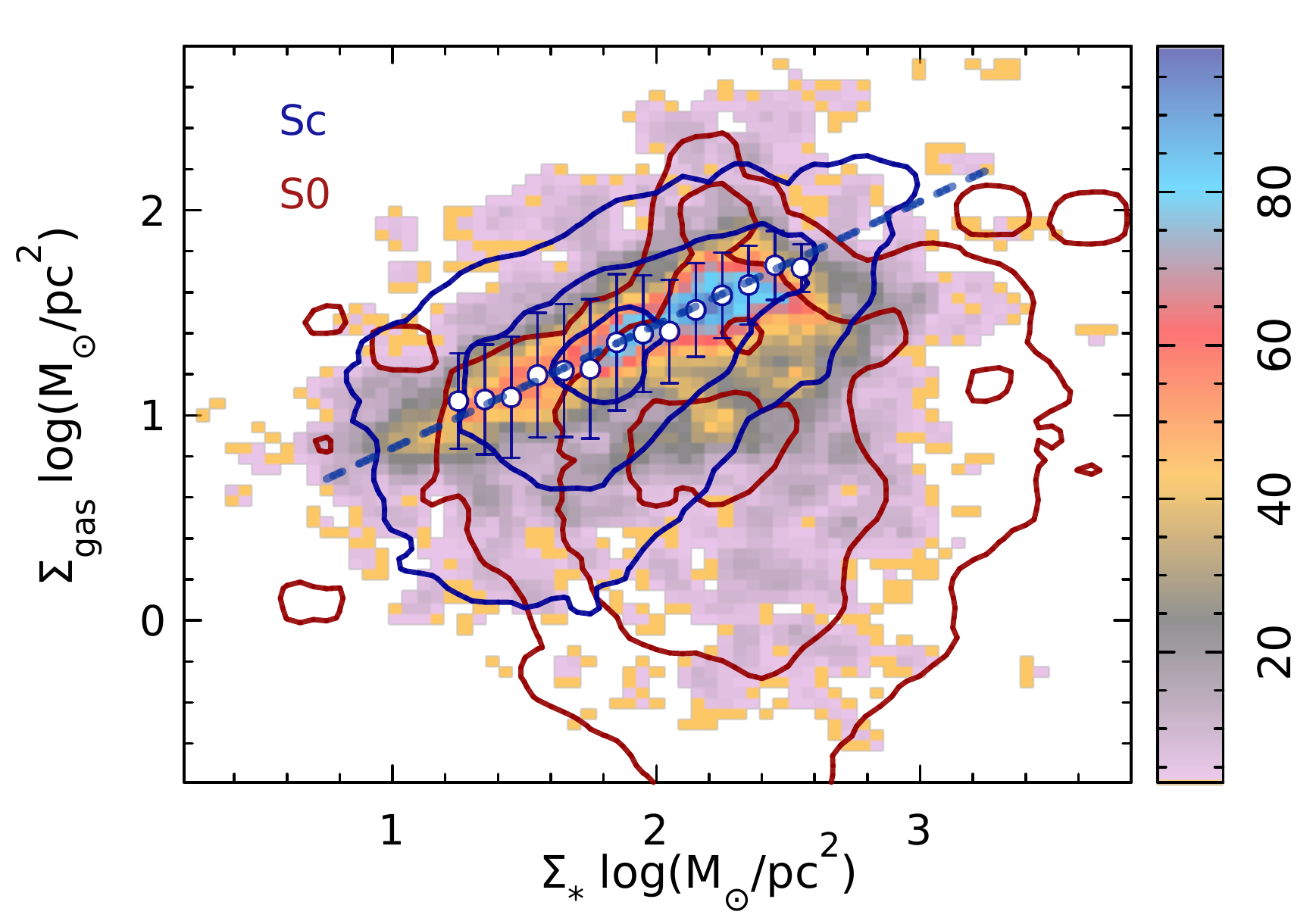}\includegraphics[width=8.5cm]{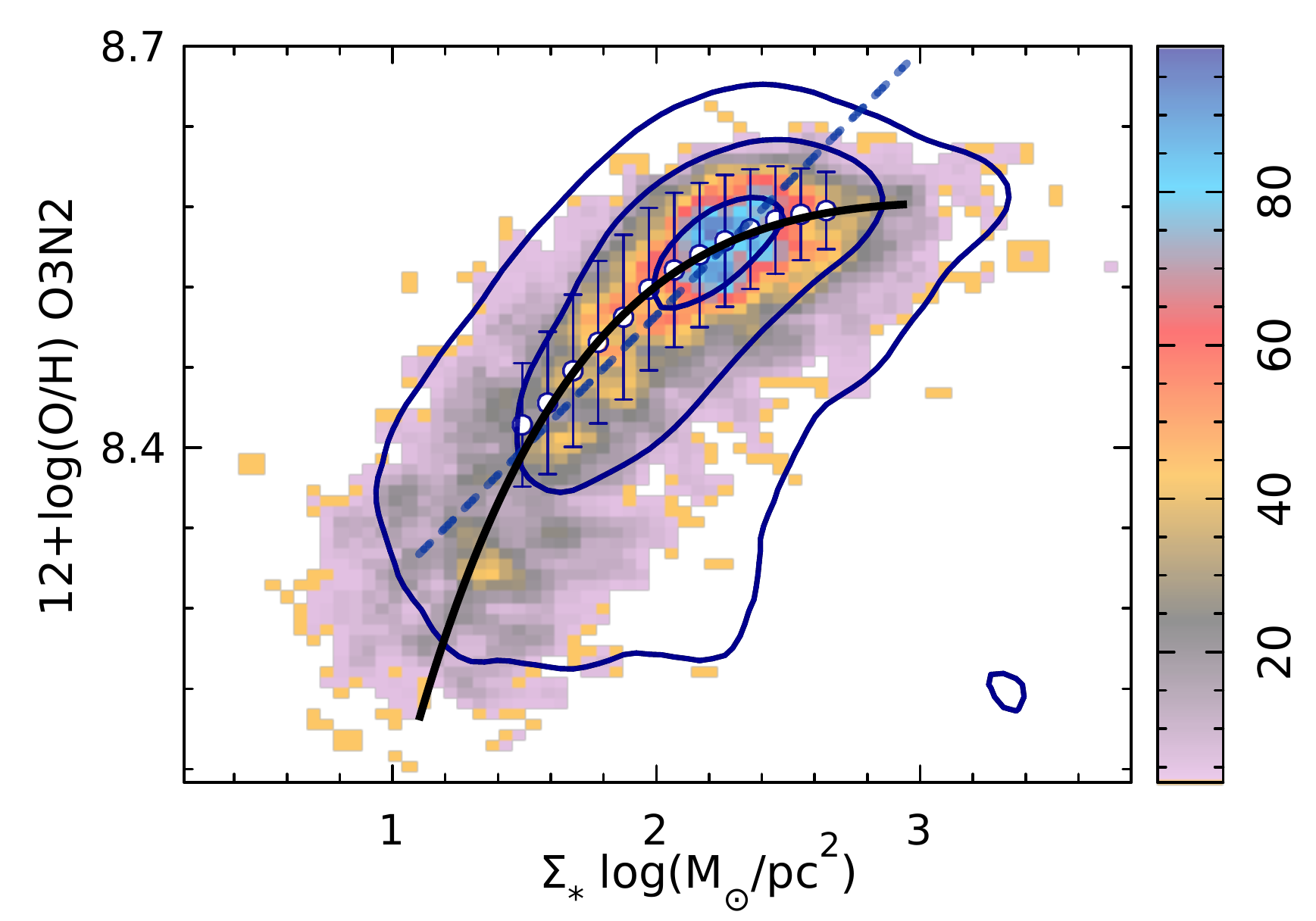}
\caption{Global and local (resolved) scaling relations compared. {\it Top-left panel:} number density distribution of galaxies in terms of mean $\Sigma_{SFR}$ against mean $\Sigma_{*}$ for entire galaxies within our compilation. The distributions are shown as color coded images, where the color bar indicates the cumulative distribution in fraction of total. In addition, we show the number density distribution of $\Sigma_{SFR}$ vs.~the $\Sigma_*$ for each individual spaxel of each individual galaxy as contour plots, for both the Sc (blue contours) and S0 (red contours) galaxies within the compilation. Each contour encircles 85\%, 50\% and 10\% of the sample, from the outermost to the innermost one. Solid white circles represent the average value of $\Sigma_{SFR}$ for bins of $\sim$0.1 dex in $\Sigma_*$, for galaxies of type Sc, and for those values encircled by the 85\% distribution contour, with errorbars corresponding to the standard deviation within each bin. The blue dashed line represents the best linear regression to these average points. {\it Top-right panel:} Similar plot for the galaxy averaged and spaxel-wise $\Sigma_{SFR}$ against $\Sigma_{gas}$ distribution. In this case, for the resolved properties we only show the contours corresponding to Sc galaxies. {\it Bottom-left panel:} Similar plot for the $\Sigma_{gas}$-$\Sigma_*$ diagram. {\it Bottom-right panel:} Similar plot for the distribution of Oxygen abundance vs.~$\Sigma_*$. In the last case, the contours correspond to the number density distribution of all the spaxels with measured Oxygen abundance for the full sample of compiled galaxies (black contours). Although in practice, it contains mostly late-type, star-forming galaxies. In this panel a best fitting curve is included as a black solid-line, following the functional form usually adopted to describe the relation between these two parameters, first proposed by \citet{sanchez13}.}
\label{fig:local}
\end{figure*}

\begin{table*}[t]
\begin{center}
\label{tab:local}
\caption[Log-log fitting to resolved relations]{Log-log fitting to resolved relations}
\begin{tabular}{llrrrr}\hline\hline
  \multicolumn{1}{c}{Relation} & \multicolumn{1}{c}{Reference} & \multicolumn{1}{c}{$\beta$} & \multicolumn{1}{c}{$\alpha$} &  \multicolumn{1}{c}{r$_c$} & \multicolumn{1}{c}{$\sigma$} \\   
\hline
rSFMS & This work &-10.35$\pm$0.03 & 0.98$\pm$0.02 & 0.96 & 0.17 \\
      & \citet{sanchez13}    &       & 0.66$\pm$0.18 & 0.66& \\
      & \citet{wuyts13}      & -8.4$^1$ & 0.95 &  & \\
      & \citet{mariana16}    & -10.19$\pm$0.33 & 0.72$\pm$0.04 & 0.63 & 0.16 \\
      & \citet{lin19}        & -11.68$\pm$0.11 & 1.19$\pm$0.01 &    & 0.25 \\
      & \citet{mariana19}    & -10.48$\pm$0.69 & 0.94$\pm$0.08 & 0.62 & 0.27 \\
\hline      
rMGMS& This work using $\Sigma_{gas}$& 0.24$\pm$0.04    & 0.60$\pm$0.02 & 0.93 & 0.08 \\
        & This work using $\Sigma'_{gas}$& -1.10$\pm$0.07   & 1.22$\pm$0.03 & 0.95 & 0.14 \\
        & \citet{lin19}        & -1.19$\pm$0.08   & 1.10$\pm$0.01 &    & 0.20 \\
        & \citet{jkbb20}       & -0.95            & 0.93          &    & 0.20 \\
\hline
rSK  & This work & -9.85$\pm$0.02   & 1.19$\pm$0.02 & 0.96 & 0.11 \\
     & Using $\Sigma'_{gas}$& -10.72$\pm$0.04   & 1.17$\pm$0.02 & 0.97 & 0.15 \\
     & \citet{bolatto17}    & -9.22            & 1.00          &      &      \\
     & \citet{lin19}        & -9.33$\pm$0.06   & 1.05$\pm$0.01 &      & 0.19 \\
\hline
rMZR & This work & 8.21$\pm$0.01    & 0.13$\pm$0.01 & 0.88 & 0.06 \\
\hline
\end{tabular}

Results for the log-log regressions between the different
parameters shown in Fig. \ref{fig:local}, with $\beta$ being
the zero-point of the relation, $\alpha$ the slope, r$_c$ the correlation
coefficient between the involved parameters and $\sigma$ the standard
deviation around the best fit relation. We include similar results 
extracted from the literature, shifted when needed to match the 
adopted units for the different quantities. We remind the reader that 
our surface density quantities are all expressed in units of $pc^2$ 
for the area. $(1)$ We should note that the \citet{wuyts13} results
are based on galaxies at $z\sim$1. Thus, the offset in $\beta$ 
reflects an evolution in the rSFMS similar to the one reported for 
the SFMS \citep[e.g.~][]{speagle14,sanchez18b}.
\end{center}
\end{table*}

\subsection{The intensive global relations}
\label{sec:intensive}

As indicated above, the global scaling relations involve extensive 
quantities (in most of the cases), relating, in many cases, the stellar 
mass (or gas mass) to other properties of galaxies, such as the SFR, 
Oxygen abundance, of both properties between themselves. As described 
before, those relations could be a pure consequence of a scaling 
between galaxies, since they involve pure extensive quantities: the 
larger and more massive a galaxy is, the higher are the values of any 
other extensive property, such as the SFR or the gas mass (at least 
for SFGs). The MZR is slightly different
in this regards, since it involves an extensive quantity (M$_*$) and
an intensive one (the Oxygen abundance). However, in many cases it is
ill-defined, since it is well known that the Oxygen abundance presents
a radial gradient in galaxies \citep[e.g.][and reference
therein]{sanchez14}. Furthermore, the Oxygen abundance is frequently 
estimated using variable physical apertures in galaxies, rather 
than a fixed aperture or a
characteristic radius \citep[see][for a recent discussion on the
topic]{sanchez19}. Thus, although the MZR involves an intensive
quantity, it is measured as an extensive one (i.e., not averaged across
the extension of galaxies). However, it is true that in this particular
case the simple scaling of the two parameters due to their extensive nature
cannot explain this relation \citep[see the discussion in the 
seminal article by][]{tremonti04}

We attempt to explore the existence of intensive global relations
(rather than extensive ones), by deriving the average surface
densities of the considered extensive parameters (M$_*$, M$_{\rm gas}$ and
SFR), and the characteristic value of the Oxygen abundance \citep[i.e.,
the value at the effective radius, following][]{sanchez13}. To that end 
we divide each extensive quantity by the effective area of each
galaxy, defined as the area within 2r$_e$, i.e.,
A$_e$=$\pi$(2r$_e$)$^2$. We choose this particular radius because
most of the IFS data explored in this review sample the galaxies up
to this galacto-centric distance. Therefore, most of the
reported extensive properties are actually aperture-limited values
corresponding to that radius. How the actual parameters were derived, 
both global and resolved, is described in detail in Sec. \ref{sec:analysis}.

We will demonstrate in the upcoming section that indeed the average 
intensive parameters derived as described before scale with the 
corresponding value at the effective radius. For this reason, 
the current results would be qualitatively similar to the ones 
found assuming a different scale-length for the galaxy. Finally, 
for the Oxygen abundance we fit its radial gradient, normalized to the
effective radius, within 0.5-2 r$_e$ with a simple linear profile
\citep[following][]{laura18}. Then, we derive the value at the
effective radius using the best fit profile for each galaxy.

Figure \ref{fig:local} shows the result of this experiment. In each
panel we show, in color-code, number density distributions of the
intensive global parameters defined before (and the characteristic 
one, in the case of the Oxygen abundance), including: (i) 
$\left< \Sigma_{\rm SFR}\right>$ versus $\left< \Sigma_{*}\right>$; 
(ii) $\left< \Sigma_{\rm SFR}\right>$ versus 
$\left< \Sigma_{\rm gas}\right>$; 
(iii) $\left< \Sigma_{\rm gas}\right>$ versus 
$\left< \Sigma_{*}\right>$ and finally 
(iv) 12+log(O/H)$_e$ against $\left< \Sigma_{*}\right>$. 
First we note that the intensive global distributions expressed in 
extensive quantities (color-coded data) follow the same trends 
as the extensive local ones (contours) reported in the literature, 
and in more detail the relations found for resolved quantities. 
The first panel ($\Sigma_{\rm SFR}$-$\Sigma_*$ diagram) corresponds 
to the SFR-M$_*$ diagram, and the distribution shows the two well 
known trends for SFGs (the SFMS) and the RGs (with a cloud of 
points well below the SFMS). 
The agreement between the global and local relations was already 
explored in detail by \citet{pan18} and \citet{mariana19}, 
showing in general good agreement. 

The second panel ($\Sigma_{\rm SFR}$-$\Sigma_{\rm gas}$ diagram) 
shows the well known SK-law for the analysed sample. Like in the 
case of the SFMS, the well known trend between the two parameters 
is visible, despite the fact that we are using a proxy to estimate 
the gas content, based on the dust attenuation \citep[][]{jkbb20}, 
as discussed in Sec. \ref{sec:ana_gas}. The third panel 
($\Sigma_{\rm gas}$-$\Sigma_{*}$ diagram) mimics the known relation
between the gas and the stellar mass, for SFGs, with a tail towards 
lower values of M$_{gas}$ ($\Sigma_{gas}$ in this case) for retired
galaxies \citep[e.g.][]{saint16,calette18,sanchez18,lacerda20}. 
Finally, the fourth panel shows the intensive version of the MZR 
diagram, thus, the distribution of the characteristic oxygen 
abundance of each galaxy (i.e., the value at r$_e$) against the 
average $\Sigma_*$. This diagram replicates the well known MZR
distribution, with an almost linear regime where the abundance 
increases with the stellar mass (here the stellar mass density) 
and a plateau at high $\Sigma_*$ values ($\sim$2 M$_\odot$ pc$^{-2}$),
where oxygen abundances reach an asymptotic value.


The main goal of this exploration is not to discuss the physical
origin of the reported relations. For that discussion we refer to
\citet{ARAA}, where the different relations are described in more
detail. In this particularly case we are just interested in showing
that the global relations, here expressed in their intensive form, 
do actually correspond to the local/resolved ones reported in the
literature. For this reason we overplotted in the different diagrams
the corresponding local/resolved distributions for the individual
spaxels within our compilation of IFS data. When required we plot 
the distributions segregated by morphology, since in many cases the
reported relations depends on the morphology (as discussed before), 
being in general different for star-forming galaxies (dominated by
late-type galaxies) than for retired galaxies (dominated by 
early-type ones) \citep[e.g.][]{lacerda20}. Panel by panel we 
can see that the resolved distributions for SFGs (represented by Sc
galaxies) follow exactly the same trends as the global intensive
properties for the $\Sigma_{\rm SFR}$-$\Sigma_*$, 
$\Sigma_{\rm SFR}$-$\Sigma_{\rm gas}$ and 
$\Sigma_{\rm gas}$-$\Sigma_*$ diagrams. Similar results are found 
when using a different estimator for the gas mass density, as the 
one described in Sec. \ref{sec:ana_gas} (labeled as $\Sigma_{gas}$).


We perform a linear regression between the different parameters 
shown in Fig. \ref{fig:local} to characterize the relation between 
them. For this analysis we derive the mode of the distribution along 
the $y-$axis in a set of bins of $\sim$0.1 dex along the $x-$axis 
for each diagram, for those values encircled by the 80\%\ density 
contour in Fig. \ref{fig:local}. The points derived this way are
represented as solid white circles in the Figure. Then we perform a
least-square regression based on a Monte-Carlo procedure, with the 
best fit parameters and their errors derived as the mean and standard
deviations of the reported parameters in each iteration. This 
procedure is very similar to the one adopted by similar recent
explorations \citep[][]{sanchez19,jkbb20}. The result of this 
procedure is included in Table \ref{tab:local}, together with 
similar values reported in the literature. In general, there is good
agreement between the current reported trends and values in the
literature. This is particular true for the $\Sigma'_{\rm gas}$ 
estimation of the gas mass density (in the relation involving this
parameter, i.e., rSK and rMgM*). The dispersion around the main trend,
characterized by the standard deviation, is also of the same order or
smaller than the one reported in the literature. In the case of the 
rMZR we included a linear regression for consistency with the other
explored relations. However, it is well known and it is clear from 
inspection of Fig. \ref{fig:local} that this functional form does 
not describe the distribution along the full range of explored 
parameters. Other functional forms, like a higher order polynomial
function \citep{rosales12} or a linear+exponential one 
\citep{sanchez13}, describe this distribution better. For this reason 
we include in the figure the best fit curve using this later model, 
with the functional form:
\begin{equation}
y=8.54+0.003\left(x-2.25\right) exp\left(5.75-x\right)
\label{eq:rMZR}
\end{equation}
\noindent where $x$ is log($\Sigma_*$) and $y$ is 12+log(O/H). Like in the 
case of the linear relations for the previous distributions, the 
trend described by this functional form is very similar to the one 
reported by recent explorations of the rMZR \citep[e.g.][]{jkbb16}.
Furthermore, similar functional forms have also been adopted to
parameterize the extensive MZR 
\citep{jkbb16,jkbb18,sanchez17a,sanchez19}.


The distributions shown in Fig. \ref{fig:local} clearly demonstrate 
that the rSFMS, rSK and rMGMS resolved relations are indeed the same
relations as the {\it intensive} versions of the SFMS, SK and 
gas-stellar ones. Even more, the individual spaxels of the RGs
(represented by S0 galaxies in here), are located off and below the
reported relations for the SFGs in the $\Sigma_{SFR}$-$\Sigma_*$ and
$\Sigma_{gas}$-$\Sigma_*$ diagrams, following exactly the same trends
as those reported for the RGs. Finally, the rMZR distribution for the 
bulk sample of galaxies follows exactly the same distribution as the
intensive version of the MZR. This distribution is always dominated by
SFGs, since they are the only ones were the oxygen abundance is
properly derived.

\begin{figure*}[t]
\includegraphics[width=8.5cm, clip, trim=15 5 7 3]{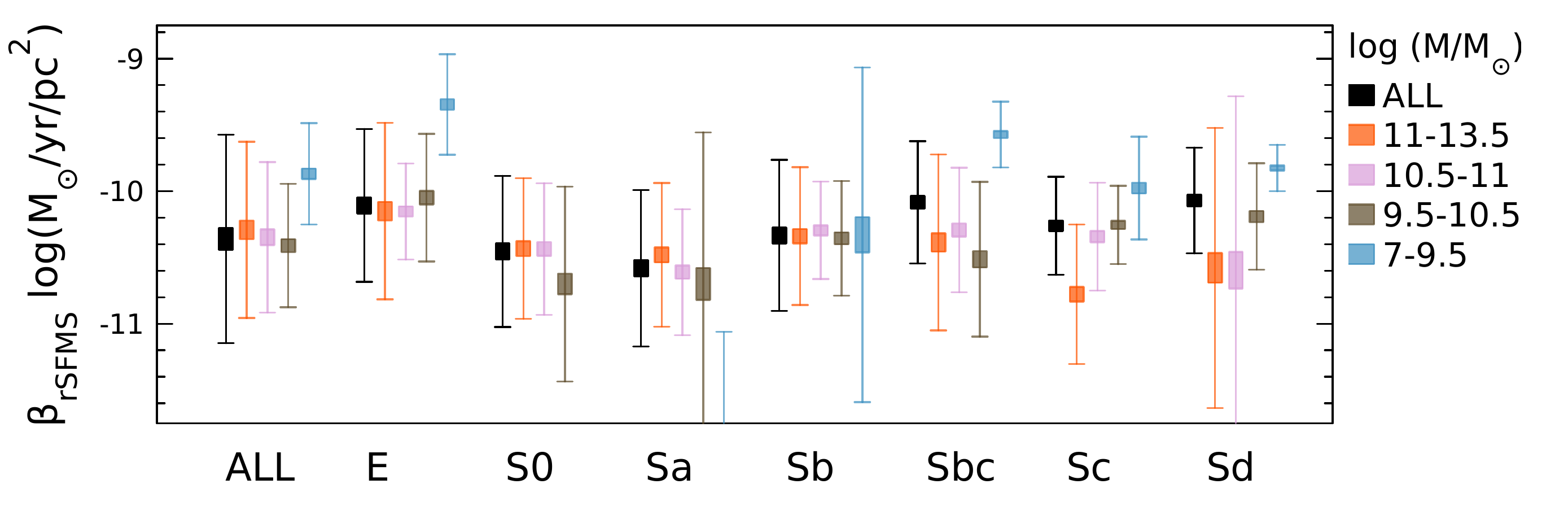}\includegraphics[width=8.5cm, clip, trim=15 5 7 3]{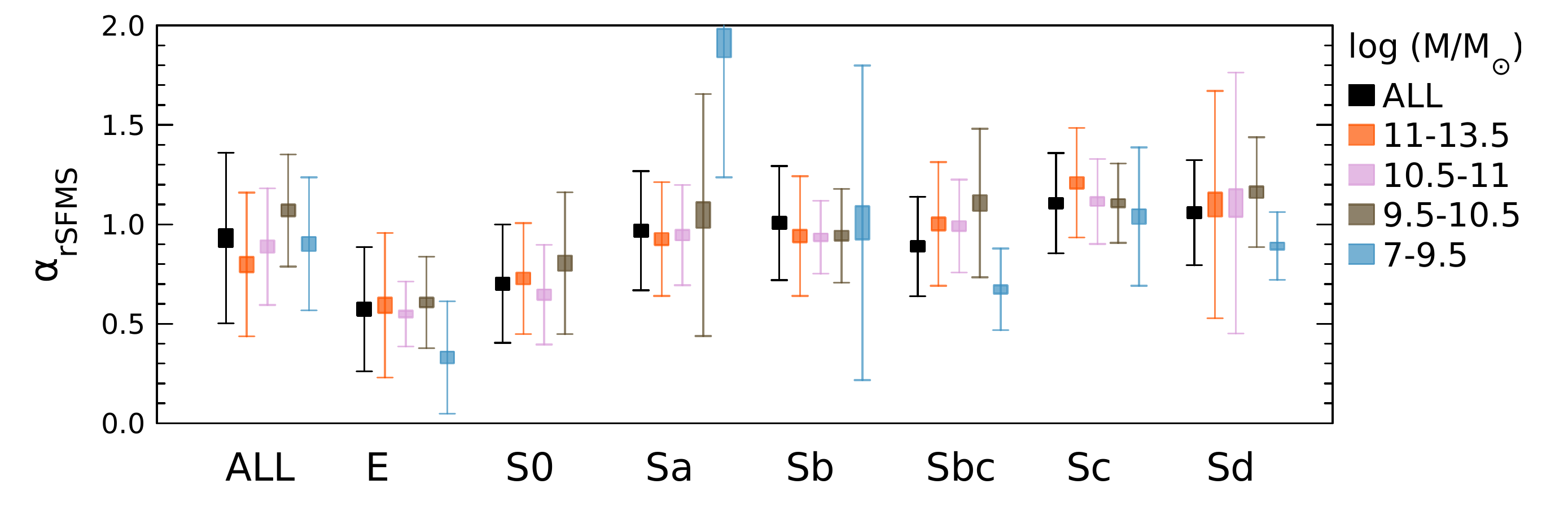}
\includegraphics[width=8.5cm, clip, trim=15 5 7 3]{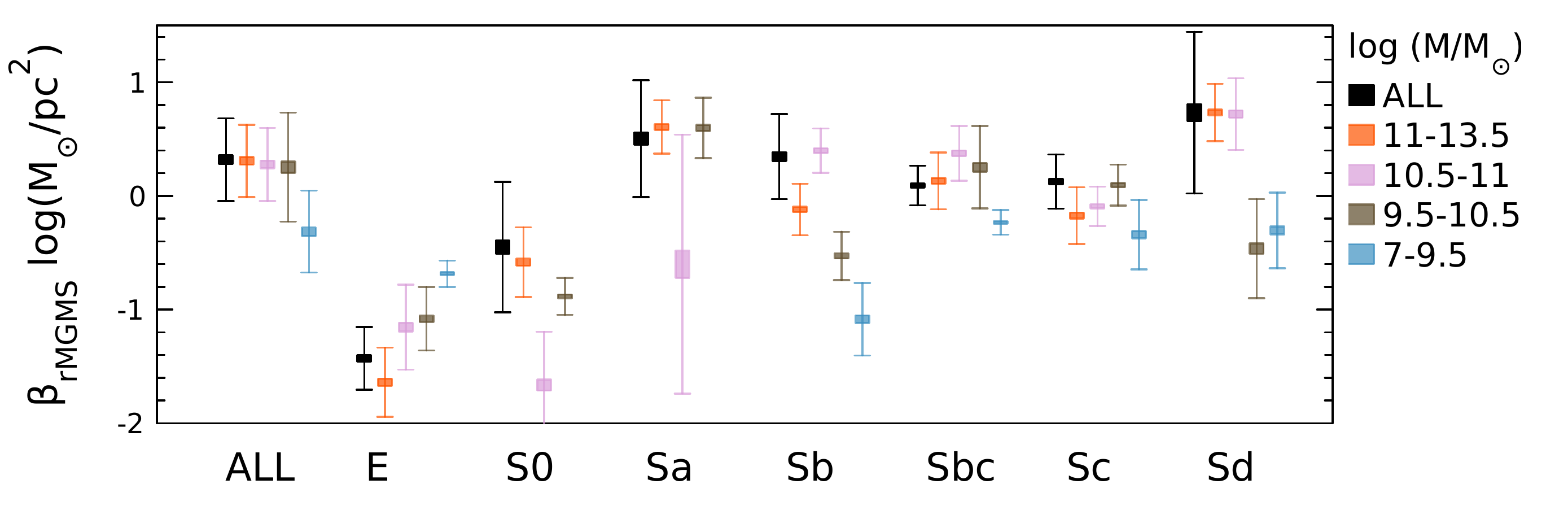}\includegraphics[width=8.5cm, clip, trim=15 5 7 3]{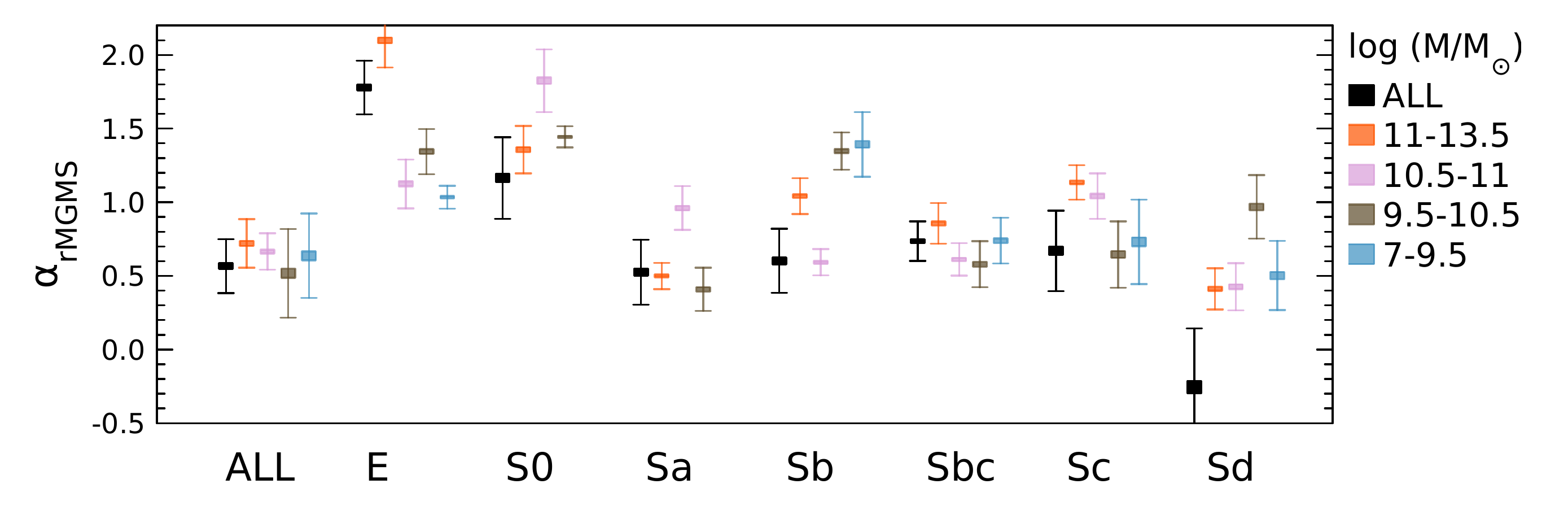}
\includegraphics[width=8.5cm, clip, trim=15 5 7 3]{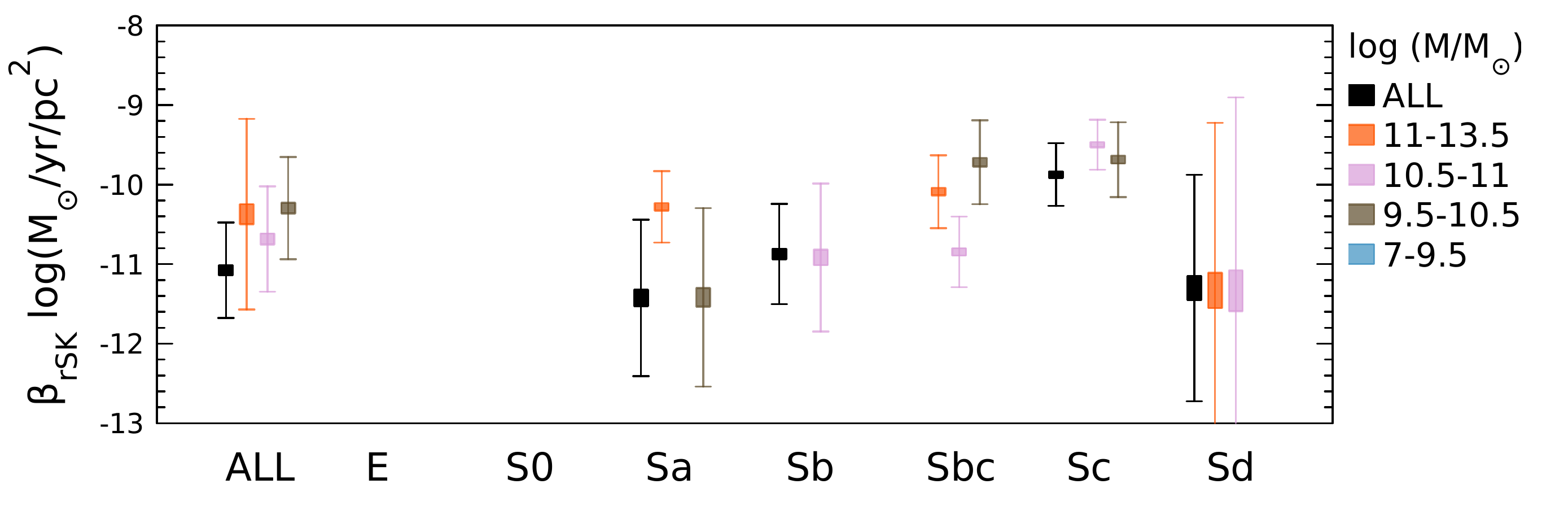}\includegraphics[width=8.5cm, clip, trim=15 5 7 3]{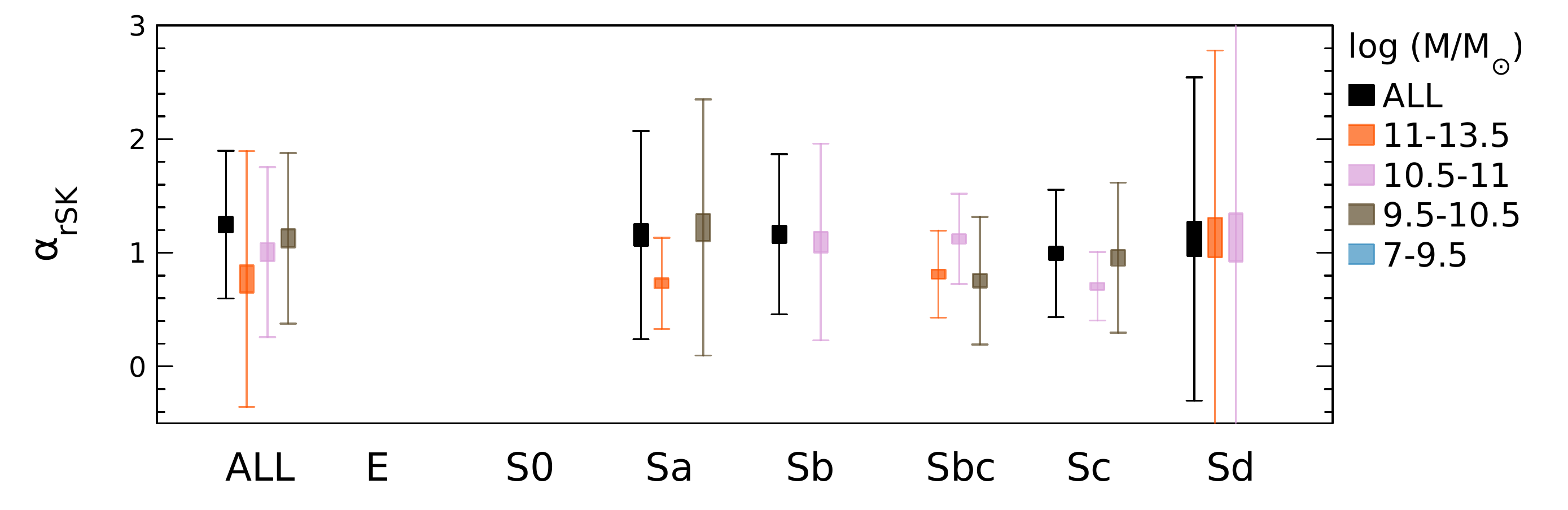}
\caption{{\it Top panel:} Zero-point of the rSFMS relation derived for
  different morphology (black symbols) and stellar mass (colors) bins. 
  For each sub-sample the mean value of the considered zero-point is
  shown, together with the standard deviation
  (shown as a rectangle) and the minimum and maximum values within the
  distribution (shown as error-bars). {\it Middle panel:} Similar
  distribution for the slope of the rMGMS. {\it Bottom panel:} Similar
  distribution for the slope of the rSFMS.}
\label{fig:res_MM}
\end{figure*}

This exercise show that the global and local/resolved
relations are indeed the same relations, covering the same ranges,
following the same distributions (and trends) and the same
morphological segregation (in general). We should note however that
selecting SFGs and RGs is not exactly the same as selecting
star-forming areas (SFAs) and retired-areas (RAs), as recently 
discussed by \citet{mariana19}. There are RAs located in (mostly 
the center) of SFGs, and also a few SFAs located in (mostly the 
outer part) of RGs
\citep[e.g.,][]{sign13,gomes16,belfiore17a}. This should surely affect the
global (intensive or extensive) relations, since they deal with average
quantities, mixing SFAs and RAs within
the considered apertures. Despite this caveat we stress the
remarkable agreement between the explored global and resolved relations.

\subsection{Universality of the local relations}
\label{sec:uni}

One relevant question regarding the local relations explored in 
the present section is their
universality. In other words, whether the resolved regions of 
different galaxies lie along the same reported relation, 
irrespective of the properties of the host galaxies (global or 
local), or, on the contrary, if part of the dispersion reported 
in those relations is the consequence of the existence of 
different relations for different families of
galaxies (or possible secondary correlations). In global 
relations this exploration has lead, for instance to the extensive 
search for a third parameter to explain (or reduce) the scatter 
in the SFMS \citep[SFE or gas fraction, e.g.][]{saintonge17} and 
the MZR relations \citep[SFR, gas fraction, e.g.][]{mann10,both13}. 
It has also lead to a debate over the need for that third parameter
\citep[e.g.][]{sanchez13,jkbb18}. Regarding the local/resolved 
relations, similar explorations were recently performed by 
\citet{jkbb16}, \citet{mariana19} and \citet{ellison20} in the 
case of the rMZR, rSFMS  and rSK (resolved SK-law), respectively. 
These find that: (i) it is not clear that a secondary parameter 
is required to explain the dispersion in the rMZR (in particular 
the $\Sigma_{\rm SFR}$); (ii) galaxies of different morphologies 
and SFE segregate both the SFMS and the rSFMS \citep[as already 
noticed by ][]{rosa16,catalan17}; and (ii) stellar mass and SFE 
segregate the rSK \citep[see also][]{bolatto17,colombo18}. 
This was already discussed in
\citet{ARAA} in detail, describing the similarities and differences
for different stellar masses and morphologies. The rMZR was not included in here since it clearly
requires a more complex parameterization. In order to present a more quantitative evaluation of the variation 
of the different local relations with global properties of galaxies 
we repeat the linear regression to the rSFMS, rMGMS and rSK 
relations discussed in the 
previous section segregated in bins of mass and morphology. 
rMZR was not included in here since it clearly
requires a more complex parameterization

We adopted the same bins as those described in \citet{ARAA}, 
including four ranges of stellar mass (M$\*=$10$^{7.5}$-10$^{9.5}$,10$^{9.5}$-10$^{10.5}$,10$^{10.5}$-10$^{11}$
and 10$^{11}$-10$^{13.5}$ M$\odot$), seven morphological types 
(E, S0, Sa, Sb, Sbc, Sc and Sd), and an additional bin including 
all morphologies, and the same stellar mass bins (labelled as $ALL$).
Figure \ref{fig:res_MM} illustrates the result of this analysis, 
showing the variation of the zero-point ($\beta$) and slope 
($\alpha$) of the linear relations against morphology and stellar 
mass. The values shown in this Figure are distributed in different 
tables available to the community \footnote{\url{http://ifs.astroscu.unam.mx/RMxAA/}}. 

The distributions confirm the results reported in the literature: 
(i) the rSFMS presents a mild change with the stellar mass, and a 
stronger variation with morphology, with SFAs of more massive 
and earlier galaxies presenting a shallower slope in the relation 
and lower values of the $\Sigma_{\rm SFR}$ for a given $\Sigma_{*}$; 
(ii) the rMGMS presents clear patterns with mass, but again, even 
stronger and clearer ones with morphology. SFAs in early-type 
galaxies present lower values of $\Sigma_{gas}$ for a given 
$\Sigma_*$ (i.e., a molecular gas deficit). An effect that is 
enhanced by the stellar mass (i.e., the gas deficit is stronger 
in more massive early-type galaxies). The relation between the 
two mass densities is steeper (larger value of $\alpha$) than 
the one reported for their counterparts in SFGs, indicating 
that the central (more massive) regions of early/massive 
galaxies still may present some molecular gas that is scarce 
in the outer regions \citep[see Fig. of 18][]{ARAA}. On the 
contrary, the SFAs of more massive SFGs present an overall higher 
$\Sigma_{gas}$ for given $\Sigma_{SFR}$ than those of less 
massive SFGs, and a shallower and relation between both parameters; 
(iii) the rSK, that has the noisiest distribution, shows a dependence 
with the mass and the morphology that is more evident in the 
zero-point (or scale) than in the slope (that seems rather 
constant, with $\alpha\sim$1). This trend indicates that the 
SFE is lower in SFAs of earlier and more massive galaxies. 

Thus, although there are local/resolved relations in galaxies 
suggesting that the SF processes are governed by physical processes 
that happen at kilo parsec scales, those relations are not 
fully universal. Indeed, most probably they are not fundamental, being the statistical effect of physical processes that happens at a much smaller scales($\sim$10 pc). They present dependencies with the global 
properties of galaxies that modulate them, indicating a clear
interconnection between local and global processes, 
as also discussed in \citet{ARAA}.

\subsection{Prevalence of local relations}
\label{sec:loca_preval}

The experiments in previous sections show that local/resolved and global
intensive relations are indeed the same relations. However, they do not
indicate which of them is prevalent in their physical origin. In 
other words, they do not show that global relations a just a purely  
integrated (averaged) version of the resolved ones. To demonstrate 
that his is the case at least mathematically, we should show that 
once there is a local/resolved relation it is inevitable to 
generate a global one, but not the contrary. We will now explore this. 

It is well known that many of the physical properties in galaxies have
a radial gradient at first order. In particular, all the properties
described in the previous section do actually present a radial decline
for all galaxy types and all stellar masses and morphologies, on
average. This is particularly true for the radial distributions of
$\Sigma_*$ and $\Sigma_{gas}$. In some particular mass/morphological
types and for some particular galaxies the radial distribution may be
flat of even inverted. This may be the case of $\Sigma_{SFR}$ and
12+log(O/H), although those are not the general/average trends. If the
local relations have a physical prevalence, this means that all radial
relations/trends are a consequence of the relation between the
explored quantities and $\Sigma_*$, and the radial dependence of this
parameter, i.e., $\Sigma_* = F(r)$. In general, the radial distribution of
$\Sigma_*$ is well represented by a log-linear dependence with the
galactocentric distance \citep[that we adopt normalized to the
effective radius, following][]{ARAA}. At first order it presents 
an almost universal exponential decline that can be parameterized 
with the following functional form:

\begin{equation}
  \Sigma_* = \Sigma_0 {\mathrm exp} \left[ -b (r/r_e) \right]
  \label{eq:loc1}
\end{equation}

\noindent where $\Sigma_0$ is the stellar mass density in the central regions,
r$_e$ is the effective radius, and $b$ is a parameter that controls
the slope of the log-linear relation (with that slope being
$b'=\frac{b}{{\mathrm ln}(10)}$).

Therefore, if a parameter $p$ (e.g., $\Sigma_{SFR}$), presents a
log-linear relation with $\Sigma_*$, in the form:

\begin{equation}
  p = c \Sigma_*^d
  \label{eq:loc2}
\end{equation}

\noindent then:

\begin{equation}
  p(r) = c \left(\Sigma_0 {\mathrm exp} \left[ -b (r/r_e) \right]\right)^d =
  c \Sigma_0^d {\mathrm exp} \left[ -d b (r/r_e) \right]
  \label{eq:loc3}
\end{equation}

\noindent or, in other form:

\begin{equation}
  p(r) =  p_0 {\mathrm exp} \left[ -b' (r/r_e) \right]
  \label{eq:loc4}
\end{equation}

\noindent where $p_0$= $c \Sigma_0^d$ and $b' = d b$. In summary, if a parameter
$p$ presents a log-linear relation with $\Sigma_*$, and this later one
presents an exponential decline with the radius, then $p$ presents
a similar dependence with the radius too.

However, the existence of a radial dependence of $\Sigma_*$ is an
empirical result, the nature of which is most probably related with the shape
of the gravitational potential, how the gas settles in that
potential (i.e., the dynamics), and how gas is transformed in
stars. But there is no
global relation between extensive or intensive properties (such as the
one explored in the previous sections) that predicts the existence of
such a radial distribution.  In other words, it could be that galaxies
follow all the four relations explored before, without presenting a
radial decline in the surface densities (or any of the explored properties). 
Thus, galaxies could equally well present flat radial
distributions (which it is unphysical for dynamical reasons) and at
the same time the global relations could still hold. Even more, it could
be that one of the parameters presents a radial decline, and another
not, without affecting the shape of the global relations.

In more detail, let $P$ be the global property corresponding to the
resolved property $p$ (e.g., like the SFR to the $\Sigma_{SFR}$). In
this particular case, we will show that it is possible that $P$
follows a global dependence with M$_*$, with the same functional form
as the local relation shown in Eq. \ref{eq:loc2}:

\begin{equation}
  P = c {\mathrm M}_*^d
  \label{eq:loc5}
\end{equation}

\noindent without fulfilling a local/resolved relation between $p$ and
$\Sigma_*$. Let's assume that $\Sigma_*$ presents a radial dependence
described by Eq. \ref{eq:loc1}, but $p$ is constant along the
extension of a galaxy. Then, no local/resolved relation would be
verified, since at different galactocentric distances $\Sigma_*$ would
present different values, while $p$ remains constant. However, if
both quantities are integrated up to a particular radius (e.g., $r<$2
r$_e$), then:

\begin{equation}
  P = \int_{0}^{2r_e} 2\pi r p \,dr = 4 \pi r_e^2 p
  \label{eq:loc6}
\end{equation}

\noindent and:

\begin{equation}
   \begin{array}{l}
    M_* = \int_{0}^{2r_e} 2\pi r  \Sigma_0 {\mathrm exp} \left[ -b (r/r_e) \right] \,dr \\\\
    \,\, = 2\pi\Sigma_0 r_e^2 b^{-2} \left[ 1 - e^{-2b}(2b+1)\right]
   \end{array}
  \label{eq:loc7}
\end{equation}

\noindent then, even in the case that all galaxies had universal $\Sigma_0$
and $p$ values, both quantities would present a similar dependence
with the effective radius, and they will present a global relation
similar to the one indicated before.

It is true that this is a very particular case, and it is totally unrealistic.
However, it illustrates that the presence of a global relation does not
guarantee the existence of a local/resolved one. The contrary, on the other
hand, is not possible. If galaxies present a local relation between
parameter $p$ and $\Sigma_*$ in the form described by Eq. \ref{eq:loc2}, then
galaxies would present a global relation as the one described by Eq. \ref{eq:loc5}.
To demonstrate so, we use the results of Eq. \ref{eq:loc7} and \ref{eq:loc4},
to show that in this case $P$ would be:

\begin{equation}
    P = 2\pi p_0 r_e^2 (b')^{-2} \left[ 1 - e^{-2b'}(2b'+1)\right]
  \label{eq:loc8}
\end{equation}

\noindent and re-calling the definition of $p_0$, and $b'$, and Eq. \ref{eq:loc7} then:

\begin{equation}
   \begin{array}{l}
  P = 2\pi c \Sigma_0^d r_e^2 D\\ \\
     \,\, = 2\pi c \left[\frac{M_*}{2\pi r_e^2 B}\right]^d r_e^2  D\\\\
     \,\, = (2\pi)^{1-d} r_e^{2-2d} B^{-d} c D M_*^d\\
   \end{array}
 \label{eq:loc9}
\end{equation}

\noindent where $$D=(db)^{-2} \left[ 1 - e^{-2db}(2db+1)\right]$$
and $$B=(b)^{-2} \left[ 1 - e^{-2b}(2b+1)\right]$$

As the effective radius r$_e$ presents an almost log-linear scaling
relation with M$_*$ \citep[e.g.][and references therein]{ARAA}, following
a similar functional form as Eq. \ref{eq:loc5}:

\begin{equation}
  r_e = \beta {\mathrm M}_*^\alpha
  \label{eq:loc10}
\end{equation}

\noindent then:

\begin{equation}
   \begin{array}{l}
  P = (2\pi)^{1-d} \beta^{\alpha(2-2d)} B^{-d} c D M_*^{d+\alpha(2-2d)}\\
   \end{array}
 \label{eq:loc11}
\end{equation}

\noindent and therefore the proposed extensive global relation
is verified (i.e., we recover naturally Eq. \ref{eq:loc5}).

Furthermore, even in the case that the relation shown in Eq. \ref{eq:loc10} is not verified,
an intensive global relation holds.
If we define the intensive global quantities
$\left< p \right> = \frac{P}{4\pi r_e^2}$ and
$\left< \Sigma_* \right> = \frac{\mathrm M_*}{4\pi r_e^2}$, as the
respective extensive ones ($P$ and M$_*$) divided by the area within
2r$_e$ (as defined in Sec. \ref{sec:intensive}). Then, based on
Eq. \ref{eq:loc7} to \ref{eq:loc9} it
is easy to demonstrate that:

\begin{equation}
   \begin{array}{l}
     \left< p \right> = c B^{-d} D {\left< \Sigma_* \right>}^d\\
   \end{array}
 \label{eq:loc12}
\end{equation}

\noindent what it is indeed a global relation between intensive quantities that
mimics the local one. In particular, it presents the same slope as the
local one, with a slightly different zero-point, when represented in a
log-log form. We should highlight here that the zero-point of this relation
is not a dimensionless quantity, and particular care should be taken
when the surface densities are expressed in different areal units (e.g., when
transforming between $pc^{-2}$ or $kpc^{-2}$, if the value of $d$ is not one).
These are the relations that match with the
local/resolved ones, as shown in Fig \ref{fig:local}. In principle,
the scale/zero-point are different by a factor $B^{-d} D$ between
local and global intensive relations. However, adopting the average
values reported for those relations, shown in Table \ref{tab:local}, we
found that the expected offset if $\sim$0.05 dex, being compatible
with zero in some cases (like the rSK and rSFMS relations).

Finally, we should stress that if local relations segregate by
morphological type (or other properties of galaxies), then global ones
should also segregate, depending on the particular effect in the
scaling of the zero-point. This effect should be stronger for the
extensive global relations, that depend on the relation between the
effective radius and the stellar mass shown in Eq. \ref{eq:loc10}. It
is known that this relation is different for early- and late-type
galaxies. Therefore, even in the case that both families present the
same local/resolved relations, their global extensive ones should
segregate just due to this effect, without involving a different physical processes.
{\bf This stresses the need to explore the global relations in their
intensive form, and not in their extensive ones, as usually done.}


\subsection{Characteristic intensive properties}
\label{sec:char}

In the previous sections we studied the existence of global
relations between galaxy parameters based on an intensive formulation, rather than an extensive one as customary in the literature. 
We used average surface densities within 2 r$_e$ in galaxies. 
In the case of the Oxygen abundance, we adopted the value at r$_e$ 
as the characteristic Oxygen abundance for a galaxy. Several 
previous studies have indeed reported that the value of many different
quantities at this particular radius are representative/characteristic
of the average quantity across the optical extent in galaxies
\citep[e.g.~oxygen abundance, stellar ages and metallicities, stellar
mass density,][]{sanchez13,rosa14,rgb17}. This is indeed a corollary of
the derivations in the previous section, as we will see here. 

We assume again that $P$ is a global extensive quantity, and it 
results from the integral of the corresponding surface density $p$ 
across the optical extent of a galaxy, as described in the previous
section. It follows a radial distribution described by
Eq. \ref{eq:loc4}. Then, following Eq. \ref{eq:loc8} and the definition 
of $\left< p \right>$, we find that:

\begin{equation}
    \left< p \right> = 0.5 p_0 (b')^{-2} \left[ 1 - e^{-2b'}(2b'+1)\right]
  \label{eq:loc13}
\end{equation}

\noindent Then defining $p_{\rm e}$ as the value of $p$ at the effective radius, it is possible to derive the relation:

\begin{equation}
    \left< p \right> =  p_{e} e^{b'} (b')^{-2} \left[ 1 - e^{-2b'}(2b'+1)\right]
  \label{eq:loc14}
\end{equation}

\noindent Thus, the average value of the surface density across the optical
extent ($r<2r_e$) is related to the value at the effective radius 
by a multiplicative constant that depends only on the slope of the 
radial gradient. Indeed, in the
case of $b'\sim$1, i.e., an exponential profile,
$\left< p \right> \sim 0.8 p_{e}$.

\section{Characteristic gradients of resolved properties}
\label{sec:grad}

So far, we have demonstrated that the existence of a
local relation between an observed property and 
$\Sigma_*$ implies that this property presents a radial gradient 
\citep[if $\Sigma_*$ presents that gradient originally, as shown
by][]{jkbb16}. The radial gradients of
all the properties explored in the previous sections do indeed exist
and were extensively discussed in \citet{ARAA}. It was shown that in
general, most of those properties present indeed a radial decline. 
This decline is frequently characterised at first order
with an exponential function (as indicated in the previous sections),
i.e.~a log-linear function.  Fitting that function to the observed
distributions it is possible to derive the corresponding value of each
considered parameter at the effective radius (that is
representative of the average value), plus the magnitude of the radial gradient/slope of the 
radial gradient, either galaxy by galaxy, or averaged by galaxy type.

\begin{figure}
  \includegraphics[width=8.5cm, clip, trim=1 5 7 3]{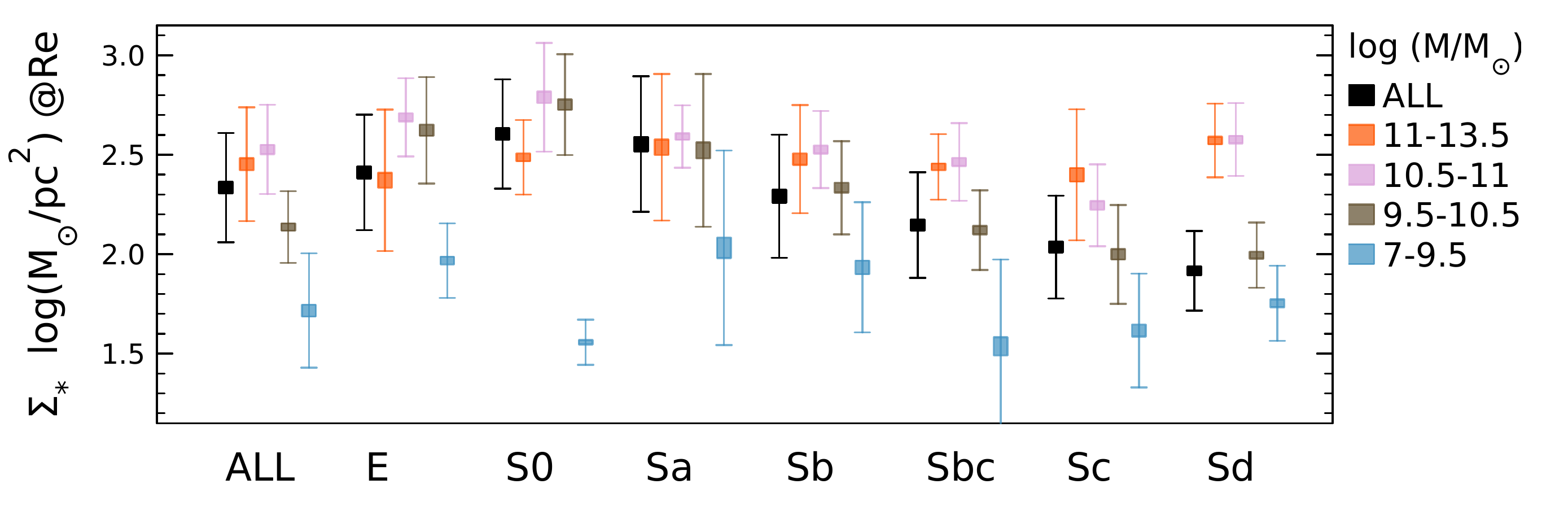}
\includegraphics[width=8.5cm, clip, trim=1 5 7 3]{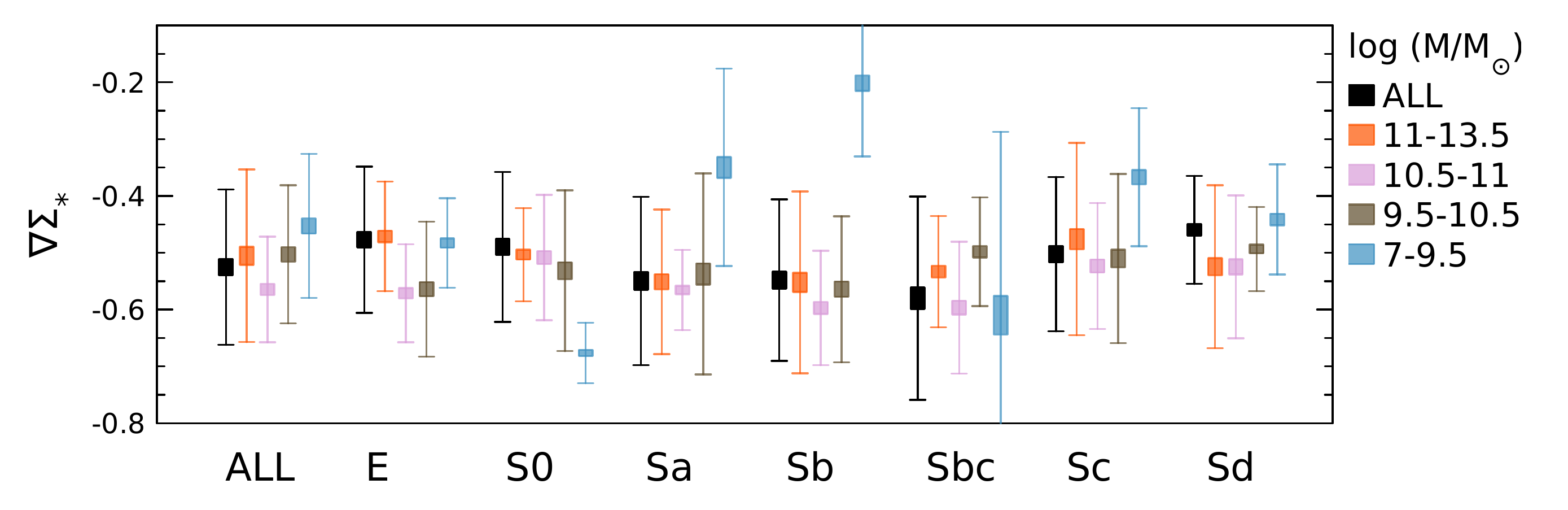}
\caption{{\it top-panel:} Characteristic value of $\Sigma_*$ at
  the effective radius for all galaxies in the sample (black symbols),
  and for different stellar masses (colors) and different
  morphologies. For each subsample we show the mean value of
  the considered parameter, together with the standard deviation
  (shown as a rectangle) and the minimum and maximum values within the
  distribution (shown as error-bars). {\it bottom-panel:} Similar
  distribution for the characteristic slope of the radial gradient of
  the $\Sigma_*$ parameter.}
\label{fig:Sigma_M}
\end{figure}

This analysis has been frequently performed to explore whether the
proposed radial distributions (and therefore the local and global
relations, as demonstrated in the previous sections) are universal
(i.e., essentially the same, not dependent on other galaxy properties)
or if they depend on the properties of the galaxies (e.g.~morphology,
mass, star-formation rate, gas content). In this particular
section we report on the log-linear fitting performed on the average
radial profiles of the properties explored by \citet{ARAA} in
different mass bins and morphological types. We use the same bins
described in that article, including four ranges of stellar masses
(M$\*=$10$^{7.5}$-10$^{9.5}$,10$^{9.5}$-10$^{10.5}$,10$^{10.5}$-10$^{11}$
and 10$^{11}$-10$^{13.5}$ M$\odot$), and seven morphological types (E,
S0, Sa, Sb, Sbc, Sc and Sc), and an additional bin including all
morphologies and the same stellar mass bins (labelled as $ALL$). 
We fit each of the radial profiles with the functional form 
described in Eq. \ref{eq:loc4}, that corresponds to the log-linear
relation:

\begin{equation}
  {\rm log} p(r) =  \beta - \alpha (r/r_e)
  \label{eq:loc15}
\end{equation}

\noindent where $\alpha$ is the slope of the gradient (related to the $b'$
parameter of Eq. \ref{eq:loc4}) and $\beta$ is the zero-point. From this
relation it is possible to derive the characteristic value (i.e., the value at $r_e$), that would be

\begin{equation}
  {\rm log} \ p_e =  \beta - \alpha
  \label{eq:loc16}
\end{equation}

\noindent In order to avoid possible resolution problems affecting the central
regions of the galaxies and to truncate to a radius covered by all IFS
data, we restricted the fitting to the radial range between 0.5 and
2.0 $r_e$. Like in the case of the local relations explored in 
Sec.~\ref{sec:uni}, we repeat the analysis for galaxy subgroups 
segregated by mass and morphology, following \citet{ARAA}. 
The results of the analysis are included in the electronically 
distributed tables.\footnote{\url{http://ifs.astroscu.unam.mx/RMxAA/}}
Figures \ref{fig:Sigma_M} to \ref{fig:OH_O3N2} present the results of
this analysis, showing the distribution of the characteristic value
(e.g., $\Sigma_{*,e}$) and the gradient slope (e.g.
$\nabla \Sigma_*$) for the different explored properties, 
in each bin of stellar mass and morphology. This way it is
possible to explore the variations of both parameters along these
galaxy properties in a more quantitative way than presented in
\citet{ARAA}, although the same conclusions are extracted from these
figures. In summary we find that:

\begin{figure}
  \includegraphics[width=8.5cm, clip, trim=15 5 7 3]{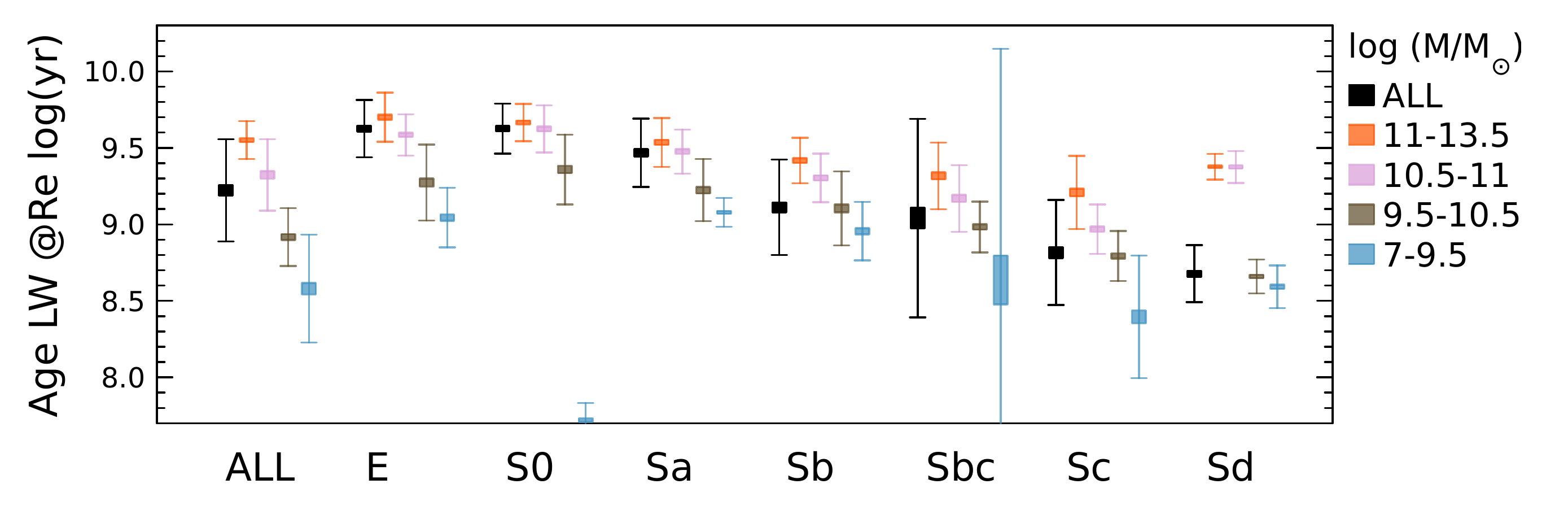}
\includegraphics[width=8.5cm, clip, trim=15 5 7 3]{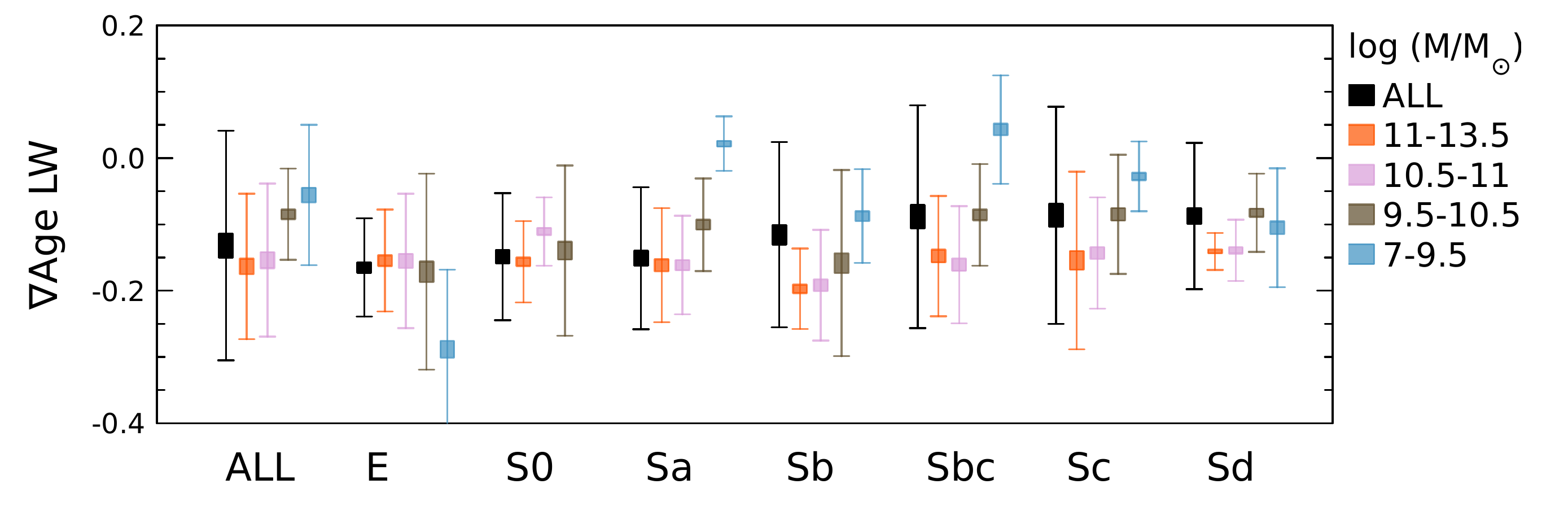}
\caption{Same as Figure \ref{fig:Sigma_M}, but for the 
luminosity-weighted ages of the stellar populations.}
\label{fig:Age_LW}
\end{figure}


{\bf Stellar mass density gradients:} Fig. \ref{fig:Sigma_M} 
  shows that the characteristic stellar
  mass density presents a clear dependence with the integrated stellar
  mass (M$_*$), as a consequence of the relation shown in
  Eq. \ref{eq:loc7}.  This way $\Sigma_{*,e}$ increases with M$_*$ in
  general, up to $\sim$10$^{11}$M$_\odot$. However, this trend is
  modulated by the morphology of the galaxy, with later types showing
  lower values of $\Sigma_{*,e}$ than earlier types for the same
  stellar mass. This trend is particularly strong for the less
  massive galaxies. Based on the results in Sec. \ref{sec:local}, this
  indicates that the radial gradient of $\Sigma_*$ should depend on
  the morphology. This is indeed appreciated in the bottom panel of
  Fig.~\ref{fig:Sigma_M}. All galaxies present a negative gradient of
  $\nabla \Sigma_*\sim -$0.5 dex. However, there are clear variations
  with the morphology, with shallower gradients for both the earliest
  (E/S0) and latest (Sc/Sd) morphological bins, but strongly modulated
  by the stellar mass. In general, the most massive Sb (and Sbc to a
  lower extent) galaxies are those presenting the sharper
  gradients. Similar results were reported by \citet{rosa14} and
  reviewed in \citet{ARAA}.

  It is worth noticing that the most massive E-type galaxies are not
  the ones with the largest $\Sigma_{*,e}$. This is a consequence of
  these galaxies not following a single exponential profile (or 
  log-linear relation) as the one shown here (i.e.~they show cores). 
  For them the parameterization adopted here is too simple.
  Finally, it is important to highlight that not all galaxy types
  are equally well represented in each mass bin, and therefore the results
  are not equally significant. In particular, the results for
  late-type galaxies in the more massive bins and for early-type ones in
  the less massive bins should be taken with care.

\begin{figure}
  \includegraphics[width=8.5cm, clip, trim=15 5 7 3]{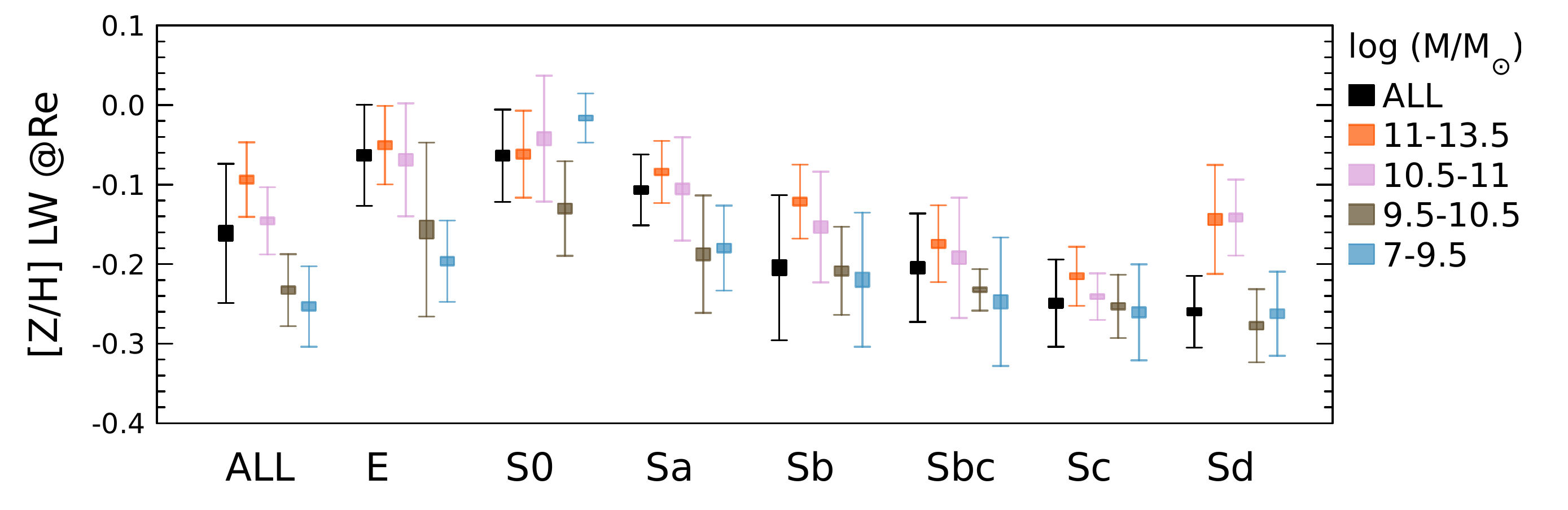}
\includegraphics[width=8.5cm, clip, trim=15 5 7 3]{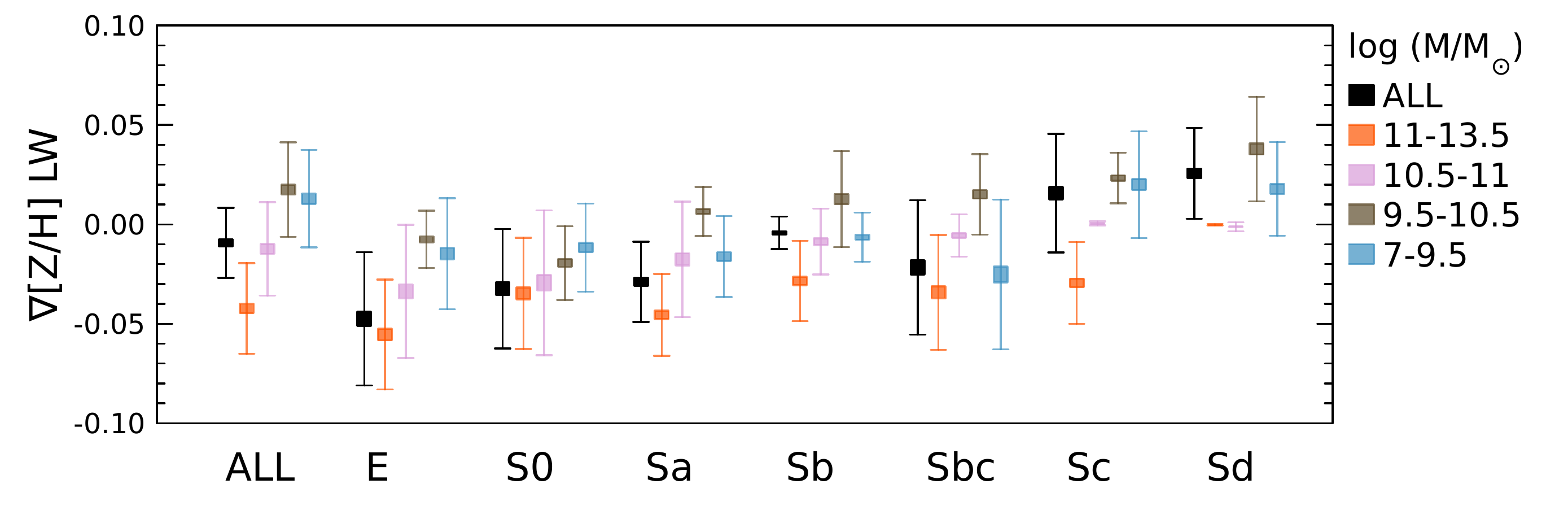}
\caption{Same as Figure \ref{fig:Sigma_M}, but for the  luminosity-weighted stellar metallicity.}
\label{fig:ZH_LW}
\end{figure}

{\bf Age and metallicity gradients: } Fig. \ref{fig:Age_LW} and
  \ref{fig:ZH_LW} show the distributions of the luminosity weighted 
  (LW) characteristic ages and metallicities of the stellar populations
  together with the slope of the radial gradients. For both parameters 
  similar trends are found, with earlier and more massive galaxies 
  showing older ($Age_e\sim$5 Gyr) and more metal rich 
  ($[Z/H]_e\sim -$0.05 dex) stellar populations, with more pronounced negative gradients in both explored quantities: 
  $\nabla_{Age}\sim -$0.2 dex/r and $\nabla_{[Z/H]}\sim -$0.04 dex/r. 
  On the other hand, later and less massive galaxies have younger
  ($Age_e\sim$0.3 Gyr) and more metal poor ($[Z/H]_e\sim$-0.3 dex) 
  stellar populations, with shallower age gradients and even flat 
  or positive metallicity gradients: $\nabla_{Age}\sim -$0.1 dex/r 
  and $\nabla_{[Z/H]}\sim $0.02 dex/r. As reviewed by \citet{ARAA} 
  this has been interpreted as a clear evidence of (i) average 
  inside-out growth of galaxies, that is more pronounced in more 
  massive and earlier types, and (ii) a change in the 
  SFHs and ChEHs from earlier to later types, with the first ones
  presenting a sharper evolution with a stronger enrichment in earlier
  times and the later ones presenting smoother evolutions with ongoing 
  enrichment processes. Similar trends are found for the mass weighted 
  ages and metallicities, although covering a narrower dynamical range 
  in the described parameters. We do not reproduce here to avoid 
  repetition, although we provide the corresponding measurements 
  in the distributed tables.

\begin{figure}
  \includegraphics[width=8.5cm, clip, trim=15 5 7 3]{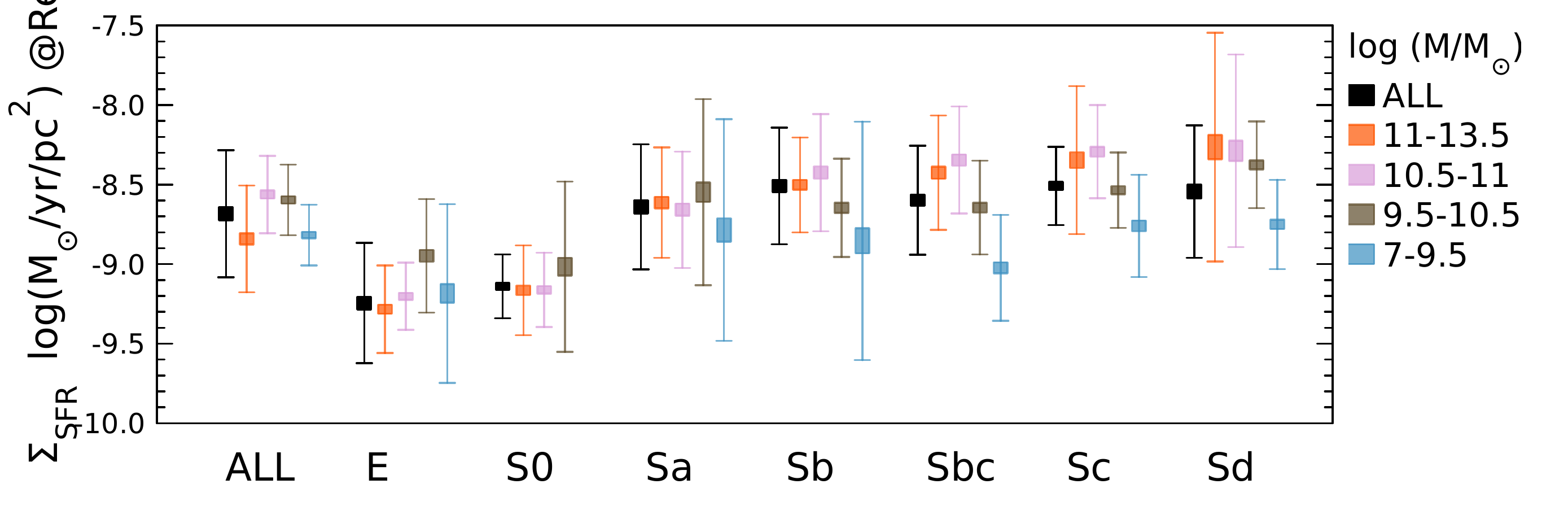}
\includegraphics[width=8.5cm, clip, trim=15 5 7 3]{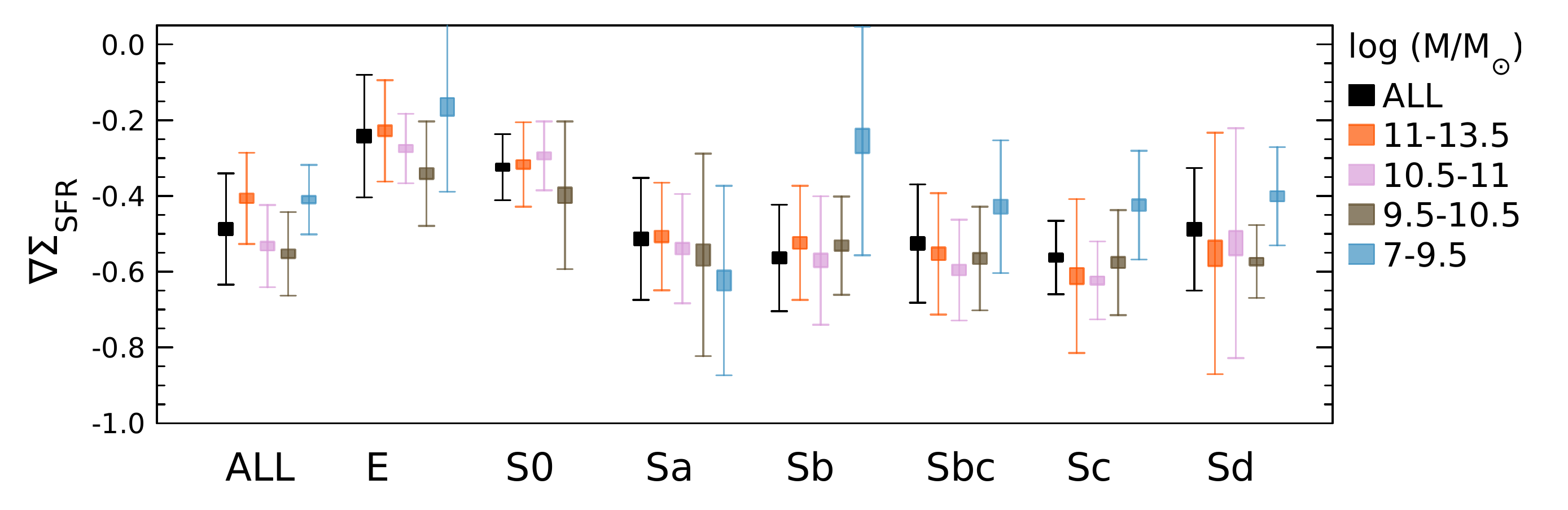}
\caption{Same as Figure \ref{fig:Sigma_M}, but for the 
star-formation rate surface density $\Sigma_{SFR}$.}
\label{fig:Sigma_SFR}
\end{figure}

{\bf Star-formation density gradients: } Fig. \ref{fig:Sigma_SFR} 
  shows the distribution of characteristic values and gradient slopes 
  of the star-formation surface densities for the different 
  morphological and stellar mass bins. As expected, early type 
  galaxies (E/S0) present lower $\Sigma_{SFR,e}$ 
  ($\sim$10$^{-9.5}$M$_\odot$yr$^{-1}$ pc$^{-2}$) at any 
  stellar mass than late type ones ($\sim$10$^{-8.3}$M$_\odot$yr$^{-1}$ 
  pc$^{-2}$).  For SFGs there is a clear trend between $\Sigma_{SFR}$ 
  and stellar mass, that is a consequence of the existence of an 
  rSFMS, as described in Sec. \ref{sec:local}, and the relation between 
  $\Sigma_{*,Re}$ and M$_*$ \citep[e.g.][]{rgb17}. 
  On the other hand earlier galaxies present 
  shallower negative radial gradient in $\Sigma_{SFR}$ 
  ($\nabla_{\Sigma_{SFR}}\sim -$0.2 dex/r) than later ones 
  ($\nabla_{\Sigma_{SFR}}\sim -$0.5 dex/r). This trend is also 
  observed with the mass, from more massive to less massive galaxies, 
  at least for E, S0 and Sa galaxies. This indicates that in these
  galaxies the SFR is not as uniformly distributed in the central 
  regions as in the case of later and less massive galaxies. This 
  is clearly a consequence of the presence of a bulge dominated by 
  retired areas, that do not form stars. The trend with the mass is
  inverted for later type galaxies (Sb to Sd), what indicates that 
  in these galaxies there is a mild increase of the SFR in the 
  outer regions (with $\nabla_{\Sigma_{SFR}}$ rising from $-$0.5 
  to $-$0.4 dex/r). This reinforces the idea that low-mass galaxies 
  may show a less strong inside-out evolution (which was shown in 
  the stellar metallicity gradients too).

\begin{figure}
  \includegraphics[width=8.5cm, clip, trim=15 5 7 3]{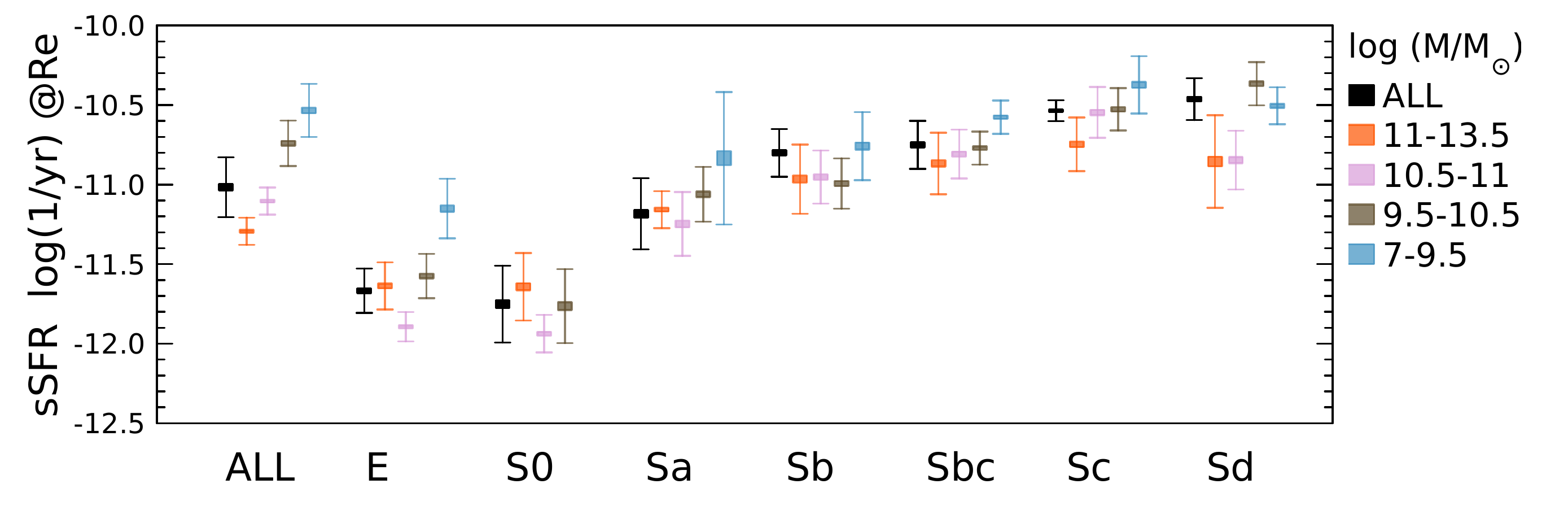}
\includegraphics[width=8.5cm, clip, trim=15 5 7 3]{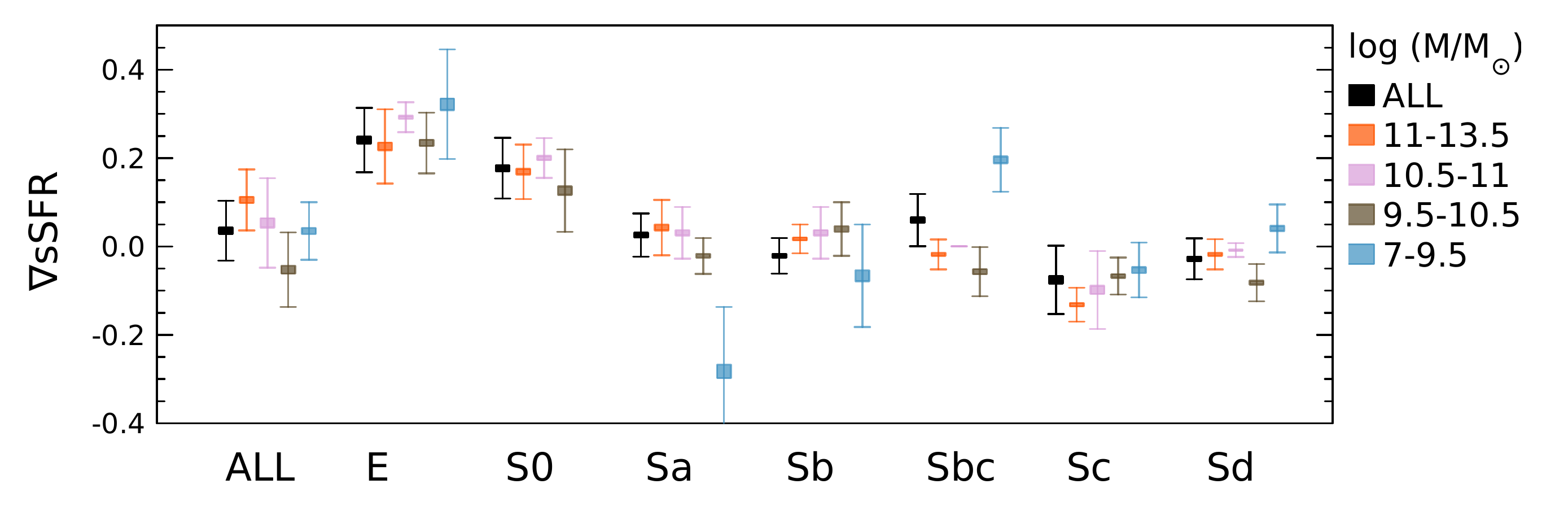}
\caption{Same as Figure \ref{fig:Sigma_M}, but for the 
specific star-formation rate.}
\label{fig:sSFR}
\end{figure}

Similar trends are reported for the radial distribution of the
specific star-formation rate, as shown in Fig. \ref{fig:sSFR}, with
clearer and sharper distributions as a consequence of the
combination of $\Sigma_{SFR}$ with $\Sigma_*$ (as
sSFR is the ratio of both quantities). Due to that the sSFR presents a
sharp segregation by morphology, with most massive early type galaxies
showing the lowest characteristic values (sSFR$_e\sim$10$^{-11.75}$
yr$^{-1}$), and a clear increase for later and less massive galaxies,
rising up to sSFR$_e\sim$10$^{-10.5}$yr$^{-1}$ for the less massive Sd
galaxies. The average radial gradient slope for the full sample is
almost zero, however there is a strong morphology and stellar mass
segregation, with positive gradients for early-type galaxies
($\nabla_{\rm sSFR}\sim$0.2 dex/r) and shallow and slightly negative
gradients for the latest type ones ($\nabla_{\rm sSFR}\sim -$0.02 dex/r).
These trends indicate that the SFR relative to the stellar mass (what
measures the sSFR) is stronger in the outer regions of early-type
galaxies than in their inner regions \citep[e.g.][]{rosa16}, 
as a consequence of both very low SFR in the inner regions and either an
inside-out dimming or a rejuvenation due to the capture of gas rich
galaxies in the outer regions of those galaxies
\citep[e.g.][]{gomes16}. On the contrary, for late-type galaxies SF 
happens at a relatively similar rate with respect to the underlying
stellar mass, what is again a direct consequence of the
already discussed rSFMS relation \citep[e.g.][]{mariana16}. The mild
but significant differences with morphology in the gradient indicates
that galaxies slightly deviate from the average rSFMS with a pattern
showing that later type ones are located above the average and earlier
spirals below, as discussed in \citet{rosa16} and \citet{mariana19}, 
and shown in the previous section.

Similar results are found when other estimators for the SFR are
adopted. In particular, in the case of the SFR derived using the
stellar population analysis described in Sec. \ref{sec:ana_ssp} 
the results
are totally compatible with the one described here. We provide 
the corresponding measurements in the distributed tables, although no
figures are provided to avoid repetition.

\begin{figure}
  \includegraphics[width=8.5cm, clip, trim=15 5 7 3]{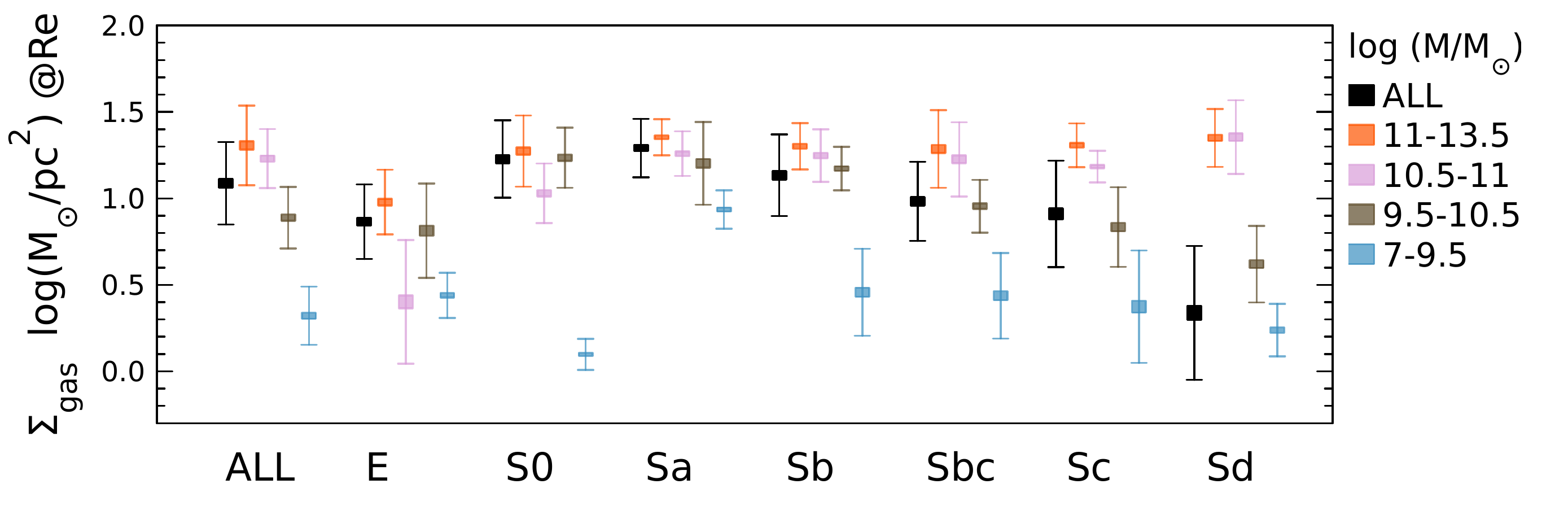}
\includegraphics[width=8.5cm, clip, trim=15 5 7 3]{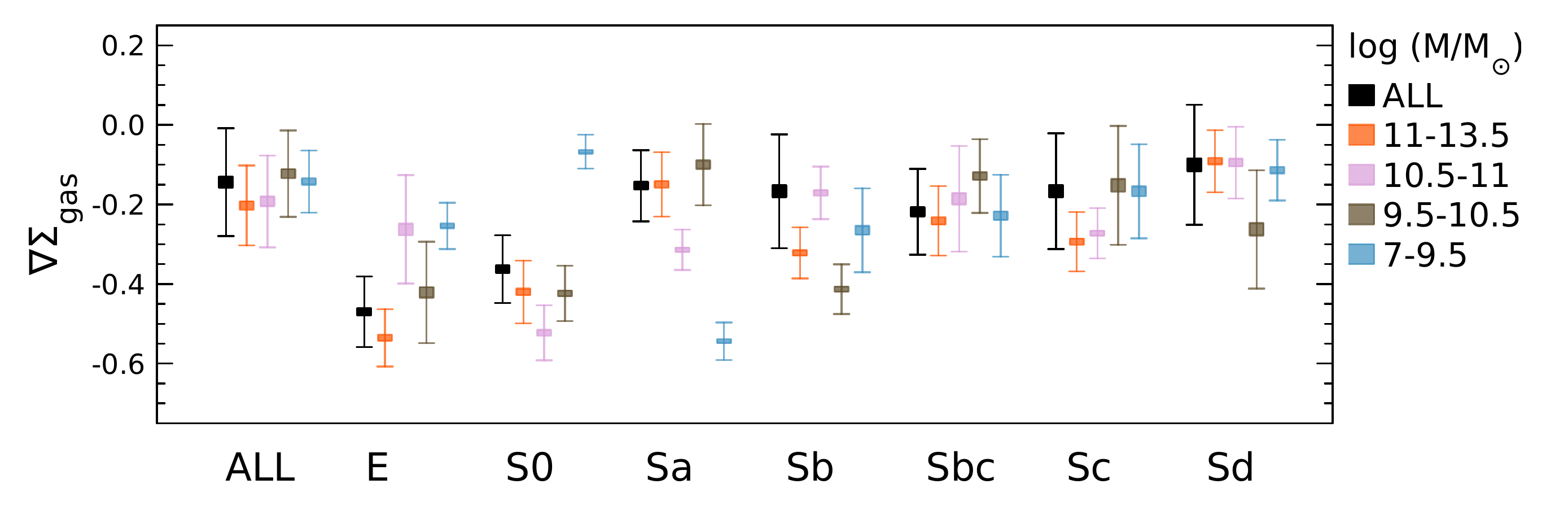}
\caption{Same as Figure \ref{fig:Sigma_M}, but for the 
molecular gas mass surface density $\Sigma_{gas}$.}
\label{fig:Sigma_gas}
\end{figure}

{\bf Gas density gradients: } Fig. \ref{fig:Sigma_gas} includes the
distributions of the characteristic values and the radial gradients
for the molecular gas mass surface density for the considered
sub-samples. We should stress out that this derivation was obtained
based on the dust-to-gas calibrator presented in \citet{jkbb19} 
(as described in Sec \ref{sec:ana_gas}). Like in the previous properties
there are clear trends, with more massive galaxies showing more gas
($\Sigma_{gas}\sim$1.2 M$_\odot$ pc$^{-2}$) than less massive ones
($\Sigma_{gas}\sim$0.2 M$_\odot$ pc$^{-2}$. Regarding morphology,
elliptical galaxies present a clear gas deficit with respect to other
early type galaxies (S0) or early-spirals (Sa, Sb), for any mass, and
for the more massive ones they present this deficit with respect to
any other galaxy of any later morphology. In average gas density
decreases from earlier to later type galaxies (besides the described
trend for pure elliptical galaxies), from ($\Sigma_{gas}\sim$1.2
M$_\odot$ pc$^{-2}$) to ($\Sigma_{gas}\sim$0.5 M$_\odot$
pc$^{-2}$). This trend is enhanced by a similar one observed in mass,
with more massive galaxies presenting larger amounts of gas 
than less massive ones. This trend is a clear consequence of the
MGMS and rMGMS relations described for SFGs between 
M$_{gas}$ ($\Sigma_{gas}$) and
M$_*$ ($\Sigma_*$) described by different authors
\citep[e.g.,][]{saint16,calette18,lin19,jkbb19}, and 
discussed in Sec.~\ref{sec:local}. The clear morphological
segregation reinforces the results, indicating that the rMGMS 
does not present a universal shape, as seen in 
Fig.~\ref{fig:res_MM} and discussed in Sec.~\ref{sec:uni}.

Regarding the slope of the radial gradient, the trends are more
complicated than the ones described so far. On average, when all
morphologies are included, the slope for late type galaxies is 
almost constant $\nabla_{\Sigma_{gas}}\sim -$0.15 dex/r,
without a clear trend with stellar mass. However, early type 
galaxies present a sharp negative gradient 
($\nabla_{\Sigma_{gas}}\sim -$0.5 dex/r. 
Furthermore, for each morphology the modulation is different: (i) less 
massive elliptical galaxies present less steep negative gradients than 
those with higher massea; (ii) S0 present shallow negative 
gradients for the lowest stellar mass ranges; (iii) Sa and Sb 
present the stronger negative gradients for the lower masses, and, 
finally, (iv) latest spirals present almost the same
gradient independently of their stellar masses.

\begin{figure}
  \includegraphics[width=8.5cm, clip, trim=15 5 7 3]{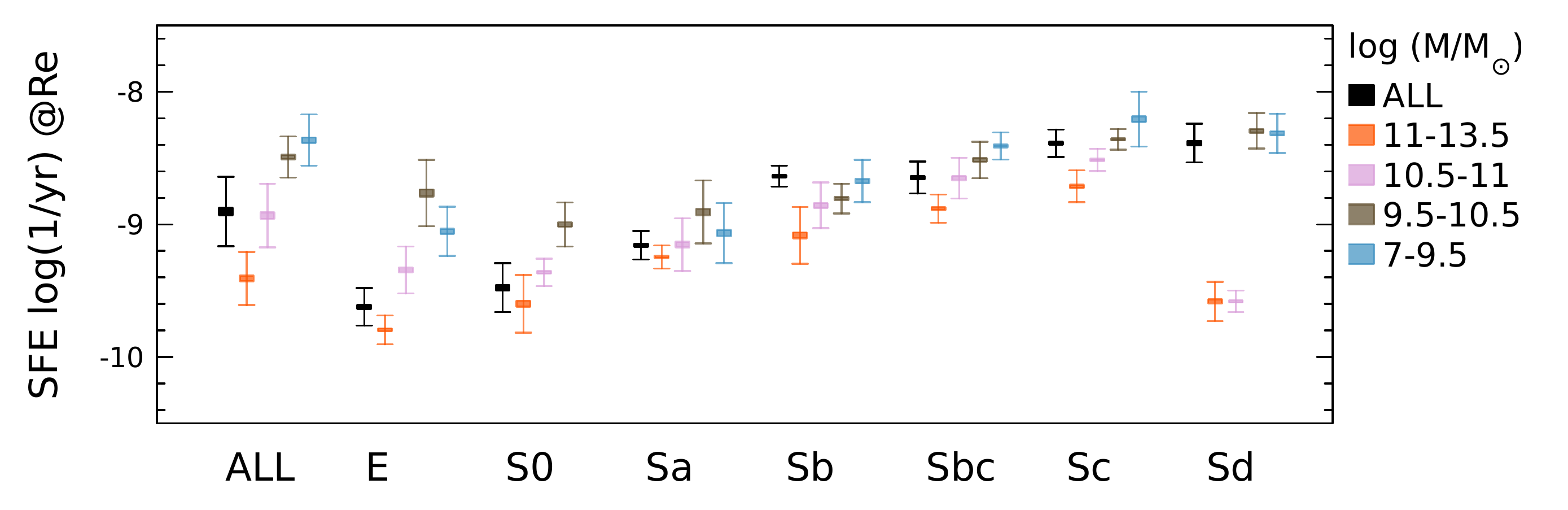}
\includegraphics[width=8.5cm, clip, trim=15 5 7 3]{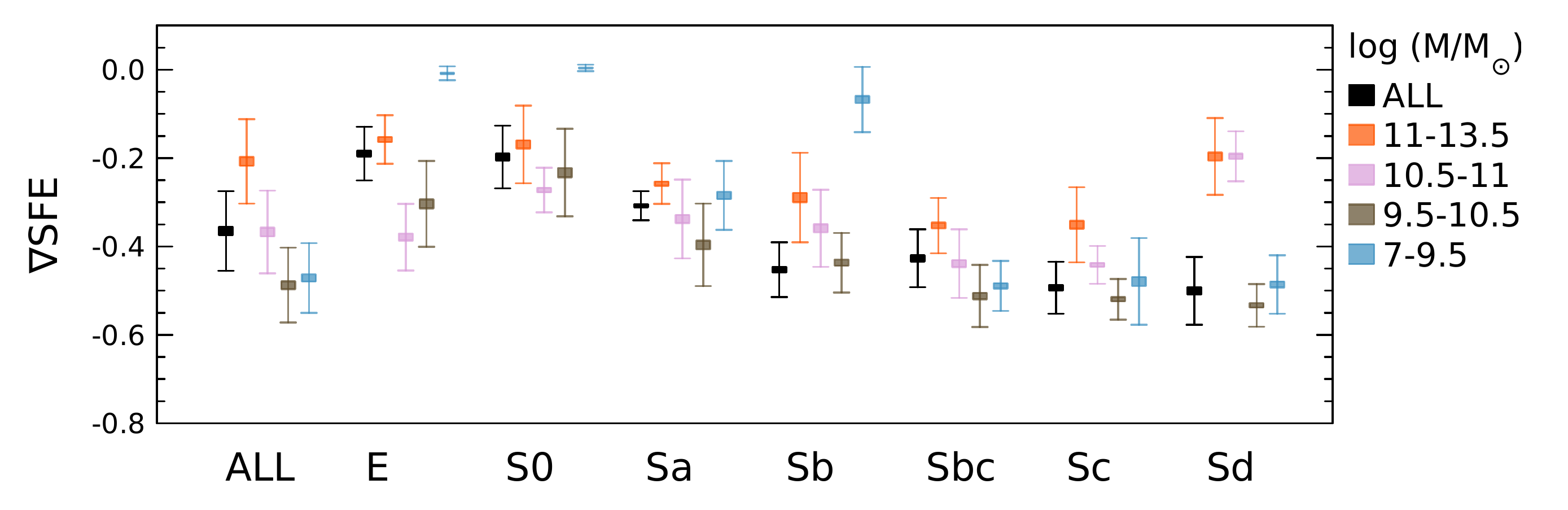}
\caption{Same as Figure \ref{fig:Sigma_M}, but for the 
star-formation efficiency.}
\label{fig:SFE}
\end{figure}

In general these results indicate that RGs present a global lack of
molecular gas, as discussed by \citet{saint16} and \citet{sanchez18}. 
Furthermore, they present a deficit that is stronger in
the inner regions than in the outer ones, which suggests that the
described inside-out quenching process \citep[e.g.][]{rosa16,belf17},
is driven by a lack of gas too. The fact that quenching seems to be
connected with the presence of a central (active) massive black-hole
has been frequently proposed in the 
literature \citep[e.g.][]{hopkins+2010}.
Recently explorations using spatially resolved IFS data support
this scenario \citep[e.g.][]{bluck19}, suggesting that AGN are 
related with inside-out quenching and the removal (or heating) of
gas that are thus prevented from settling as molecular clouds. 
However, there is no general consensus in this regard, since other 
recent contributions
\citep[e.g.~][]{ellison20} indicate that star-formation efficiency
could be the primary driver for halting of the SF rather than a real
lack of gas. We should be cautious in this regards since most
estimations of the spatially resolved molecular gas density, based on CO
observations, are biased towards late-type galaxies \citep[e.g., EDGE,
ALMAQUEST,][]{bolatto17, lin19}, while the current estimations,
although indirect, also covers early-types (in particular pure
ellipticals), where the strongest drop in $\Sigma_{gas}$ is reported.
This unsettled result highlights the need for a large CO mapping survey
over an unbiased sample already covered by an IFS-GS with a similar spatial
resolution (and deep enough) to shed light to this fundamental question.


{\bf Star-formation efficiency gradients: } Fig. \ref{fig:SFE} shows
similar plots as the ones described before for the star-formation
efficiency (SFE=$\Sigma_{SFR}$/$\Sigma_{gas}$), i.e., the inverse of the
depletion time $\tau_{dep}$. The average value found for all galaxies 
is  SFE$\sim$10$^{-9}$ yr$^{-1}$, thus, $\tau_{dep}\sim$ 1 Gyr), a value
slightly lower than the typical value reported of $\sim$2 Gyr in other
studies using {\it direct} gas estimations based on CO observations
\citep[e.g.][]{bolatto17,utomo17,colombo18}. This difference is most
probably due to a zero-point effect in the adopted dust-to-gas 
calibration (see Table \ref{tab:local}.
Despite this offset, as reviewed in \citet{ARAA}, the trends
with mass and morphology are similar to the ones described in the
literature \citep[e.g.][]{colombo18}: earliest and most massive galaxies
have lower values of SFE$\sim$10$^{-9.8}$ yr$^{-1}$ (larger
$\tau_{dep} \sim$ 6 Gyr) than later-type one, with values as high as
SFE$\sim$10$^{-8.25}$ yr$^{-1}$ (i.e., lower $\tau_{dep} \sim$ 0.2 Gyr).

In addition to this global dependence with galaxy mass and morphology,
there are also evident radial differences. On average the SFE presents a
gradient with a negative slope of $\nabla_{SFE}\sim$-0.4 dex/r. However,
the gradient is shallower, the more massive and earlier is a galaxy
($\nabla_{SFE} -$0.2 dex/r), and steeper the less massive and later type
it is ($\nabla_{SFE} -$0.5 dex/r). There are some significant outliers
in this trend, that correspond to galaxy sub-groups with a very small 
number of galaxies in our sample (such as low mass early-types). 
In summary, as indicated
by different studies there is no single SFE ($\tau_{dep}$) among 
galaxies and within galaxies. Different scenarios to explain this
variation that affects in different way galaxies of different masses,
morphologies and galactocentric distances have been proposed: (i) the 
effect of the orbital or dynamical time ($\tau_{dyn}$), that relates the
gravitational instability (e.g., spiral arms) with an increase
in the SFE \citep{silk97,elme97}; (ii) the self-regulation of the
star-formation that increases the local velocity dispersion 
(e.g.~the local pressure) and decreases the SFE \citep{silk97}; 
(iii) the stabilization of molecular clouds that decreases the 
SFE (or even quenches the SF) due to the presence of warm/hot 
orbits associated with a bulge \citep{martig09}; (iv) the
differential/local gravitational potential that may affect the 
SFE \citep[e.g.][]{saintonge2011} ; (v) the metal content of the 
ISM that affects the cooling; or (vi) a combination of all of 
them \citep[e.g.][]{dey19}. 

\begin{figure}
  \includegraphics[width=8.5cm, clip, trim=15 5 7 1]{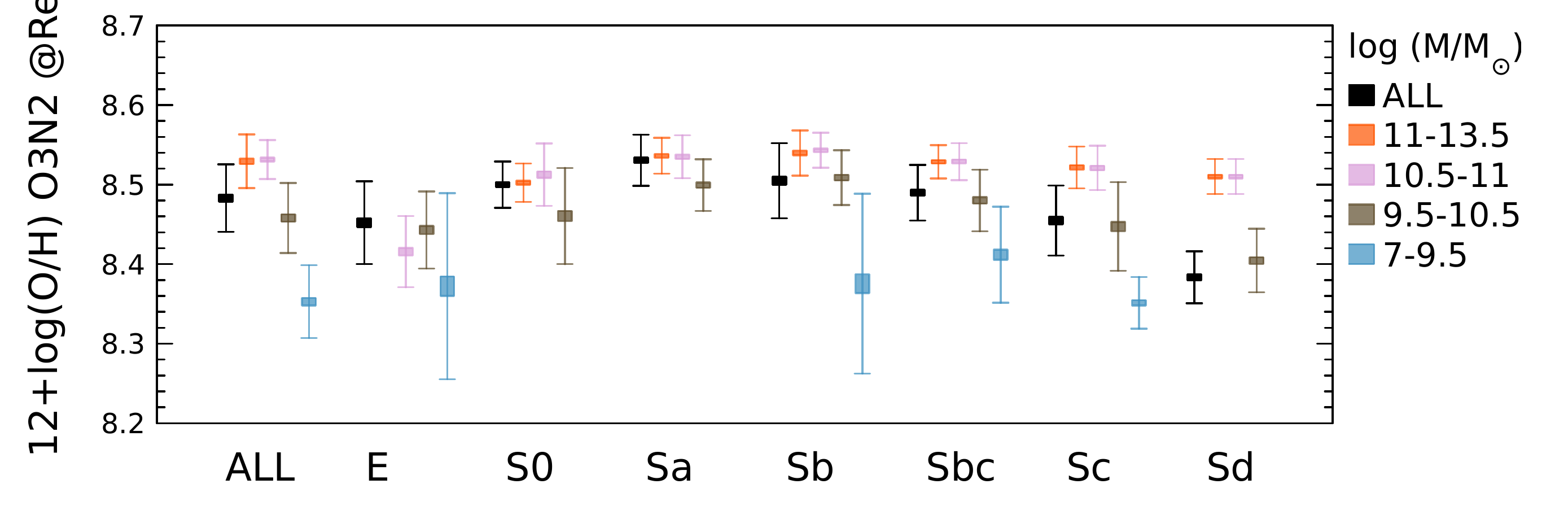}
\includegraphics[width=8.5cm, clip, trim=15 5 7 1]{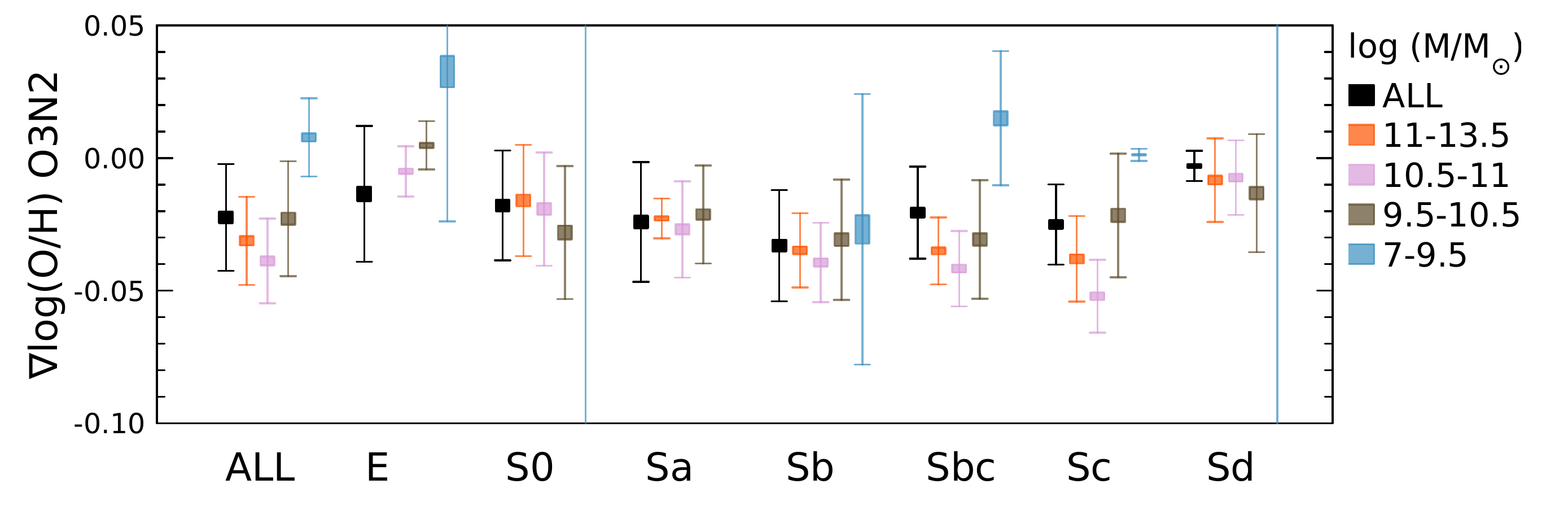}
\caption{Same as Figure \ref{fig:Sigma_M}, but for the 
gas phase oxygen abundance derived using the O3N2 calibrator.}
\label{fig:OH_O3N2}
\end{figure}

{\bf Oxygen abundance gradients: } Fig. \ref{fig:OH_O3N2} shows the
distribution of the characteristic oxygen abundance and the slope of
the corresponding radial gradient for the different stellar masses and
morphologies explored in this section. We adopted the O3N2
calibrator proposed by \citet{marino13}, what imposes certain
restrictions/biases regarding the absolute values of the abundances
and the range of values covered with each galaxy
\citep[see][]{kewley08,sanchez19}. However most of the
qualitative results are independent of the adopted calibrator 
\citep[in particular regarding the abundance gradient,
e.g.~][]{laura16,laura18}. Another
caveat to note is that oxygen abundance is only derived for those
galaxies that present SF ionized regions populating the explored
spatial range well enough to derive a reliable gradient. Therefore,
the number of galaxies in the most early-type subgroups are really 
small.

On average, galaxies in the nearby Universe have an oxygen abundance 
of 12+log(O/H)$\sim$8.5 dex, however, this is strongly modulated by
the stellar mass due to the well known MZR \citep[][]{tremonti04}, and
the resolved version discussed in Sec. \ref{sec:local}. The most
massive galaxies present a plateau in the oxygen abundance around
$\sim$8.5 dex, while the less massive ones present values as low as
$\sim$8.3 dex. We remind the reader that the actual absolute values may
change from calibrator to calibrator. Furthermore, a trend with
morphology is apparent. However, in most of the cases the
abundance is very similar among galaxies of the same mass, despite
their morphology. Thus, the morphology trend seems to be induced by
the mass-morphology and MZR relations, rather than being a real
secondary trend. Only for the less massive and latest type galaxies
there is some appreciable difference induced by the morphology and
not only by the stellar mass.

Regarding the abundance gradient it was reported that late-type
galaxies of mass M$_*>$10$^{9.5}$ M$\odot$ present a very
similar value for their slope \citep[][]{sanchez13}, that for the
explored calibrator on that study has a value of 
$\nabla_{\rm log(O/H)}\sim -$1 dex/r. This
result was confirmed by more recent explorations based on different
datasets \citep[e.g.][]{laura16,laura18}. At lower
stellar masses the abundance gradient becomes shallower
\citep[e.g.][]{belf17}. In general these results are reproduced in
Fig. \ref{fig:OH_O3N2}, although with a different characteristic
abundance gradient, being $\nabla_{\rm log(O/H)}\sim -$0.03 dex 
(due to the use of a
different calibrator). Little to no morphology dependence is
appreciate in the average slope (when all masses are
considered), from E to Sc galaxies. A significantly shallower slope is
detected only for Sd galaxies, with a slope near to zero.  Curiously,
the reported mass dependence seems to be present only when the most
late type galaxies (Sc and Sd) are considered, and for Elliptical
galaxies (a very limited number of objects). Therefore, it is not clear
if the reported mass dependency is actually due to the mass or to the
morphology, an issue that should be explored in more detail in the future.

\section{Discussion and Conclusions}
\label{sec:summary}

In this manuscript we have reviewed the most recent results based on 
IFS extracted from recent galaxy surveys following \citet{ARAA}, but 
including additional details that clarify some of the conclusions 
presented in that review. We made use of the large collection of 
galaxies observed with IFS selected for this previous study. All of 
them were analyzed using the same tool \citep[{\sc Pipe3D},][]{pipe3d}, 
in order to homogenize as much as possible the compiled data. Based 
on those parameters we explored different spatially resolved and 
integrated properties of galaxies.

First, we explored the local/resolved nature of the dominant ionization
processes in galaxies. Based on this analysis we demonstrate that the 
use of integrated or aperture limited parameters of the emission lines 
(line flux intensities and ratios) may produce significant errors in 
the interpretation of the ionizing sources in galaxies. We show the 
main trends of the integrated and resolved line ratios with galaxy 
morphology and masses, illustrating them in detail with a few galaxies 
observed with the IFU that provides the best spatial resolution so far 
for a wide FoV (i.e.~MUSE). Based on that analysis we update the 
practical scheme presented in \citet{ARAA} to classify the ionization 
sources at the spatial resolution considered in this review ($\sim$1 kpc).
Our main conclusion is that the location within the classical diagnostic
diagrams is not enough to distinguish between the dominant ionizing 
sources in galaxies. The inclusion of a third
parameter, like the EW(H$\alpha$), partially mitigates the intrinsic 
degeneracy in the observed line ratios. The additional use of (i) 
the knowledge of the composition of the underlying stellar population, 
(ii) the spatial distribution and shape of the ionized gas, and (iii) 
a knowledge of the gas kinematics, substantially improves our ability 
to distinguish between different ionizing sources. However, all this 
is hampered by the spatial and spectral resolution of the adopted data.

Then, we explored in a more quantitative way the global and local 
relations that seem to result from the star-formation processes in galaxies 
(and regions within galaxies) already reviewed in \citet{ARAA}: the 
SFMS, MGMS, SK and MZR relations. We revisited the most recent results 
in this regards and re-evaluated them using our extensive dataset. 
The main conclusions of this exploration are that: (i) there are 
resolved/local versions of the global relations that are verified 
at kiloparsec scales for SFAs; (ii) those relations are similar in 
shape to global ones. This is particularly true for the intensive 
versions of those relations, that totally overlap with the resolved/local 
ones; (iii) global and local relations are verified for star-forming 
galaxies (and areas/regions within galaxies), but not for retired galaxies 
(areas). Thus, these relations are tightly related to the 
star-formation activity. Retired galaxies (areas) follow different 
trends that are more loose relations or clouds; (iv) global and local 
relations are not fully universal: i.e., they are similar but not 
exactly the same for different galaxy types and stellar masses; 
(iv) global relations can be derived from local ones with less 
assumptions than the other way around. Indeed, just considering 
that more massive galaxies are in general more extended (i.e.~the 
existence of a M$_*$-R$_e$ relation), the existence of a local 
relation implies the presence of a global one. The contrary is not 
true: i.e.~it is possible to have global relations without the 
need for local ones. 

All these results suggest that local relations have a physical 
prevalence over global ones and the star-formation is governed by 
processes that leave clear imprints at kiloparsec scales rather than galaxy 
wide. In fact, these processes are likely originated at much smaller scales. They involve most probably the self-regulation 
introduced by feedback that modulates the gravitationally driven 
trend of molecular clouds to collapse and form new stars. The rSFMS 
and rSK relations, and the dispersion around them, are most probably 
the result of the bouncing effect of those two processes working 
one against each other. This way, the MGMS relation would highlight 
the ability of the local gravitational potential to hold both 
components bound to the system. Under this interpretation is 
is naturally explained that the rSFMS and rMGMS evolve along 
cosmic times: (i) galaxies (and regions within them) in the past 
present higher SFRs at a given M* (or $\Sigma_*$) and (ii) galaxies 
(and regions within them) present larger amounts of molecular gas.
The expected evolution is demonstrated for the rSFMS just comparing the
relation reported at $z\sim0$ \citep{sanchez13} with the one reported 
at $z\sim0.7$ \citep{wuyts13}, showing a similar evolution as the 
one found for the SFMS \citep[e.g.][]{speagle14}. Under this 
scenario the rSK (SK) relation should present a weaker evolution 
along cosmological times. The same scenario would explain the 
morphological segregation of the rSFMS and MGMS relations, since 
later-type galaxies (and SFAs within them) would present a 
smoother evolution of the rSFMS as a consequence of a smoother 
star-formation history \citep[e.g.][]{lopfer18}.

This way the dependence or modulation of local/resolved relations 
with global properties of galaxies is naturally explained. The 
dynamical stage of a galaxy influences the way the star formation 
happens (or is halted) in different locations, modifying the local 
relations. The presence of a bulge seems to stabilize the molecular 
clouds \citep{martig09} and it would be the reason behind the 
morphological segregation of the SFMS \citep[e.g.]{catalan17} and 
rSFMS \citep[e.g.][]{rosa16,mariana19}. On the contrary, local 
effects, like the over-pressure produced by the pass of a spiral 
arm, may enhance the SFR locally, producing local fluctuations 
in the local relations \citep[e.g.][]{laura17,laura19}. Another 
interpretation is that the SFMS (and the rSFMS) holds only for 
the disk (or regions within disk) of galaxies (regions with 
cold orbits), and the morphological segregation is a consequence 
of the increase of M$_*$ ($\Sigma_*$) due to the inclusion of 
stars in warm/hot orbits. This is supported by the recent results 
by \citet{jairo19}, that shows that SFMS holds for the disk component 
of galaxies once they are spatially decomposed.

Additional global processes, like galactic winds or the presence 
of an AGN can alter those relations. Physically, it is known that 
galaxy-wide winds are important for setting galaxy metallicities and 
even baryonic fractions.
In particular, galactic winds are usually claimed to be responsible 
for the shape of the MZR. The linear rising phase of this relation, 
observed at M$_*<$10$^{10}$M$_\odot$, is easily interpreted as a 
natural consequence of the internal enrichment processes in galaxies
\citep[e.g.][]{pily07}. However, the flat regime at high mass 
requires a certain amount of gas inflow/outflow, or a dependence of the IMF with the metallicity \citep[as reported by ][]{navarro15}, with galaxies 
exhibiting an equilibrium between the inflow of pristine gas, the 
internal enrichment, and the outflow of metal rich material
\citep[e.g.][]{tremonti04,belf16a}. The same scenario could 
explain the rMZR. Outflows and inflows may produce a differential 
effect between the central and the outer regions in galaxies. 
Recent explorations indicate that outflows are indeed needed to 
shape the rMZR \citep[e.g.][]{jkbb18}, when considered as local 
effects not as global ones. However, gas leaking has to be 
compensated by gas accretion \citep[e.g.][]{salmeida19}, which 
seems to have a much stronger influence in the flattening of the 
rMZR than outflows \citep[Fig. 11][]{jkbb18}. Once the rMZR is 
shaped, the global MZR emerges as pure integral, as demonstrated 
in Sec. \ref{sec:local}.

On the other hand, it seems that AGN may be more relevant in 
the quenching of star-formation activity. The energy injection 
by these powerful sources is the primary candidate to explain the 
halt of star formation either by removal or heating of gas
\citep[e.g.][]{hopkins+2010}. The location of AGN hosts in the 
green valley region of galaxies in different diagrams (like the 
CMD or the SFR-M$_*$), reinforces this perception
\citep[e.g.][]{kauff03,sanchez04,schawinski+2014,lacerda20}. 
Nevertheless,  quenching is not an instantaneous and coherent 
process galaxy wide. It is a local process too, evolving from the 
inside-out \citep{rosa16,belfiore17a}, with AGN hosts being in 
transition between SFGs and RGs in this regards too \citep{sanchez18}. 
Recent explorations suggest that the presence of a central 
massive black hole is directly connected with the global and local 
quenching \citep{bluck19}. Thus, again, although the presence 
of an AGN could be considered as a global process, its effects 
in the processes and relations are local.

Finally, to be efficient, both processes, galactic winds and AGN, 
require that the injected kinetic energy equals or overpasses the 
local escape velocity. 
For this reason their effect was much stronger in earlier cosmological 
times, when galaxies were less massive and had larger SFRs and they 
hosted AGN more frequently. Maybe for this reason their effect seem 
less evident in the relations and patterns explored in this 
review, which corresponds to those of nearby galaxies. For instance, 
the metal redistribution usually associated
with galactic outflows should have a limited effect in todays 
massive galaxies. It is known that massive galaxies present a 
strong metal enrichment in their early cosmological times
\citep[e.g.][Camps-Fari\~na in prep.]{vale09, walcher15}, in agreement 
with their known SFHs \citep[e.g.][]{panter07,thomas10}. At the same 
time, they show Oxygen abundances \citep[e.g.][]{sanchez13} and 
stellar metallicity gradients \citep[e.g.][]{rosa14} that agree 
with a local downsizing and inside-out growth \citep[e.g.][]{eperez13}. 
The gradients in their stellar and ionized gas properties, like the 
ones explored in Sec. \ref{sec:grad}, are therefore a fossil 
record of the early evolutionary phases in galaxies. Thus, even 
in the period of more violent SFRs and therefore stronger and more 
frequent outflows, the metal redistribution induced was not strong 
enough to blur the observed gradient. It is expected that nowadays, 
when outflows are more scarce and less energetic
\citep{ho18,carlos19,carlos20}, their effect is even less prelevant 
in this regards, at least for massive galaxies. 

On the other hand, low mass and late-type galaxies could be more 
strongly affected by SF driven outflows. Their SFHs, global or 
resolved, present a smoother cosmological evolution, being still 
in the rising phase for the lowest and latest type galaxies
\citep[e.g.][]{lopfer18}, with smaller inside-out differences. As 
a consequence, in those galaxies the signatures of inside-out formation 
(e.g.~radial gradients in the explored properties, Sec. \ref{sec:grad}), 
are less evident. This is particularly true for the Oxygen abundance 
and stellar metallicity distributions. As already discussed in 
\citet{ARAA} those galaxies present a flat or even inverse gradient 
in these properties, even potentially implying an outside-in growth 
phase. In general, it can be considered that their overall evolution 
is delayed with respect to that of more massive galaxies with 
earlier morphological types. 

The influence of galaxy environment on the evolution of galaxies and 
the observed global and local patterns has been discussed very little 
in this review \citep[and in][]{ARAA}. The main reason is that galaxy 
samples currently observed by IFS-GS are still not large enough or 
will require a more detailed reevaluation to distinguish the 
environmental effects. In this regards, the results of the SAMI 
project (which has by design a large sub-sample of cluster galaxies) 
and the GASP survey \citep[][that explores the gas stripping in 
galaxies entering in clusters]{GASP}, would be of a particular importance.
The comparison between the properties of field and cluster members, 
or central and satellite galaxies, in large samples, like the one 
provided by MaNGA, would also provide more insight on this particular 
problem. However, the current number of these analyses is too limited 
and they are too recent to make firm conclusions in this regards.

In summary, we reviewed here our current understanding of the 
interconnection between the local and global properties of the 
ionized gas and stellar populations in galaxies in the nearby 
universe. In particular, we discuss the ionization processes 
and the star formation and metal enrichment cycle in galaxies. 
The main conclusion, in the same line as the ones presented in 
\citet{ARAA}, is that these processes are governed mostly by 
local physical processes that are modulated by global ones. 
Whether the influence of those global properties is through 
local ones or not would be a matter of exploration for the next years.





\section*{ACKNOWLEDGMENTS}

\update{We acknowledge the anonymous referee for reading this manuscript and
helping us to improve its content.}

We thank the sharing of ideas with the IA-MaNGA team, in particular with Prf. V. Avila-Reese, and the help with the morphological analysis of the MaNGA dataset by Dr. H. Hernandez-Toledo. The enthusiasm and hard working of Dr. H. Ibarra-Medel and Dr. Mariana Cano Diaz.

We are grateful for the support of a CONACYT grant CB-285080 and FC-2016-01-1916, and funding from the PAPIIT-DGAPA-IN100519 (UNAM) project.

Parts of this research was conducted by the Australian Research Council Centre of Excellence for All Sky Astrophysics in 3 Dimensions (ASTRO 3D), through project number CE170100013.

The SAMI Galaxy Survey is based on observations made at the Anglo-Australian Telescope. The Sydney-AAO Multi-object Integral field spectrograph (SAMI) was developed jointly by the University of Sydney and the Australian Astronomical Observatory. The SAMI input catalogue is based on data taken from the Sloan Digital Sky Survey, the GAMA Survey and the VST ATLAS Survey. The SAMI Galaxy Survey is supported by the Australian Research Council Centre of Excellence for All Sky Astrophysics in 3 Dimensions (ASTRO 3D), through project number CE170100013, the Australian Research Council Centre of Excellence for All-sky Astrophysics (CAASTRO), through project number CE110001020, and other participating institutions. The SAMI Galaxy Survey website is http://sami-survey.org/.

This project makes use of the MaNGA-Pipe3D dataproducts. We thank the IA-UNAM MaNGA team for creating this catalogue, and the ConaCyt-180125 project for supporting them

Funding for the Sloan Digital Sky Survey IV has been provided by
the Alfred P. Sloan Foundation, the U.S. Department of Energy Office of
Science, and the Participating Institutions. SDSS-IV acknowledges
support and resources from the Center for High-Performance Computing at
the University of Utah. The SDSS web site is www.sdss.org.

SDSS-IV is managed by the Astrophysical Research Consortium for the 
Participating Institutions of the SDSS Collaboration including the 
Brazilian Participation Group, the Carnegie Institution for Science, 
Carnegie Mellon University, the Chilean Participation Group, the French Participation Group, Harvard-Smithsonian Center for Astrophysics, 
Instituto de Astrof\'isica de Canarias, The Johns Hopkins University, 
Kavli Institute for the Physics and Mathematics of the Universe (IPMU) / 
University of Tokyo, Lawrence Berkeley National Laboratory, 
Leibniz Institut f\"ur Astrophysik Potsdam (AIP),  
Max-Planck-Institut f\"ur Astronomie (MPIA Heidelberg), 
Max-Planck-Institut f\"ur Astrophysik (MPA Garching), 
Max-Planck-Institut f\"ur Extraterrestrische Physik (MPE), 
National Astronomical Observatories of China, New Mexico State University, 
New York University, University of Notre Dame, 
Observat\'ario Nacional / MCTI, The Ohio State University, 
Pennsylvania State University, Shanghai Astronomical Observatory, 
United Kingdom Participation Group,
Universidad Nacional Aut\'onoma de M\'exico, University of Arizona, 
University of Colorado Boulder, University of Oxford, University of Portsmouth, 
University of Utah, University of Virginia, University of Washington, University of Wisconsin, 
Vanderbilt University, and Yale University.

This study uses data provided by the Calar Alto Legacy Integral Field Area (CALIFA) survey (http://califa.caha.es/).

Based on observations collected at the Centro Astron\'omico Hispano Alem\'an (CAHA) at Calar Alto, operated jointly by the Max-Planck-Institut fur Astronomie and the Instituto de Astrof\'\i sica de Andaluc\'\i a (CSIC).

%



\bibliographystyle{ar-style2}
\bibliography{ARAA,extra,carlos}


\end{document}